\documentclass[aps,prx,twocolumn,showpacs,psfig,superscriptaddress,longbibliography]{revtex4-2}

\usepackage{listings}
\usepackage{tikz}
\usepackage{times}
\usepackage{graphicx}
\usepackage{float}
\usepackage{latexsym,amsmath,amssymb,bm,euscript}
\usepackage{color}
\usepackage{subfigure}
\usepackage{epstopdf}
\usepackage[colorlinks=true,linkcolor=blue,citecolor=blue]{hyperref}
\usepackage{soul}
\usepackage[normalem]{ulem}
\usepackage{mathrsfs}
\usepackage{amssymb}
\usepackage{algpseudocode}
\usepackage{algorithm}
\usepackage{amsmath}
\usepackage{braket}
\usepackage{booktabs}
\usepackage{chngcntr}
\usepackage{lettrine}
\usepackage{bbding}
\usepackage{xspace}
\usepackage{textcomp}
\usepackage{textcase}
\usepackage{setspace}
\usepackage[T1]{fontenc}
\usepackage{float}
\usepackage{graphicx}
\usepackage{multirow}
\usepackage{mathtools}
\usepackage{threeparttable}
\usetikzlibrary{shapes}

\def\para{\ensuremath{/\kern -0.8em /}\xspace}

\def\beqn{\begin{eqnarray}}
\def\eeqn{\end{eqnarray}}
\def\beq{\begin{equation}}
\def\eeq{\end{equation}}

\newcommand{\Beq}{\begin{eqnarray*} }
\newcommand{\Eeq}{\end{eqnarray*} }
\newcommand{\Bmat}{\left(\begin{matrix}}
\newcommand{\Emat}{\end{matrix}\right)}

\graphicspath{{../Fig/}}

\begin{document}
\title{Thermal Tensor Network Approach for Spin-Lattice Relaxation in Quantum Magnets}
	
\author{Ning Xi}
\thanks{These authors contributed equally to this work.}
\affiliation{CAS Key Laboratory of Theoretical Physics, Institute of
Theoretical Physics, Chinese Academy of Sciences, Beijing 100190, China}
\affiliation{School of Physics, Zhejiang University, Hangzhou, 310058, China}

\author{Yuan Gao}
\thanks{These authors contributed equally to this work.}
\affiliation{Peng Huanwu Collaborative Center for Research and Education,
School of Physics, Beihang University, Beijing 100191, China}
\affiliation{CAS Key Laboratory of Theoretical Physics, Institute of
Theoretical Physics, Chinese Academy of Sciences, Beijing 100190, China}

\author{Chengchen Li}
\affiliation{Department of Physics and Beijing Key Laboratory of Opto-electronic
Functional Materials and Micro-nano Devices, Renmin University of China,
Beijing 100872, China}

\author{Shuang Liang}
\affiliation{Institute of Physics, Chinese Academy of Sciences, Beijing 100190, China}
\affiliation{Yangtze River Delta Physics Research Center, Liyang, Jiangsu 213300, China}

\author{Rong Yu}
\email{rong.yu@ruc.edu.cn}
\affiliation{Department of Physics and Beijing Key Laboratory of Opto-electronic
Functional Materials and Micro-nano Devices, Renmin University of China,
Beijing 100872, China}
\affiliation{Key Laboratory of Quantum State Construction and Manipulation 
(Ministry of Education), Renmin University of China, Beijing, 100872, China}

\author{Xiaoqun Wang}
\email{xiaoqunwang@zju.edu.cn}
\affiliation{School of Physics, Zhejiang University, Hangzhou, 310058, China}

\author{Wei Li}
\email{w.li@itp.ac.cn}
\affiliation{CAS Key Laboratory of Theoretical Physics, Institute of
Theoretical Physics, Chinese Academy of Sciences, Beijing 100190, China}
\affiliation{Peng Huanwu Collaborative Center for Research and Education,
School of Physics, Beihang University, Beijing 100191, China}

\begin{abstract}
Low-dimensional quantum magnets, particularly those with strong spin frustration, 
are characterized by their notable spin fluctuations. Nuclear magnetic resonance 
(NMR) serves as a sensitive probe of low-energy fluctuations that 
offers valuable insight into rich magnetic phases and emergent phenomena 
in quantum magnets. Although experimentally accessible, the numerical 
simulation of NMR relaxation rates, specifically the spin-lattice relaxation 
rate $1/T_1$, remains a significant challenge. Analytical continuation based 
on Monte Carlo calculations are hampered by the notorious negative sign 
for frustrated systems, and the real-time simulations incur significant costs 
to capture low-energy fluctuations. Here we propose computing the relaxation 
rate using thermal tensor networks (TTNs), which provides a streamlined 
approach by calculating its imaginary-time proxy. 
We showcase the accuracy and versatility of our methodology by applying 
it to one-dimensional spin chains and two-dimensional lattices, where we 
find that the critical exponents $\eta$ and $z\nu$ can be extracted from the 
low-temperature scalings of the simulated $1/T_1$ near quantum critical points. 
Our results  also provide insights into the low-dimensional and frustrated 
magnetic materials, elucidating universal scaling behaviors in the Ising chain 
compound CoNb$_2$O$_6$ and revealing the renormalized classical behaviors 
in the triangular-lattice antiferromagnet Ba$_8$CoNb$_6$O$_{24}$. We apply 
the approach to effective model of the family of frustrated magnets AYbCh$_2$
(A = Na, K, Cs, and Ch = O, S, Se), 
and find dramatic changes from spin ordered to the proposed quantum spin 
liquid (QSL) phase. 
Overall, with high reliability and accuracy, the TTN methodology offers a 
systematic strategy for studying the intricate dynamics observed across a 
broad spectrum of quantum magnets and related fields.
\end{abstract}

\date{\today}
\maketitle

\section{Introduction}
Frustrated quantum magnets offer an ideal material platform for investigating 
rich many-body phases and phenomena. They are characterized by unconventional 
magnetically ordering such as spin supersolidity \cite{Xiang2024NBCP} and 
fractional magnetization plateau state \cite{shangguan2023}, and disordered 
spin states like the quantum spin liquid (QSL) that remain strongly fluctuating 
even down to zero temperature. The QSL does not show any symmetry 
breaking, yet has intrinsic topological order and hosts fractional excitations, 
beyond the Landau theory of phases of matter~\cite{Anderson1973,Balents2010,
Zhou2017,Broholm2020QSL}.

The studies of frustrated magnets demand an integration of advanced 
many-body approaches with experimental methodologies, in particular 
spectroscopic measurements at very low temperature. Amongst various 
dynamical measurements, nuclear magnetic resonance (NMR) constitutes 
an exceptionally sensitive probe of low-energy excitations. The NMR relaxation 
rates, including the spin-lattice relaxation rate $1/T_1$ and the spin-echo 
decay rate $T_{\rm 2G}$, offer the advantage of capturing contributions of 
very low energy excitations and from various momenta across the Brillouin 
zone (BZ). 

NMR has been widely used in the investigations of low-dimensional 
quantum magnets~\cite{Takigawa1996SpinChain,Sachdev1994,Sachdev2019,
Sandvik1995,Sandvik1995_2D,TMRGT1_2000,CoNbO2014PRX,Cui2019SCVO,
E8_2021,TMGO_NMR,Zheng2017PRL,Cui2023DQCP}. 
In particular, it serves as a very powerful and rather unique technique 
in the studies of highly frustrated QSL candidate materials.
For example, in spin-chain compounds the observed scalings of relaxation 
rates $1/T_1$ reveal the Tomonaga-Luttinger liquid (TLL) behaviors
\cite{Sachdev1994,Takigawa1996SpinChain} and Ising quantum criticality
\cite{Sachdev2019,CoNbO2014PRX,Cui2019SCVO,E8_2021}. 
For 2D quantum magnets, the temperature dependence of $1/T_1$ has also 
been employed to detect the Kosterlitz-Thouless phase in triangular-lattice 
quantum antiferromagnet TmMgGaO$_4$~\cite{TMGO_Theo,TMGO_NMR}, 
possible quantum spin liquid in the Kitaev candidate $\alpha$-RuCl$_3$
\cite{Zheng2017PRL}, and the proximate deconfined quantum critical point 
(QCP) in the Shastry-Sutherland magnet SrCu$_2$(BO$_3$)$_2$
\cite{Wang2023SSM,Cui2023DQCP}, among others. 

From the theoretical side, except for few exactly solvable systems
\cite{Sachdev1994,Sachdev2019,Yang2022,Barzykin_2001}, 
calculations of the spin-lattice relaxation rate $1/T_1$ with precision 
pose significant challenges. Specifically, by taking the strictly local 
hyperfine form factor~\cite{Sandvik1995,Sandvik1995_2D}, the 
spin-lattice relaxation rate can evaluated as
\begin{equation}
\label{Eq:T1Def}
1/T_1 \propto \mathcal{S}^{ }(\omega=0) \simeq \lim\limits_{\omega \rightarrow 0} 
T\sum_{\bf q} \sum_{\alpha=x,y,z} \chi_{\alpha \alpha}^{\prime \prime}({\bf q}, 
\omega)/\omega,
\end{equation}
where $\chi^{\prime \prime}_{\alpha \alpha}({\bf q}, \omega)$ is the 
imaginary part of the dynamical susceptibility of spin component $\alpha$ 
and at momentum $\boldsymbol{q}$, with $\mathcal{S}^{ }(\omega=0)$ the 
static spin structure factor.
To compute the spin-lattice relaxation~\cite{Realtime_2016,Realtime2_2016,
Capponi_2019,RMP_2021}, one first needs to prepare the low-temperature 
equilibrium state as an initial step for dynamic simulations. Following this, 
real-time evolution should be performed to long time scales, to provide 
sufficient energy resolution into the low-energy spectroscopy. Both steps 
present significant challenges in many-body calculations.

Alternatively, in quantum Monte Carlo (QMC) calculations there have been 
attempts to calculate $1/T_1$ from imaginary-time correlation functions 
by using numerical analytical continuation~\cite{TMRGT1_2000,
Sandvik1995,Sandvik1995_2D,MCT1_2018,Tang2020}. 
A large collection of algorithms have been developed to deal with the 
somewhat ill-posed analytical continuation problem, including maximal 
entropy~\cite{TMRGT1_2000,Huscroft_2000}, stochastic analytic continuation
\cite{ShaoH2017}, stochastic pole expansion~\cite{huang2023stochastic,
huang2023reconstructing}, and Nevanlinna analytical continuation
\cite{huang2023reconstructing,Nevanlinna_2021}, etc. However, these 
approaches necessitate the careful selection of a specific scheme
and the meticulous adjustment of hyperparameters, to capture the 
essential features inherent in the spectral function. 
One way to avoid numerical analytical continuation is resorting to 
the imaginary-time proxy (ITP). In early years ITP has been exploited 
to simulate the attractive electron Hubbard model~\cite{Scalettar1992}, 
and recently also to quasi-1D spin chain systems~\cite{MCT1_2018,FYC_2020},
within QMC calculations. However, this approach were hindered 
when applied to the frustrated magnets and fermion lattices 
away from half filling, by the notorious sign problem in QMC.

Thermal tensor networks (TTN) and their renormalization group methods
provide an efficient and accurate approach for simulating the quantum lattice
systems down to low temperature~\cite{Bursill1996DMRG,Wang1997TMRG,
Zwolak2004,Feiguin2005,White2009METTS,Li2011,Czarnik2012PEPS,
Czarnik2015PEPS,Chen2017,Dong2017,Chen2018,Kshetrimayum2019annealing,
tanTRG2023}. In particular, the matrix product operator (MPO) based 
approaches [see Fig.~\ref{Fig1}(a)], including the linearized tensor
renormalization group (TRG)~\cite{Li2011,Dong2017}, exponential TRG
\cite{Chen2017,Chen2018}, and the tangent-space TRG~\cite{tanTRG2023}
have been developed to simulate large-scale quantum spin systems
\cite{Li2019,Chen2019} and the doped Hubbard-type models
\cite{Chen2021SLU,Chen2022tbg,tanTRG2023} down to low temperatures. 
Recently, they have been applied to study the TLLs in 1D spin chains
\cite{LiuSpin1Chain,YuCPL2021}, triangular-lattice antiferromagnets
\cite{Li2020TMGO,TMGO_NMR,Liu2022,Gao2022NBCP,Xiang2024NBCP}, 
and the honeycomb Kitaev magnets~\cite{HLi2020PRR,Li2021NC,
Li2023KitaevPRB,Zhou2023NC}, among others.

In this study, we demonstrate that the cutting-edge TTN methodology serves 
as a potent tool for assessing the spin-lattice relaxation rates, specifically $1/T_1$. 
This is achieved by calculating the ITPs at zero frequency: first-order proxy 
$\mathcal{S}_{1}(\omega=0)$ or second-order $\mathcal{S}_{2}(\omega=0)$, 
as defined in the subsequent analysis. Note that the TTN is free of sign 
problem, and this approach is thus capable of studying $1/T_1$ in the 
frustrated quantum magnets. With this approach, we reveal the TLL 
scalings of the 1D XXZ Heisenberg antiferromagnetic chain and the 
quantum critical behavior of the 2D transverse-field Ising model by 
accurately extracting $\eta$ and $z\nu$ critical exponents from 
$\mathcal{S}_{1,2}(\omega= 0)$. Moreover, we extract the site- and 
momentum-resolved spin excitation gaps of spin-1 Heisenberg chain, 
providing an accurate estimate of the renowned Haldane gap. For the 
square-lattice Heisenberg model, the temperature scale of renormalized 
classical regime is obtained. 

We apply this approach also to analyze the NMR measurements on realistic 
quantum magnets, including the Ising chain compound CoNbO$_{6}$ and 
triangular lattice Heisenberg antiferromagnet Ba$_8$CoNb$_6$O$_{24}$. 
For the Ising chain case, inspired by the model calculation results we perfectly 
collapse the experimental data and accurately determine the critical fields and 
exponents $\eta$, $z\nu$ in CoNbO$_{6}$. With triangular-lattice Heisenberg 
model calculations, we obtain similar behaviors in $1/T_1$ as observed in the 
compound Ba$_8$CoNb$_6$O$_{24}$, where two temperature scales can be 
identified. The triangular-lattice QSL candidate AYbCh$_2$ family has raised
great interest recently, whose coupling can be tuned by substituting the alkli 
elements A=Na, K, Cs, and Ch=O, S, Se~\cite{Liu2018ARX,Ranjith2019NYS,
Dai2021NYS,Scheie2024KYS}. Our simulated $1/T_1$ results of the effective 
$J_1$-$J_2$ triangular-lattice model pave the way for the NMR probe of possible 
QSL in this highly frustrated material family.

The rest of the paper is organized as follows. In Sec.~\ref{Sec:MM} 
we introduce the TTN approach and low-dimensional lattice models 
used in the present work. Section~\ref{Sec:Rslt} presents our main 
results of 1D and 2D lattice models, with a focus on the universal 
scalings in $1/T_1$ of spin-1/2 Heisenberg chain and estimate of 
spin gap in spin-1 Haldane chain. The square-lattice transverse-field 
Ising and Heisenberg models are also studied.
In Sec.~\ref{Sec:NMR} we compare our model calculations with 
NMR measurements on realistic compounds, including the spin-chain 
material CoNb$_2$O$_6$ and the triangular-lattice antiferromagnet 
Ba$_8$CoSb$_6$O$_{24}$. Section~\ref{Sec:Discus} is devoted to 
the summary and outlook.

\section{Models and Methods}
\label{Sec:MM}

\subsection{Quantum spin models}
In this work, we consider two types of spin models defined on several 1D 
and 2D lattices, which include the Heisenberg antiferromagnet (HAF) and 
transverse-field Ising (TFI) models. The Hamiltonian of spin-$S$ HAF model 
with XXZ anisotropy reads
\begin{equation}
H_{\rm XXZ} = \sum_{\langle i,j \rangle}  J_{xy} (S_i^x S_{j}^x + S_i^y S_{j}^y)
+ J_z S_i^z S_{j}^z,
\label{Eq:XXZ}
\end{equation}
where $J_{xy}$ ($J_z$) represents the coupling between nearest neighboring 
(NN) sites within $xy$-plane (along $z$-axis). In the present study we take $J_{xy} 
\equiv 1$ as the energy scale, and the isotropic HAF model corresponds to 
$J_z=J_{xy}=1$. We consider the $S=1/2$ and $S=1$ HAF chains in 1D, 
as well as the $S=1/2$ HAF model on the square and triangular lattices in 2D.

The Hamiltonian of TFI models on both 1D chain and 2D square lattice can be 
written as
\begin{equation}
H_{\rm Ising} = -J \sum_{\langle i,j \rangle} S_i^z S_j^z - h \sum_i S_i^x,
\label{Eq:QIsing}
\end{equation}
where $J\equiv1$ is the energy scale of the ferromagnetic TFI model, 
and $h$ is the transverse field tuning quantum fluctuations. The critical 
fields in the two models are $h_c=0.5$ (for the 1D chain) and $h_c \simeq 
1.522$~\cite{Deng2002} (for the square lattice), which separate the spin 
ordered and paramagnetic disordered phases.

\subsection{Thermal tensor network renormalization}
In the calculations below, we employ the linearized TRG (LTRG)~\cite{Li2011} 
for 1D and tangent-space TRG (tanTRG)~\cite{tanTRG2023} for 2D lattices 
to obtain the equilibrium density operator $\rho(\beta)$ and estimate the
imaginary-time proxies $\mathcal{S}_{1,2}$. In both approaches, 
we perform imaginary-time evolutions of the density matrix $\rho(\beta)$ following 
the equation $d\rho/d\beta = -H\rho$, where $\beta$ is the inverse 
temperature, and the initial state $\rho(\beta=0)\equiv I$. 
For 1D HAF and TFI spin chains, the finite-$T$ density matrices $\rho{(\beta)}$ 
has an efficient matrix product operator (MPO) representation as shown in 
Fig.~\ref{Fig1}(a), which can be obtained by LTRG~\cite{Li2011,Dong2017} 
(see also Appendix~\ref{App:LTRG}).
In practice, we exploit 6th-order Trotter-Suzuki decomposition ~\cite{trotter2010}
so that the discretization error is sufficiently small in our calculations. In the LTRG
process, bond dimension of the MPO $\rho(\beta)$ generically increases as the
inverse temperature $\beta $ increases, and we retain bond dimension 
up to $\chi=600$ that produces fully converged data in 1D systems.

For the 2D HAF and TFI models, we exploit the tanTRG approach firstly 
invented to deal with 2D many-electron problems~\cite{tanTRG2023,Qu2023dwave} 
and here also applicable to quantum spin models~\cite{Wang2023SSM}.
To tackle the finite-size 2D lattices, we map them into quasi-1D system 
with long-range interactions, and represent the density matrix $\rho{(\beta)}$ 
also as an MPO [c.f., Fig.~\ref{Fig1}(c)]. 
Using techniques from the time-dependent variational principle~\cite{TDVP2011,
GeometryTDVP2020}, we integrate the evolution equation restricted within the 
MPO manifold of a given bond dimension. In practice, cylindrical geometries 
are considered with widths up to $W=12$ for the square-lattice TFI, $W=8$ for 
the square-lattice HAF, and $W=6$ for the triangular-lattice HAF models. The 
retained bond dimension is up to $\chi^*=4000$ multiplets (equivalently $\chi
\simeq 16,000$ individual states) for the 2D models.

\subsection{Imaginary-time proxies $\mathcal{S}_{1,2}(0)$ for $1/T_1$}
With the density matrix $\rho(\beta)$ obtained, we further compute the 
imaginary-time spin correlators
\begin{equation}\label{SSco}
\langle S_i^\alpha(\tau)S_j^\alpha \rangle =
\frac{1}{\mathcal{Z}(\beta)}{\rm Tr}[e^{(-\beta+\tau)\hat{H}}
S_{i}^{\alpha} e^{(-\tau)\hat{H}}S_j^{\alpha}],
\end{equation}
where $\mathcal{Z}(\beta)={\rm Tr} [\rho(\beta)]$ is the partition function. 
The spin correlator is related to the imaginary dynamical susceptibility 
$\chi_{\alpha \alpha}^{\prime \prime}(\boldsymbol{q}, \omega)$ via 
\begin{equation}\label{s12a}
\left\langle S_{-\boldsymbol{q}}^\alpha(\tau) S_{\boldsymbol{q}}^\alpha(0)\right\rangle
=\frac{1}{2 \pi} \int d \omega \frac{e^{-\omega \tau}}{1-e^{-\omega \beta}}
\chi_{\alpha \alpha}^{\prime \prime}(\boldsymbol{q}, \omega).
\end{equation}
One can calculate $1/T_1^{\alpha \alpha}$ by first extracting 
$\chi_{\alpha \alpha}^{\prime \prime}(\boldsymbol{q}, \omega)$ 
from Eq.~\eqref{s12a} by using numerical analytical continuation, 
then plug it into Eq.~\eqref{Eq:T1Def}.
There are a variety of specific techniques invented to effectively extract 
the dynamical information, each with its own advantages and limitations
\cite{TMRGT1_2000,Sandvik1995,Sandvik1995_2D,MCT1_2018,Tang2020,
Huscroft_2000,ShaoH2017,huang2023stochastic,huang2023reconstructing,
Nevanlinna_2021}. 

\begin{figure}[]
\includegraphics[width=0.8\linewidth]{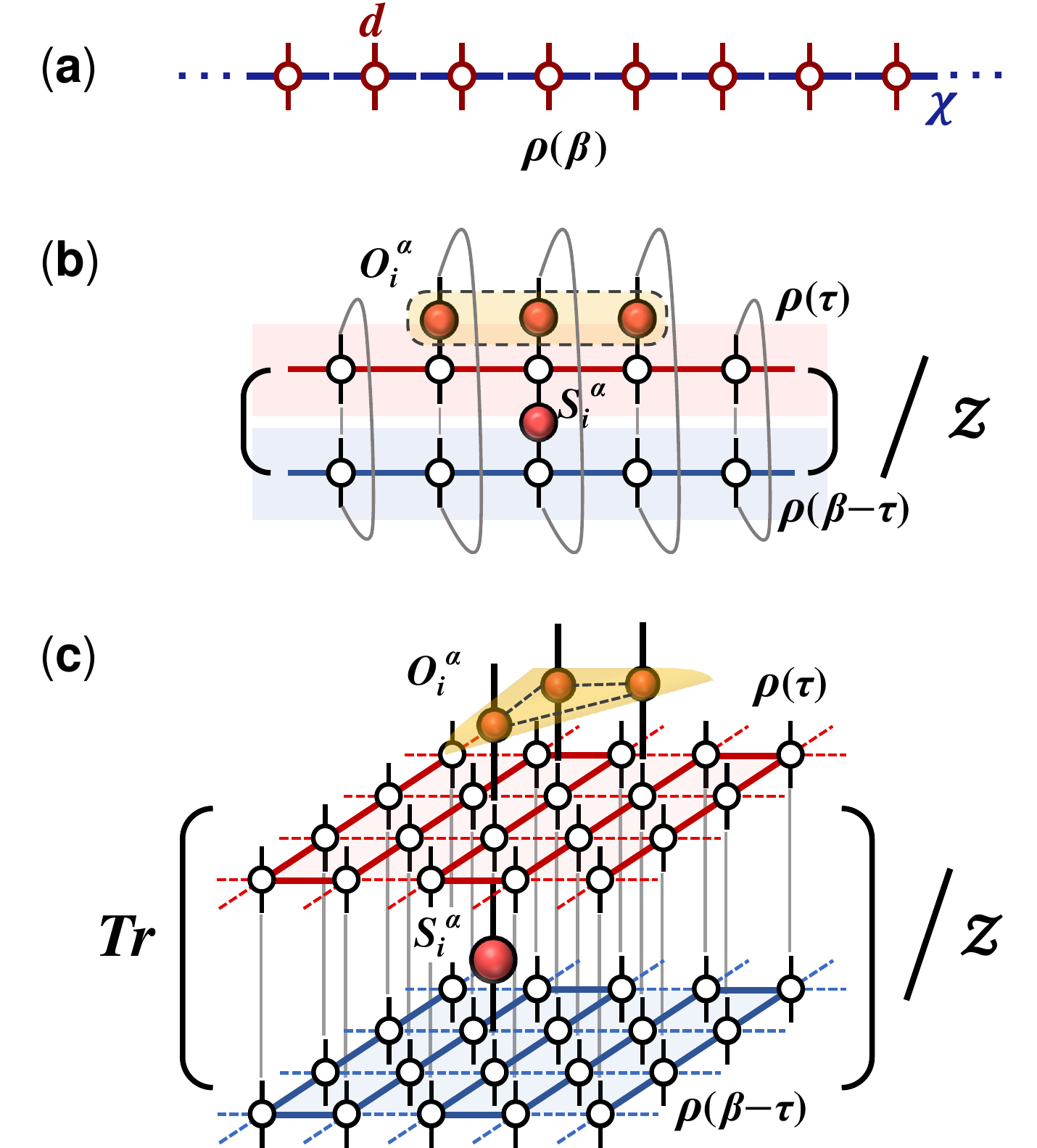}
\caption{(a) The MPO representation of density matrix $\rho(\beta)$
of 1D quantum spin chains. The geometric bond dimension is 
denoted as $\chi$ and the local Hilbert space dimension as $d$.
The tensor network evaluation of the imaginary-time proxies
$\mathcal{S}_{1,2}(\omega=0)$ are illustrated for (b) 1D chain 
and (c) 2D lattice mapped into quasi-1D system. $\mathcal{Z} 
\equiv \rm{Tr}[\rho(\beta)]$ is the partition function, and the ITPs 
$\mathcal{S}_1(0)$ and $\mathcal{S}_2(0)$ can be computed 
by tracing the bilayer tensor network with local operators 
$S_i^\alpha$ and $O_i^\alpha$ (see definition in the main text).}
\label{Fig1}
\end{figure}
 
Alternatively, here we resort to an efficient and convenient proxy 
that extracts the low-energy dynamics from imaginary-time correlator 
in the large-$\beta$ limit. Taking $\tau = \beta/2$ in Eq.~\eqref{s12a}, 
we arrive at
\begin{equation}\label{beta22}
\left\langle S_{-\boldsymbol{q}}^\alpha(\beta / 2) S_{\boldsymbol{q}}^\alpha(0)\right\rangle
=\frac{1}{2 \pi} \int d \omega \frac{\chi_{\alpha \alpha}^{\prime \prime}(\boldsymbol{q}, \omega)}{2 \sinh (\beta\omega/2)}.
\end{equation}
Assuming $\chi_{\alpha \alpha}^{\prime \prime}(\boldsymbol{q}, \omega)$ 
is analytic at $\omega=0$, we conduct Taylor expansion
\begin{equation}
\label{Taylor}
\begin{split}
\chi^{\prime \prime}_{\alpha \alpha}(\boldsymbol{q}, \omega) &= \omega\left.\sum_{n=0}^{\infty} \frac{\omega^{2 n}}{(2 n+1) !}
\frac{\mathrm{d}^n \chi^{\prime \prime}_{\alpha \alpha}(\boldsymbol{q}, \omega)}
{\mathrm{d} \omega^n}\right|_{\omega=0} \\
&= \omega \sum_{n} f^{\alpha \alpha}_{n}(\boldsymbol{q},\omega),
\end{split}
\end{equation}
where only odd terms in $\omega$ are allowed. 
Substitute the expansion in Eq.~\eqref{Taylor} into Eq.~\eqref{beta22}, we retain
only the leading order in $\omega$
\begin{equation} \label{S1}
\begin{split}
\left\langle S_{-\boldsymbol{q}}^\alpha(\beta / 2) S_{\boldsymbol{q}}^\alpha(0)\right\rangle
& \approx \frac{1}{4 \pi}\left[(\frac{2}{\beta})^{2}f^{\alpha \alpha}_0(\boldsymbol{q},0) \int_{0}^{\infty} d \tilde{\omega}
\frac{\tilde{\omega}}{\sinh(\tilde{\omega})}\right ].\\
&\overset{\beta \to \infty}{\approx} \frac{\pi}{2\beta^2} f^{\alpha \alpha}_0(\boldsymbol{q}, 0),
\end{split}
\end{equation}
where the dimensionless parameter $\tilde{\omega}=\beta\omega / 2$. 

On the other hand, by substituting the expansion of $\chi''(\bf{q}, \omega)$ 
into Eq.~\eqref{Eq:T1Def}, we retain the lead order term and obtain
\begin{equation}
\label{Eq:T1Expand}
\mathcal{S}^{ }(\omega=0) \simeq  \frac{1}{\beta}
\sum_{\boldsymbol{q}} \sum_\alpha f^{\alpha \alpha}_0(\boldsymbol{q},0).
\end{equation}
By comparing Eq.~\eqref{S1} with Eq.~\eqref{Eq:T1Expand}, we arrive at
\begin{equation}\label{Eq:PS1} 
\mathcal{S}^{ }_1(\omega = 0) \simeq  \beta \sum_{\bf q} \sum_\alpha \langle
S_{-\boldsymbol{q}}^\alpha(\beta/2)S_{\boldsymbol{q}}^\alpha \rangle 
\end{equation}
where $\mathcal{S}_1(\omega=0)$ is introduced as a proxy for $1/T_1$ 
(up to a constant factor) in the large $\beta$ limit. Equation~\eqref{Eq:PS1} 
shows that the dynamical spin structure, and also the spin-lattice relaxation
rate, can be obtained conveniently via the imaginary-time correlation.

Below we compute $\mathcal{S}_1(0)$ for various systems at low temperature 
$T\lesssim J$. For gapped systems, it is obvious that $\mathcal{S}^{\alpha \alpha}_1(\boldsymbol{q},\omega=0) \equiv \beta \langle S_{-\boldsymbol{q}}^\alpha(\beta/2)S_{\boldsymbol{q}}^\alpha \rangle$ decay exponentially following 
{$\exp[-\Delta_S({\bf q})/T]$, with $T = 2/\beta$} and $\Delta_S({\bf q})$ the 
spin excitation gap at momentum ${\bf q}$. One remark is that Eq.~\eqref{Taylor} 
relies on the assumption that $\chi^{\prime \prime} ({\bf q}, \omega)$ is analytical 
at $\omega=0$. In certain gapped systems, $\chi^{\prime \prime}({\bf q}, \omega)$  
may have a pole and hence become singular at $\omega=0$. For those systems, 
we can single out the pole, calculate its contribution to $1/T_1$ separately, 
and treat the regular part as described above.  

Along a similar line, we can define the second-order proxy, $\mathcal{S}_2(0)$ 
to $1/T_1$, by taking derivatives of the correlator $\langle 
S_{-\boldsymbol{q}}^\alpha(\tau) \, S_{\boldsymbol{q}}^\alpha \rangle$ 
with respect to the imaginary time $\tau$ in Eq.~(\ref{s12a}), namely,
\begin{equation}\label{s22a}
\partial_{\tau}^2 \left\langle S_{-\boldsymbol{q}}^\alpha(\tau)
S_{\boldsymbol{q}}^\alpha(0)\right\rangle=\frac{1}{2 \pi}
\int d \omega \frac{\omega^{2}e^{-\omega \tau}}{1-e^{-\omega \beta}}
\chi_{\alpha \alpha}^{\prime \prime}(\boldsymbol{q}, \omega).
\end{equation}
Once again we take $\tau=\beta/2$, insert the Taylor expansion of 
$\chi_{\alpha \alpha}^{\prime \prime}(\boldsymbol{q}, \omega)$ in 
Eq.~\eqref{Taylor}, and retain the leading order in $\omega$, i.e.,
\begin{equation}\label{S2}
\partial_{\tau}^2\langle S_{-\boldsymbol{q}}^\alpha(\tau)
S_{\boldsymbol{q}}^\alpha \rangle|_{\tau=\beta/2}\overset{\beta \to \infty}{\approx}
\frac{\pi^3}{\beta^4} f_0(\boldsymbol{q}, 0).
\end{equation}
With Eq.~\eqref{Eq:T1Expand} we arrive at the expression of the 
second-order proxy,
\begin{equation}
\mathcal{S}^{ }_2(\omega=0) 
\equiv \beta^3\sum_{\boldsymbol{q}} \sum_{\alpha} \partial_{\tau}^2
\langle S_{-\boldsymbol{q}}^\alpha (\tau) S_{\boldsymbol{q}}^\alpha \rangle|_{\tau=\beta/2} 
\sim 1/T_1^{}.
\end{equation}
In practice,  the second-order derivative can be evaluated following
\begin{equation}\label{partS2}
\begin{aligned}
&\partial_{\tau}^2\langle S_j^\alpha(\tau)S_j^\alpha \rangle|_{\tau=\tau_{0}} \\
& =\frac{1}{\mathcal{Z}}{\rm Tr} \big ( e^{(-\beta+\tau_0){H}}
[{H},[{H},S_{j}]]e^{(-\tau_0){H}}S_j \big ) \\
& =\langle O_j^\alpha(\tau_0)S_j^\alpha \rangle,
\end{aligned}
\end{equation}
where $O^\alpha_{j}=[{H},[{H},S_{j}^\alpha]]$ is the commutator, and thus 
we have $\mathcal{S}^{\alpha \alpha}_2(\omega=0)=\beta^3\sum_{j} \langle 
O_j^\alpha(\beta/2)S_j^\alpha \rangle$. 

As shown in Figs.~\ref{Fig1}(b,c), $\mathcal{S}_1(0)$ and $\mathcal{S}_2(0)$ 
can be evaluated as imaginary-time correlations. They are obtained by contracting 
the two MPOs, each representing the density matrix $\rho(\beta/2)$, with the local 
operators $S_i^\alpha$ and $O_i^\alpha$ defined above. The ITPs $\mathcal{S}^{}_{1(2)}(0)$ 
offer an expedient and fast way to estimate the finite-temperature and low-frequency 
dynamics, achieving a computational efficiency akin to traditional thermodynamic 
calculations performed within TTN framework.

\begin{figure}[]
\includegraphics[width=0.76\linewidth]{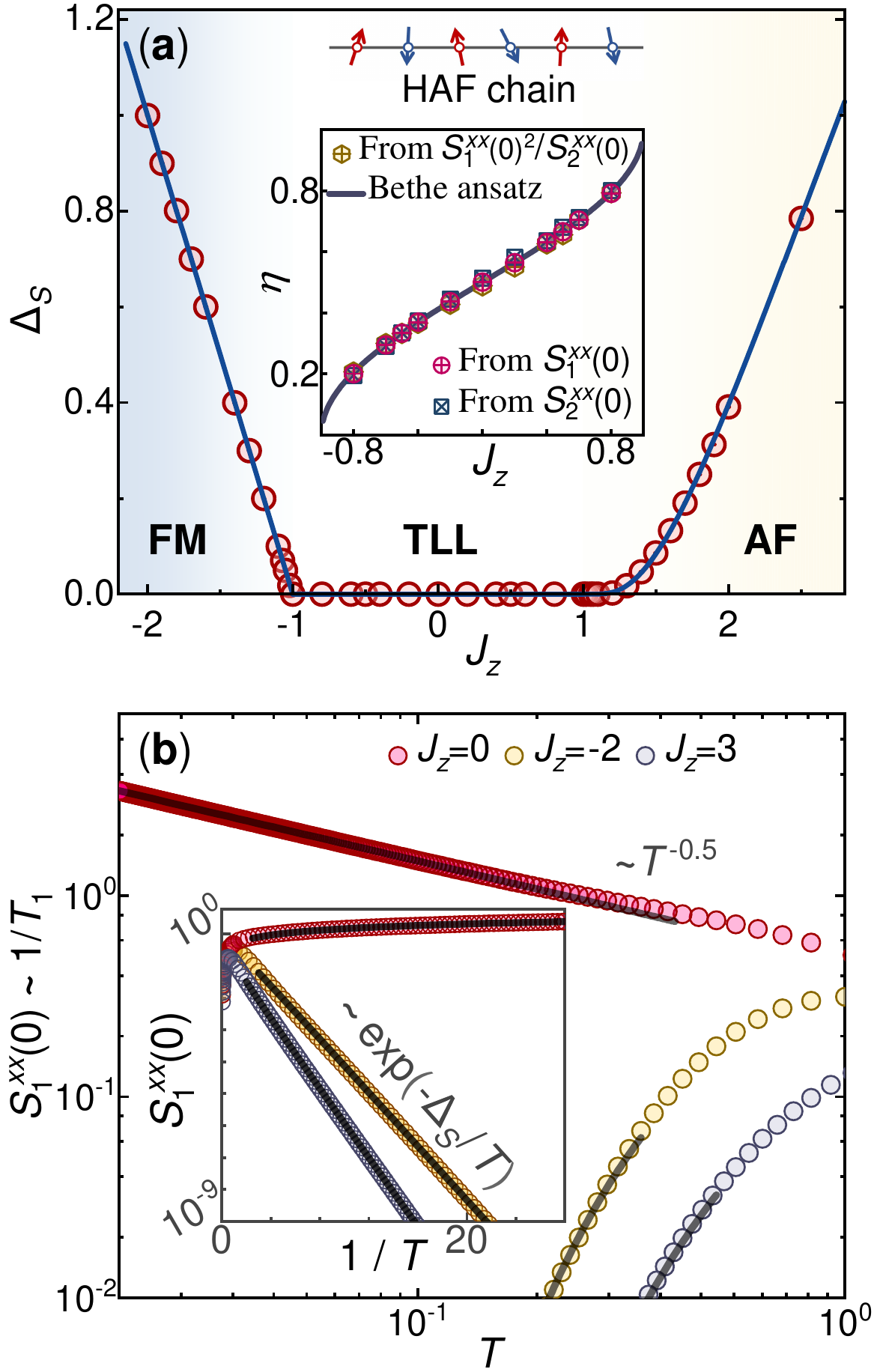}
\caption{
(a) shows the results of spin-1/2 HAF chain with varying $-2 \leq J_z \leq 3$
and fixed $J_{xy}\equiv 1$. In both the gapped FM and AF phases, the 
spin gaps $\Delta_{\rm S}$ are calculated from the exponential decay of 
$\mathcal{S}_{1}^{xx}(0)$ at low temperatures. The inset demonstrates 
the exponent $\eta$ fitted in the TLL phase using three distinct proxies
(see Appendix~\ref{App:B}), 
all of which align perfectly with the analytical results (purple line) from 
Bethe ansatz. This validates the precision and reliability of our method.
(b) The calculated $\mathcal{S}^{xx}_{1}(0)\sim 1/T_{1}$ data and 
their low-$T$ scalings in the spin-$1/2$ HAF chain with $J_z=0, -2, 3$. 
In the FM ($J_z=-2$) and AF ($J_z=3$) phases, $\mathcal{S}^{xx}_1(0)$
decays exponentially as further elaborated in the inset; for the XY case 
($J_z=0$), $\mathcal{S}_{1}^{xx}(0)$ manifests an algebraic scaling 
$T^{\eta-1}$ with the anomalous dimension $\eta=1/2$.
}
\label{Fig2}
\end{figure}

\section{Results}
\label{Sec:Rslt}

\subsection{Spin-lattice relaxation in the TLL phase of spin chains}
We consider the prototypical spin-1/2 HAF chain with an XXZ anisotropy
[c.f., Eq.~(\ref{Eq:XXZ})], and estimate the spin-lattice relaxation 
rate $1/T_1$ by computing the proxies $\mathcal{S}_{1(2)}(0)$. 
As the anisotropy $J_z$ varies from -2 to 3, the system changes 
from the easy-axis ferromagnetic (FM), through the gapless TLL 
phase, to the Ising AF phases.

In Fig.~\ref{Fig2}(a), we show the results of spin excitation gaps $\Delta_{\rm S}$
extracted from $\mathcal{S}_{1}^{xx}(0)$, where the gap scales linearly 
versus $|J_z|$, namely, $\Delta_{\rm S}=-J_z-1$ in the FM phase; 
while it follows an exponential behavior, i.e., $\Delta_{\rm S} \sim 
\exp{(-\xi/\sqrt{J_z-1})}$ in the AF phase~\cite{fradkin_2013} ($\xi$ is a 
constant, see Appendix~\ref{App:B}), indicating the occurrence of a 
Berezinskii-Kosterlitz-Thouless (BKT) transition at $J_z=1$. 
In Fig.~\ref{Fig2}(b), we show the details of temperature scalings 
in $\mathcal{S}_{1}^{xx}(0)$, which decay exponentially as 
{$\exp{(-\Delta_{\rm S}/T)}$} in both the easy-axis FM and AF 
phases with $|J_z| > 1$. 

In the intermediate TLL phase with $|J_z| < 1$, we find algebraic behaviors 
$\mathcal{S}_{1}^{xx}(0) \sim T^{(\eta+d-2)/z}$~\cite{Cui2023DQCP,Hong2021}, 
where $d$ is the spatial dimension, $z$ is the dynamical exponent, and $\eta$ 
represents the anomalous dimension that relates to the Luttinger parameter via 
$K=1/4\eta$. In particular, in Fig.~\ref{Fig2}(b) we find $\mathcal{S}_{1}^{xx}(0) 
\sim T^{-1/2}$ in the $J_z=0$ case, consistent with the expectation of $z=1$ and 
$\eta=1/2$ for the XY chain. Tuning the coupling away from $J_z=0$, we estimate 
the exponents $\eta$ from the scalings of $\mathcal{S}^{xx}_{1,2}(0)$, and the ratio 
$\mathcal{S}^{xx}_2(0)^2/S^{xx}_1(0)$ from each parameter $J_z$ (see more details 
in Appendix~\ref{App:B}). The results are collected in the inset of Fig.~\ref{Fig2}(a),
which agree excellently with the analytical results from Bethe ansatz~\cite{Luther1975}.

\begin{figure}[]
\includegraphics[width=0.86\linewidth]{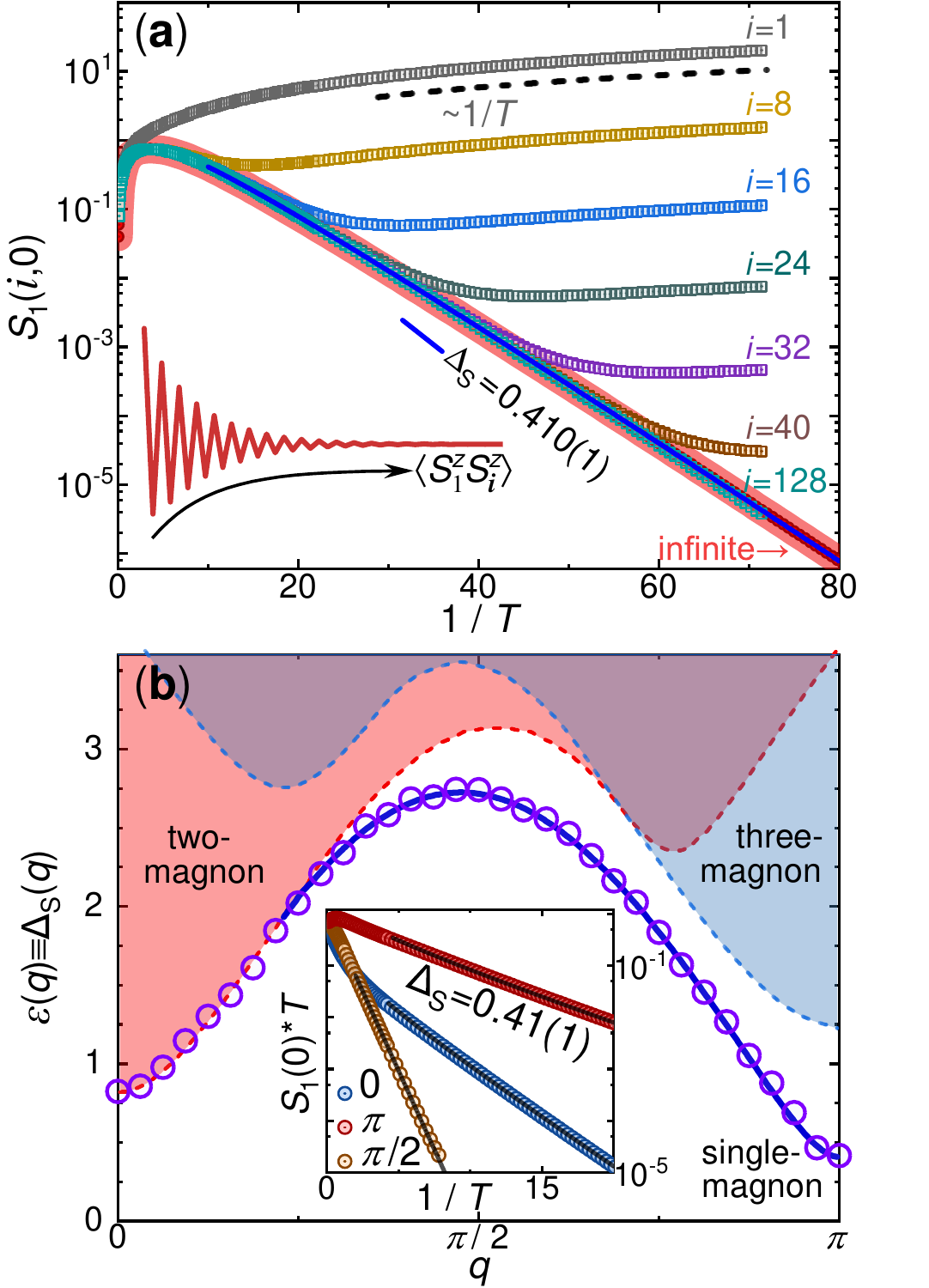}
\caption{
(a) The site-resolved results $\mathcal{S}_{1}(i,0)$ 
measured at various sites $i$
of the $L=256$ chain, plotted with the results obtained 
directly in the thermodynamic limit (red squares labeled
as ``infinite''). The $1/T$ scaling for $i$ near the ends of
the chain reflects the zero edge modes. In the bulk 
of the chain it decays exponentially with (inverse) temperature,
from which the Haldane gap $\Delta_S \simeq 0.410(1)$ 
can be accurately fitted.
(b) The momentum-resolved spin gap $\Delta_{\rm S}(q)$ (open purple 
circle) for the spin-1 HAF chain estimated from the exponential decay of
$T \mathcal{S}_1(q, \omega=0) \sim \exp{[-\Delta_S(q)/T]}$ (see the inset). 
The Haldane gap can also be estimated at momentum $q=\pi$, 
where $\Delta_S(\pi) \simeq 0.41(1)$. The
single-magnon dispersion (blue solid line), two-magnon continua (red 
shaded range), and three-magnon continua (blue shaded range) are 
adapted from Ref.~\cite{White1993Spin1}.
}
\label{Fig3}
\end{figure}

\subsection{Topological edge mode and the spin gap in the spin-1 Haldane chain}
Initially proposed theoretically~\cite{Haldane1983,AKLT1987} and subsequently 
verified via high-precision DMRG calculations~\cite{White1993Spin1,DMRGLec}, 
the spin-1 HAF chain described by Eq.~(\ref{Eq:XXZ}) with $J_{xy}=J_z=1$ exhibits 
a gapped Haldane phase. This phase has been proposed to host the symmetry-protected 
topological order~\cite{Gu2009tensor}, and supports free spin-1/2 edge states that 
reside at both open boundaries of the spin chain. The topological edge mode 
exponential decays as moving into the bulk of the system~\cite{White1993Spin1}. 
In Fig.~\ref{Fig3} we show the simulated $\mathcal{S}_1(0) \sim 1/T_1$ of the 
$S=1$ HAF chain. The spatial and momentum distribution of $\mathcal{S}_1(0)$ 
can be obtained, from which the Haldane gap and topological edge modes are extracted.

\begin{figure*}[]
\includegraphics[width=0.95\linewidth]{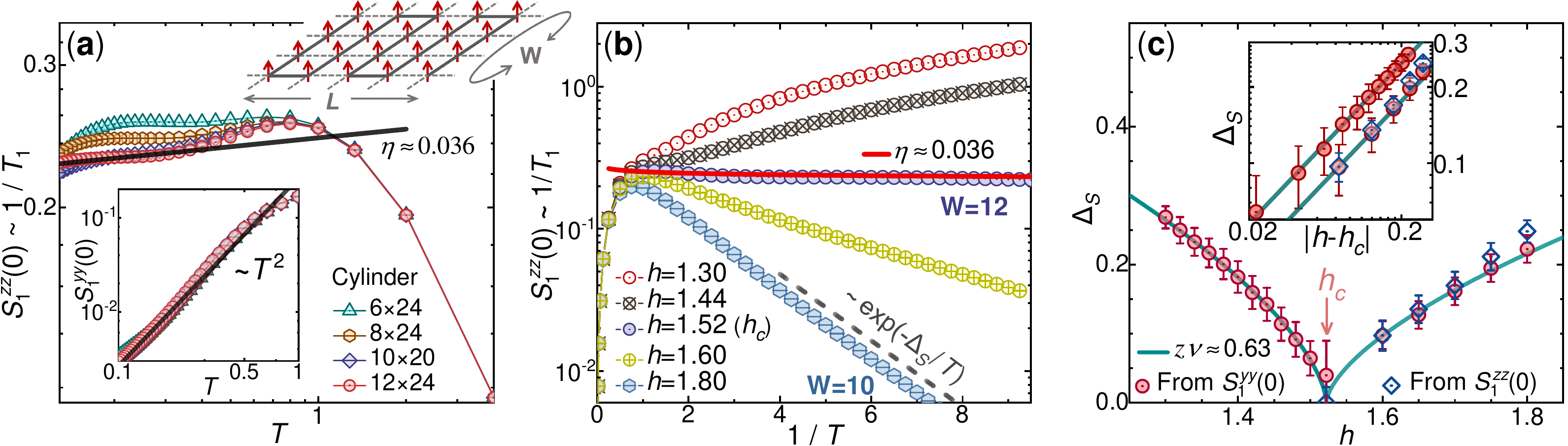}
\caption{
(a) Simulations of spin-lattice relaxation rate in the square-lattice TFI model
defined on cylinders (see upper right inset) with different system widths up to
$W=12$. The low-temperature scaling of $1/T_{1}\sim \mathcal{S}_{1}^{zz}(0)$ 
at the QCP follow the scaling $T^\eta$ (with $\eta \simeq 0.036$) as indicated by
the solid line. $\mathcal{S}_{1}^{yy}(0)$ is also computed and found to scale as 
$T^2$ (bottom left inset).
(b) The behaviors of $\mathcal{S}_1(0)$ for two gapped phases, ordered for 
$h< h_c \simeq 1.52$ and quantum disordered for $h > h_c$, are in sharp contrast 
to that at the QCP $h=h_c$. We perform the calculations on $10\times20$ cylinder 
for gapped phase and show the results at QCP on the largest system size $12\times24$.
(c) Red circles represent estimated spin gaps $\Delta_S$ from $\mathcal{S}_1^{yy}(0)$ 
(for both $h<h_c$ and $h>h_c$) and the blue diamonds are from $\mathcal{S}_1^{zz}(0)$ 
for $h>h_c$. We find $\Delta_{\rm S} \propto\left| h-h_{c} \right|^{z\nu}$ near the QCP, 
with exponent $z\nu\simeq0.63$ as indicated by the blue lines. The inset presents a
log-log plot of the data, which more clearly illustrates this universal algebraic scaling.
}
\label{Fig4}
\end{figure*}

To reveal the edge modes, we introduce the site-resolved relaxation rate
$\mathcal{S}_1^{}(i, \omega=0) = \beta \langle S_i (\beta/2) S_i(0) \rangle$,
where $i$ label the site in the spin chain of length $L$. In Fig.~\ref{Fig4}(a) 
we show results at different site $i$ of the spin-1 HAF chain (with open boundary 
conditions), and find the spins become ``asymptotically free'' as approaching 
both ends of the chain. This is revealed by the scaling $\mathcal{S}_{1}(i\ll L/2, 
\omega=0)$ at low temperature, which diverges as $1/T$ and reflects the 
existence of zero mode at the boundary. On the other hand, as $i$ moves 
into the bulk, the calculated $\mathcal{S}_{1}(i \approx L/2, \omega=0)$ join 
the exponential decaying behavior as in an infinite chain. The fitted spin gap
from the exponential scaling is $\Delta_{\rm S} \simeq 0.410(1)$ [c.f., Fig.~\ref{Fig3}(a)], 
which is in excellent agreement with the DMRG results of the Haldane gap
\cite{White1993Spin1}.

The momentum dependence of the spin excitation gap $\Delta_{\rm S}({q})$ can 
also be extracted by fitting the exponential scaling $\mathcal{S}_1({q}, \omega=0)
\sim 1/T \exp{[-\Delta_{\rm S}({q})/T]}$, which are shown in Fig.~\ref{Fig4}(b). 
For $q \gtrsim \pi/4$, $\Delta_S({q})$ reproduces the single-magnon dispersion
\cite{White1993Spin1}, while for $q\lesssim \pi/4$ $\Delta_S(q)$ follows the 
lower boundary of the the two-magnon continuum.
The two-magnon bound state has lower energy than that of the single-magnon 
excitation, and $\mathcal{S}_1(0)$ always captures the lowest excitation gap 
of the system.

\subsection{Universal scalings near the (2+1)D Ising QCP}
Below we show that our approach can be used to study the universal scalings 
near the QCP in 2D lattice models. In Fig.~\ref{Fig4} we exploit state-of-the-art 
$\tan$TRG approach~\cite{tanTRG2023} to simulate the square-lattice TFI 
model [c.f., Eq.~\eqref{Eq:QIsing}]
on cylinder geometries with widths up to $W=12$ [c.f., inset in Fig.~\ref{Fig4}(a)], 
and study the scaling behaviors of $1/T_1$ near the Ising QCP. The transverse 
field introduces quantum fluctuations that melt the magnetic order at 
$h_c\simeq 1.52$~\cite{Deng2002}.

In Fig.~\ref{Fig4}(a), we find that as system size increases $\mathcal{S}^{zz}_1(0)$
gradually falls into the universal scaling $\mathcal{S}^{zz}_1(0) \sim T^{\eta}$ 
with $\eta \simeq 0.036$, consistent with the (2+1)D Ising universality class. 
We also find that $1/T_1$ exhibits strong anisotropic behaviors, and show 
$\mathcal{S}_{1}^{yy}(0)$ in the lower left inset of Fig.~\ref{Fig4}(a), which
reveals a different algebraic scaling $\mathcal{S}_{1}^{yy}(0) \sim T^2$. 
In Fig.~\ref{Fig4}(b), the results of $\mathcal{S}^{zz}_1(0)$ are shown in 
both gapped phases, where we find it decays exponentially for the ordered 
phase with $h<h_c$ while diverges in the paramagnetic phase for $h>h_c$,
in sharp contrast to the scaling behaviors right at QCP.

\begin{figure}
\includegraphics[width=0.8\linewidth]{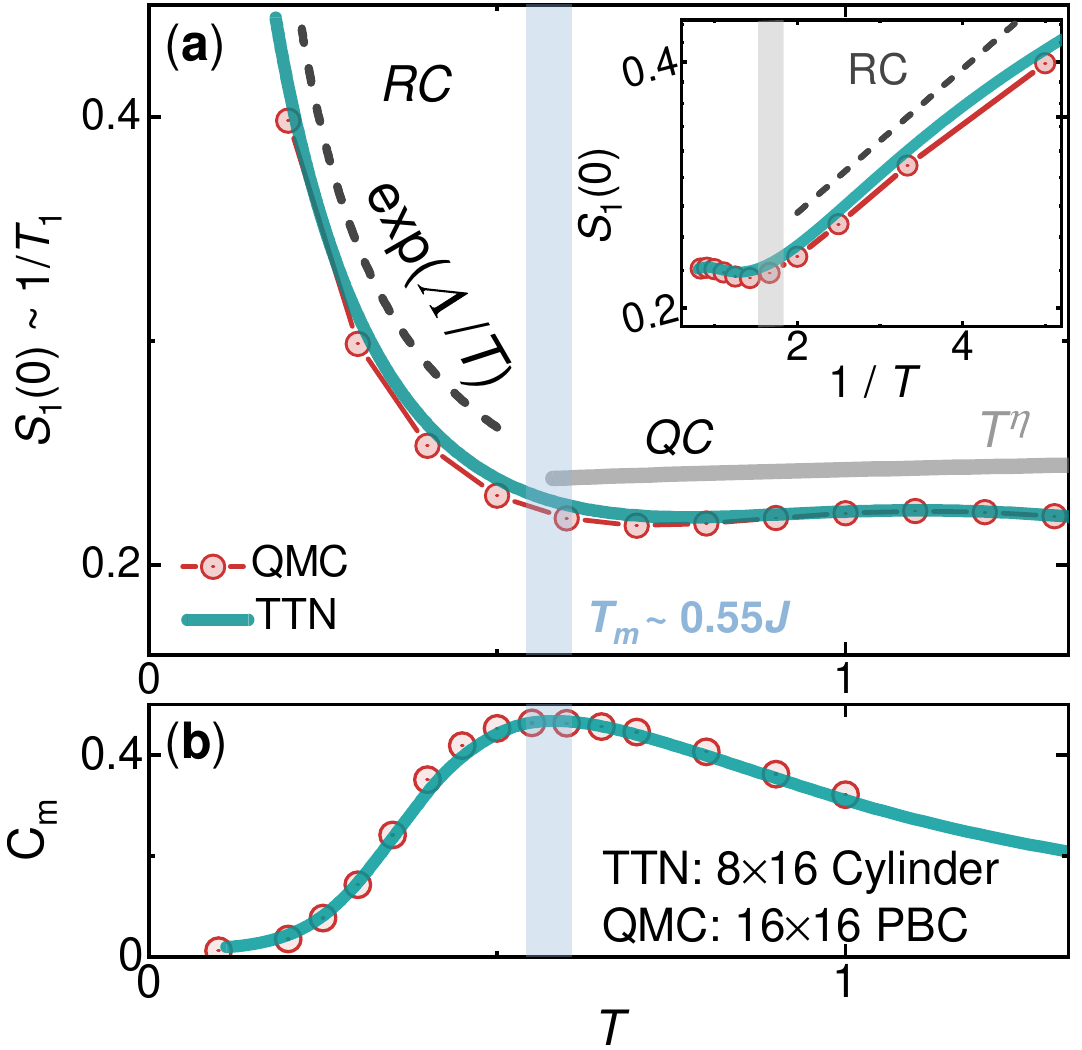}
\caption{
(a) The calculated $\mathcal{S}_{1}(0)$ for the SLHAF model. 
Results are shown for a width-8 cylinder (length $L=$16) using 
TTN, represented by the cyan line, and a 16$\times$16 square 
lattice (with both PBC) using QMC, denoted by the red solid point. 
(b) shows the specific heat data 
calculated by TTN (cyan line) and QMC (red markers). exponential 
diverging behavior $1/T_1 \sim \exp{(\Lambda/T)}$ in the RC regime.
The dashed and solid gray lines are guides to eyes, illustrating 
the exponential divergence in the RC region characterized by 
$\Lambda\simeq 0.2$ and the quantum-critical (QC) behavior associated 
with the 3D Heisenberg exponent of $\eta\simeq$0.035, respectively. 
The shaded area denotes the crossover temperature scales 
$T_m\simeq 0.55 J$.
}
\label{Fig5}
\end{figure}

Based on the $\mathcal{S}_1(0)$ data obtained in the two gapped phases, 
we further extract the spin gap $\Delta_{\rm S}$ from the exponential decaying 
behaviors [c.f., Appendix~\ref{App:B}]. The so-determined spin gaps are 
collected and shown in Fig.~\ref{Fig4}(c), and we find the results fall into an 
algebraic scaling $\Delta_{\rm S} \propto \left| h-h_{c} \right|^{z\nu}$ near the 
Ising QCP, with the fitted exponent $z\nu \simeq 0.63$ again consistent with 
the (2+1)D Ising universality class.

\begin{figure*}
\includegraphics[width=0.95\linewidth]{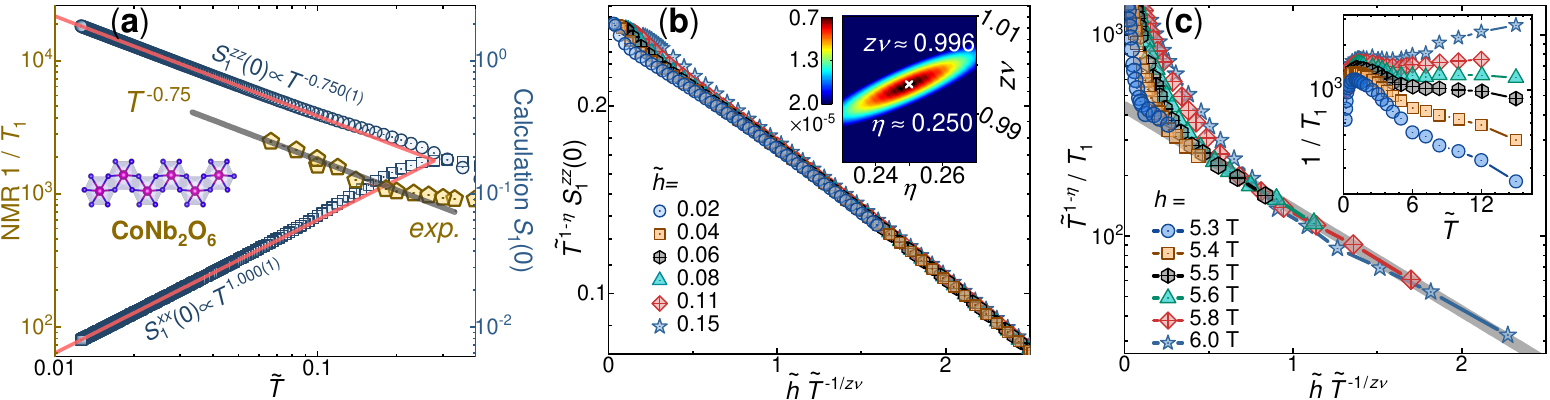}
\caption{
(a) shows the calculated $\mathcal{S}_{1}^{xx,zz}$ of the TFI chain near 
QCP ($\tilde{h}= 0$) and makes a comparison to the experimental $1/T_1$ 
data. The dimensionless parameters are defined as $\tilde{h}=(h-h_c)/h_c$ 
and $\tilde{T}=T/J$, with $J\simeq30$~K and $h_{c}\simeq5.2$~T for 
CoNb$_2$O$_6$. The navy markers ($\odot$ for $S_{1}^{xx}$ and $\boxdot$ 
for $S_{1}^{zz}$) are from the model calculations, and the brown pentagons 
are NMR measurements on CoNb$_2$O$_6$~\cite{CoNbO2014PRX}.
$\mathcal{S}_{1}^{zz}(0) \propto T^{-0.75}$ dominates the scaling behaviors
at low temperature and is in excellent agreement with NMR experiments.
(b) reveals the universal scaling by collapsing the calculated 
$\mathcal{S}^{zz}_1(0)$ data near the QCP, from which we find the critical 
exponents are in excellent agreement with (1+1)D Ising universality class.
The inset shows the contour map of the root mean square error (RMSE) 
in the $z\nu$-$\eta$ parameter space, where the minimum is located at 
$z\nu\simeq 0.996$ and $\eta\simeq 0.250$. Following this line, we collapse
the experimental $1/T_1$ data with the critical exponents of (1+1)D Ising 
universality class ($z\nu=1, \eta=1/4$) in panel (c). The original experimental 
data adapted from Ref.~\cite{CoNbO2014PRX} are shown in the inset.
}
\label{Fig6}
\end{figure*}

\subsection{Renormalized classical behaviors on the square lattice}

According to the Mermin-Wagner theorem, the square-lattice HAF (SLHAF) system 
remains disordered at all finite temperatures. However, a crossover occurs at certain 
characteristic temperature scale $T_m$, below which the system exhibits the 
renormalized classical (RC) behaviors. Within the RC regime, the correlation length 
and spin-lattice relaxation $1/T_1$ exhibit exponential divergence as temperature 
lowers~\cite{Sandvik1995_2D,Chubukov1994SL}. 

In Fig.~\ref{Fig5}, we simulate such RC behavior through $\mathcal{S}_1(0)$ 
with TTN and QMC calculations, where the former is obtained on the 8$\times$16 
cylinder and the latter on a larger, 16$\times$16 lattice with periodic boundary 
conditions (PBCs). The excellent agreement reveals that $\mathcal{S}_{1}(0)$ 
results are stable on different lattice geometries, and both results increase rapidly
below about $T_m \sim 0.55 J$. As shown in the inset of Fig.~\ref{Fig5}(a), the 
simulated $\mathcal{S}_{1}(0)$ follows an exponential diverging scaling 
$\exp{(\Lambda/T)}$ in the RC regime, below a possible QC regime at intermediate temperature. 

In Fig.~\ref{Fig5}(b), we present the results of the specific heat $C_m$
obtained from TTN and QMC calculations. 
Despite the differences in their respective geometries ($8\times16$ cylinder
vs. $16\times16$ torus), these results again demonstrate very good agreement. 
In the specific heat $C_m$ there exhibits a rounded peak at approximately 
$T_m \simeq 0.55 J$, which, together with the rapid increasing behaviors 
of $\mathcal{S}_1(0)$ at lower temperature, indicates the crossover into 
RC regime.

\section{Comparisons to Experimental measurements on Quantum Magnets}
\label{Sec:NMR}

In this section, we switch from theoretical model calculations to the domain 
of realistic quantum magnets. In past years, NMR measurements have 
been widely applied in the studies of spin dynamics in quantum magnets. 
However, theoretical analyses and comparisons to measured NMR data, 
particularly for frustrated magnets, remain relatively scarce. Below, 
we focus on two compounds, namely, the quasi-1D spin chain compound 
CoNb$_2$O$_6$, the triangular lattice antiferromagnet 
Ba$_8$CoNb$_6$O$_{24}$, and $J_1$-$J_2$ QSL candidates
AReX$_2$~\cite{Ranjith2019NYS,Scheie2024KYS}. With comprehensive 
comparisons to NMR experiments, we analyze these experimental results 
in the light of theoretical model calculations. 

\subsection{Universal scaling in the Ising chain CoNb$_2$O$_6$}
As a prototypical quantum spin system, the TFI chain compound 
CoNb$_2$O$_6$ has been shown to host intriguing critical phenomena 
driven by transverse fields~\cite{Coldea2010,E8_2021,Cui2019SCVO,
WangZ2018,Sachdev2019,Fava2020,Morris2021}. 
Apart from the possible complex spin-spin couplings in the realistic 
compound~\cite{Morris2021,Fava2020,Moore2011}, it is believed 
that the critical behaviors near the QCP are nevertheless well 
described by the (1+1)D Ising universality class, and thus can 
be studied through simulating the TFI chain model. Various experiments, 
including the low-temperature thermodynamics~\cite{Liang2015,Xu2022}, 
inelastic neutron scattering~\cite{Coldea2010,woodland2023tuning}, 
terahertz spectroscopy~\cite{WangZ2018,Morris2021}, and NMR
\cite{CoNbO2014PRX}, etc., have been conducted to study the phenomena 
near QCP in spin-chain compound CoNb$_2$O$_6$. In Fig.~\ref{Fig6},
we apply our TTN approach to study the TFI chain and compare the 
results to NMR measurements.

In Fig.~\ref{Fig6}(a), we show the low-temperature scaling behaviors of two
spin-resolved proxies $\mathcal{S}_{1}^{xx}(0)$ and $\mathcal{S}_{1}^{zz}(0)$
at $h_{x}^c=0.5$. The former follows an algebraic decay with 
$\mathcal{S}_1^{xx}(0) \sim T$, while the latter diverges as 
$\mathcal{S}_1^{zz}(0)\sim T^{\eta-1}$ with $\eta\simeq1/4$ and thus 
dominates the low-temperature behaviors. 
Therefore, our results well explain the observed $1/T_1\sim T^{-3/4}$ 
scaling obsderved in NMR experiments [also shown in Fig.~\ref{Fig6}(a)].
Furthermore, we also calculate the $\mathcal{S}_1^{zz}(0)$ in the vicinity 
of QCP (in the $\tilde{h}>0$ side) and find the data fall into a universal function 
with properly chosen parameters and exponents. In Fig.~\ref{Fig6}(b), 
we plot $\tilde{T} \mathcal{S}_1^{zz}(0) = f(\tilde{h} \tilde{T}^{-1/z\nu})$, 
where $\tilde{h}=(h-h_c)/h_c$ and $\tilde{T}=T/J$, with the spin coupling 
$J\simeq30$~K and the critical field $h_{c}\simeq5.2$~T. Through optimal 
data-collapsing process, represented by the minimal RMSE, we determine 
critical exponents $z\nu\simeq 0.996$ and $\eta\simeq0.25$ as depicted 
in the inset of the contour map, in excellent agreement with  the $(1+1)$D 
Ising universality class. In light of this theoretical calculation, we effectively 
``condense'' the experimental data with the corresponding universal scaling 
function, as presented in Fig.~\ref{Fig6}(c).

\begin{figure}
\includegraphics[width=0.8\linewidth]{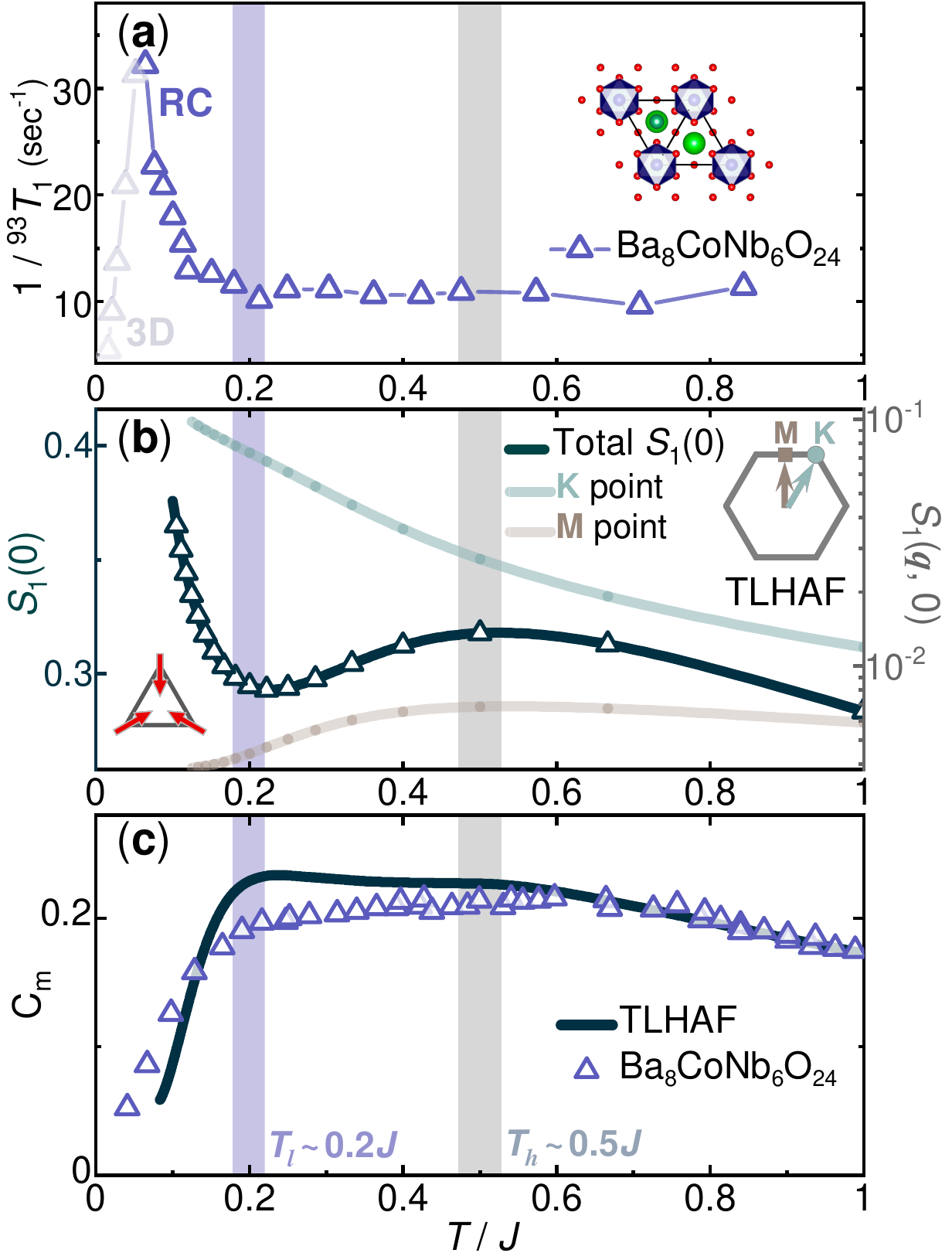}
\caption{
(a) shows the experimental measurements of the spin-lattice relaxation 
$1/T_1$ on a triangular-lattice compound Ba$_8$CoNb$_6$O$_{24}$
\cite{BCNO2018}, with the layered structure shown in the inset. (b) shows 
the calculated $\mathcal{S}_1(0)$ of the TLHAF model on a YC $6\times12$ 
cylinder, including the $\mathcal{S}_1(0)$ averaged on all ${\bf q}$ points 
and that located at ${\bf q}=(\frac{2\pi}{3}, \frac{2\sqrt{3}\pi}{3})$({\bf K} point) 
and $(0,\frac{2\sqrt{3}\pi}{3})$ ({\bf M} point).
(c) The lines represent calculated specific heat data 
and the triangle markers are the experimental data. The two shadow areas 
denote two temperature scales $T_h\simeq 0.5 J$ and $T_l\simeq 0.2 J$.
}
\label{Fig7}
\end{figure}

\subsection{NMR probe of the incipient order in Ba$_8$CoNb$_6$O$_{24}$}
Regarding the 2D antiferromagnets, it has been proposed theoretically that 
both square and triangular lattice Heisenberg models exhibit the RC behaviors 
at low temperature~\cite{Chubukov1994SL,Chubukov1994TL}, with an exponential 
divergence in the NMR relaxation rate $1/T_1$. Previous numerical studies 
have confirmed the RC behavior on the square lattice~\cite{Sandvik1995_2D}, 
as also shown in Fig.~\ref{Fig5} of the present study. However, whether the 
frustrated triangular-lattice system exhibits RC behavior and corresponding 
incipient magnetic order at low temperature remained unresolved.
 
To be specific, the triangular lattice 
HAF (TLHAF) magnet constitutes a prototypical frustrated system
\cite{Chubukov1991,Starykh2015} that has been originally proposed
to host a QSL~\cite{Anderson1973}. Subsequent investigations utilizing 
precise numerical methods have uncovered that the ground state of the 
system features a coplanar 120$^\circ$ magnetic order~\cite{Bernu1992,
Capriotti1999,White2007}, which precludes the existence of a QSL state 
at absolute zero temperature. Nevertheless, from a finite-temperature 
perspective the system remains disordered and has a large magnetic 
entropy till low temperature~\cite{Elstner1993,Kulagin2013}. This 
thermodynamics anomaly has been noticed in early days~\cite{Elstner1993} 
and confirmed more recently by diagrammatic Monte Carlo calculations
\cite{Kulagin2013}. The situation is dramatically different for the SLHAF 
system studied above (c.f., Fig.~\ref{Fig5}), the predicted RC behavior in 
TLHAF~\cite{Chubukov1994TL} has not been found in early numeric 
studies~\cite{Elstner1993,Kulagin2013}.

Our TTN approach can also be applied to study the frustrated TLHAF 
quantum magnets, with the results shown in Fig.~\ref{Fig7}. Compared 
to those of the SLHAF in Fig.~\ref{Fig5}, where the $\mathcal{S}_1(0)$ 
data increase exponentially in the RC regime, in Fig.~\ref{Fig7}(b) 
the $\mathcal{S}_1(0)$ results of TLHAF remain flat, even slightly 
suppressed, for $T \lesssim T_h \simeq 0.5 J$. The $\mathcal{S}_1(0)$ 
data are found to be significantly enhanced only below $T_l\simeq 0.2 J$. 
With momentum resolution, in Fig.~\ref{Fig5}(b) we further show that 
$\mathcal{S}_1({\bf q=K}, 0)$ increases monotonically as temperature 
lowers and becomes vert prominent for $T \lesssim T_l$.
This signals the onset of long-sought RC behaviors associated
with three-sublattice order in the TLHAF system~\cite{Chen2019}, 
and we find the spin-lattice relaxation $1/T_1$ provides a highly sensitive 
probe for the incipient order in both frustrated and unfrustrated HAF 
systems. Besides, the $\mathcal{S}_1(0)$ results exhibit a maximum 
at the higher temperature scale $T_{h}$, which may be ascribed to 
the enhancement of $\mathcal{S}_1({\bf q=M}, 0)$ due to activation 
of roton-like excitations~\cite{Chen2019} in the TLHAF system. 

This theoretical insight significantly enhances our understanding of 
the recently studied TLHAF materials~\cite{Doi2004,Zhou2012,
Susuki2013,Ma2016,Ito2017,BCNO2017,BCNO2018}. In particular, 
the cobaltate compound 
Ba$_8$CoNb$_6$O$_{24}$, as reported in Refs.~\cite{BCNO2017,
BCNO2018}, is considered an ideal realization of the TLHAF model 
due to its weak interlayer coupling and isotropic Heisenberg spin 
exchange. The NMR measurements have been conducted down to 
low temperature, and in Fig.~\ref{Fig7}(a) we show the experimental 
results adapted from Ref.~\cite{BCNO2018}, where the $1/T_1$ data 
of Ba$_8$CoNb$_6$O$_{24}$ level off even below $T_h\simeq 0.55 J$ 
and fall into the RC scaling below $T_l \simeq 0.2 J$,
in striking similarity with model calculations shown in Fig.~\ref{Fig7}(b).
In addition, the existence of two temperature scales, $T_h$ and $T_l$,
in TLHAF is also witnessed by the magnetic specific heat $C_m$ shown 
in Fig.~\ref{Fig7}(c). Remarkably, the experimental $C_m$ curve exhibits 
a hump at $T_h$ and a shoulder at about $T_l $, in consistent with our 
TTN calculations on cylinders even with limited width~\cite{Chen2018}.

\begin{figure}
\includegraphics[width=\linewidth]{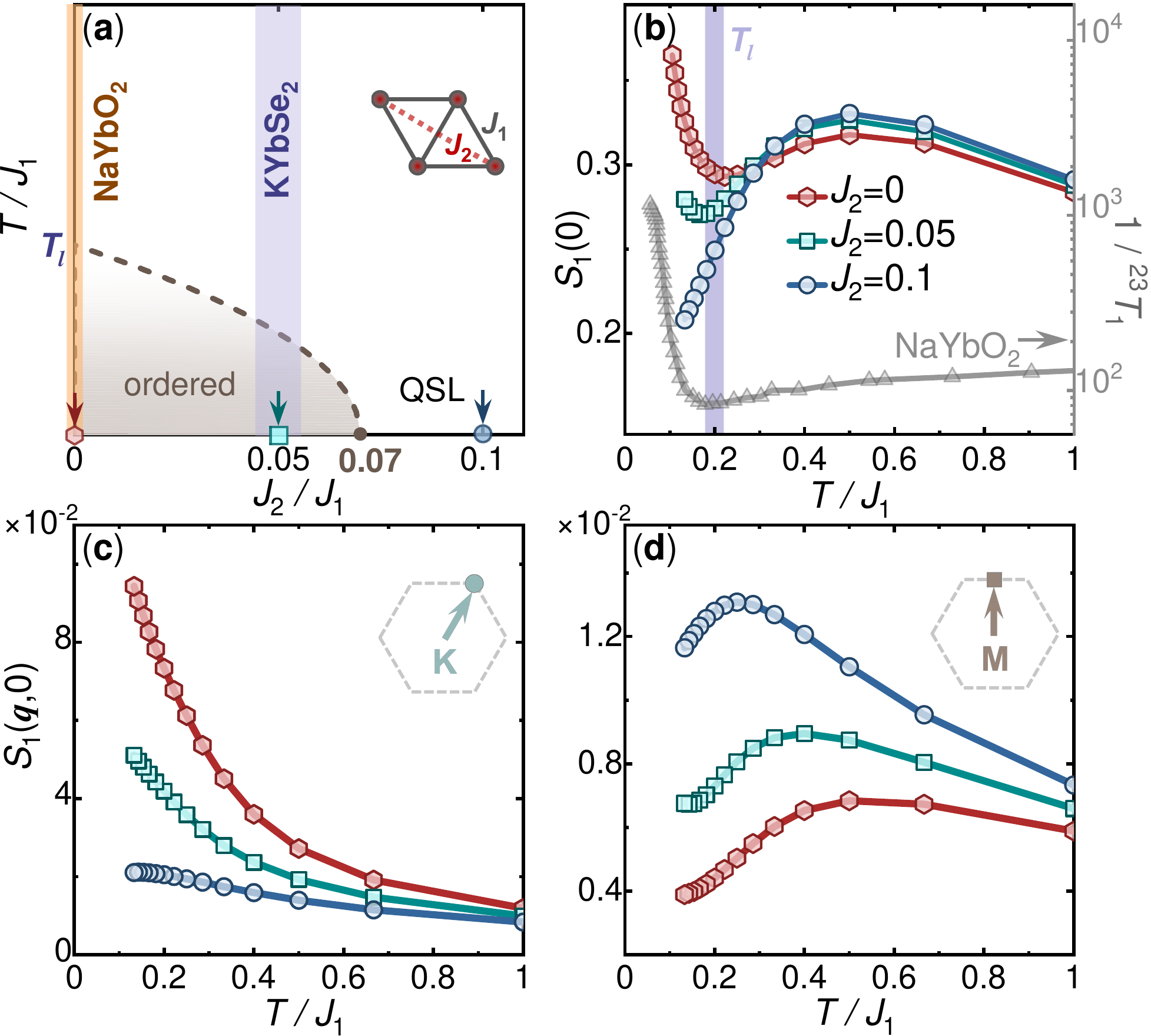}
\caption{
(a) Sketch of $J_1$-$J_2$ TLHAF phase diagram, where the 
two considered compounds NaYbO$_2$ and KYbSe$_2$, as
well as the three simulated parameters $J_2/J_1=0, 0.05, 0.1$, 
are also indicated. A quantum phase transition from the ordered 
phase to QSL occurs at $J_2/J_1\simeq0.07$.  
The inset shows the nearest-neighbor interaction $J_1$ and 
the next-nearest-neighbor interaction $J_2$ in a triangular lattice. 
The range of coupling ratio $J_2/J_1$  for KYbSb$_2$ is
adapted from the estimation in Ref.~\cite{Scheie2024KYS}, 
and $J_{\rm NYO} \simeq 5.5$~K of NaYbO$_2$ from Ref.~\cite{Wilson2019NYO}.
(b) shows the calculated $\mathcal{S}_1(0)$ of the $J_1$-$J_2$ TLHAF 
model on a YC $6\times12$ cylinder, and compares the results to the 
experimental $1/T_1$ data measured on NaYbO$_2$~\cite{Ranjith2019NYO}.
(c) and (d) present the calculated $\mathcal{S}_{1}({\bf q}, 0)$  at the {\bf K}
and {\bf M} points in the BZ (see insets), respectively.
}
\label{Fig8}
\end{figure}

\subsection{Spin-lattice relaxation of the QSL candidates AYbCh$_2$}
The triangular-lattice compound AYbCh$_2$ (with A=Na, K, Cs, and 
Ch = O, S, Se) has garnered significant attention in recent studies as 
a promising candidate for hosting QSL or proximate QSL on the triangular lattice
\cite{Liu2018ARX,Wilson2019NYO,Ranjith2019NYS,Dai2021NYS,Scheie2024KYS}.
These compounds are believed to be described by an effective $J_1$-$J_2$
model on the triangular lattice, where $J_1$ represents nearest-neighboring 
spin coupling, and $J_2$ is the next-nearest coupling [see Fig.~\ref{Fig8}(a)].
There are proposals using periodic table to tune the ratio $J_2/J_1$ in the 
compounds. For example, as shown in Fig.~\ref{Fig8}(a), NaYbO$_2$ and
KYbSe$_2$ can have different coupling ratios that locate them in different 
regimes (ideally $J_1$ vs. proximate $J_1$-$J_2$ QSL) of the phase diagram.

To examine the NMR signatures of possible QSL, in Fig.~\ref{Fig8}(b) 
we compute $\mathcal{S}_1(0)$ for different coupling ratios $J_2/J_1$, 
and find it shows distinctive behaviors in the ordered phase with 
$J_2/J_1\lesssim 0.07$ and QSL regime with $0.07 \lesssim 
J_2/J_1 \lesssim 0.15$~\cite{Zhu2015,Hu2015}. In Fig.~\ref{Fig8}(b),
we show the NMR results of the compound NaYbO$_2$, which 
corresponds to $J_2/J_1 \simeq 0$~\cite{Wilson2019NYO}, 
and find the relaxation rate $1/T_1$ increases rapidly at low 
temperature~\cite{Ranjith2019NYO}.
By comparing experimental results of NaYbO$_2$ and model calculations 
in Fig.~\ref{Fig8}(b), we find the measured spin-lattice relaxation $1/T_1$ 
suggest the presence of incipient order rather than QSL behaviors in 
NaYbO$_2$, akin to Ba$_8$CoNb$_6$O$_{24}$ analyzed in Fig.~\ref{Fig7}.
The characteristic temperature where $1/T1$ starts to increase is about 
$T/J_{\rm NYO} \approx 0.2$ for NaYbO$_2$, very close to that of pure TLHAF 
model and indeed suggesting a negligible $J_2$ in the compound. 

We further increase $J_2$ coupling and find the low-$T$ values of 
$\mathcal{S}_1(0)$ gets suppressed and the incipient order characteristic 
temperature also decreased. For $J_2/J_1=0.05$ that corresponds to 
KYbSe$_2$~\cite{Scheie2024KYS}, we find the low-temperature increase
in NMR relaxation rate is only modest, suggesting the proximity to QSL.
For $J_2=0.1$ case well located in the QSL phase, we observe decreasing 
$\mathcal{S}_1(0)$ as temperature lowers, following approximately an 
algebraic scaling that reflects a gapless nature of the QSL~\cite{Iqbal2016,
Hu2019}. This echoes the previous $1/T_1$ measurements on possible Dirac QSL materials~\cite{Itou2010}.
 To clarify the contributions from different $\mathbf{q}$ points in 
the BZ, we show in Fig.~\ref{Fig8}(c,d) the simulated $\mathcal{S}_1({\bf q}, 0)$ 
with ${\bf q = K, M}$. In Fig.~\ref{Fig8}(c), we observe that $\mathcal{S}_1({\bf K}, 0)$ 
gets suppressed at low temperature and the divergence behaviors due to 
incipient order no longer exist. On the other hand, the relaxation rate 
$\mathcal{S}_{1}({\bf M}, 0)$ gets enhanced in Fig.~\ref{Fig8}(d), reflecting 
softening of the spin excitations at ${\bf M}$ point. This is consistent 
with previous dynamical study of $J_1$-$J_2$ TLHAF~\cite{Sherman2023}, where both K and M points are shown to be gapless. 
Therefore, our findings 
not only present a numerical approach for assessing $\mathcal{S}_{1,2}(0)$ 
and recognizing dynamical trait of geometric frustration, but also underscore 
the potential of NMR relaxation as a highly sensitive diagnostic tool for probing 
the QSL state in the AYbCh$_2$ family.

\section{Discusions and outlook}
\label{Sec:Discus}
Essential for identifying exotic quantum spin states and emergent 
phenomena, low-energy excitations and spin fluctuations can be 
sensitively probed by spectroscopic techniques like NMR. However, 
numerical simulations of the temperature-dependent dynamical fluctuations, 
such as those characterized by the low-frequency dynamical spin structure 
$\mathcal{S}(0) \sim 1/T_1$ vs. $T$, remained challenging to compute.

In this work, we demonstrate that the TTN --- a powerful approach for calculating 
equilibrium properties at finite temperature~\cite{Bursill1996DMRG,Wang1997TMRG,
Feiguin2005,Li2011,Czarnik2012PEPS,White2009METTS,Chen2018,
Kshetrimayum2019annealing,tanTRG2023} --- can also simulate the NMR 
relaxation rate $1/T_1$, with spatial, momentum, and spin resolutions. The 
approach can be used to compute proxies $\mathcal{S}_{1,2}(\omega=0)$ 
with both high accuracy and controllability. In addition to spin-lattice relaxation 
$1/T_1$, our TTN approach can also compute the spin-spin relaxation rate 
$1/T_{2 \rm G}$, leveraging the calculations of imaginary-time correlators.
Our method enables extract extractions of the critical exponents ($\eta$ and 
$z\nu$), recognition of topological edge modes, incipient orders, and the 
possible QSL phase, among others, without the need for the somewhat 
ill-posed analytical continuation or other complex dynamical approaches.

It thus offers us a very useful tool for analyzing the NMR experiments on 
realistic compounds, particularly the QSL candidates with high degree of 
spin frustration. For the spin-chain compound CoNb$_2$O$_6$, the 
NMR results show an universal scaling $\mathcal{S}_1^{zz}(\omega=0) \sim 
1/T_1 \propto T^{-3/4}$ and determined critical exponent $z\nu \simeq 1$
through the data-collapse analysis inspired by the numerical results. For the 
TLHAF compound Ba$_8$CoNb$_6$O$_{24}$, we find the low-temperature 
scale for incipient spin order is evident from the $1/T_1$ measurements. 
The AYbSe$_2$ family on the $J_1$-$J_2$ triangular lattice has been 
intensively studied in experiments, and our calculations provide valuable 
insights into their spin dynamics and useful guidance for exploring possible 
QSL in these compounds.

\setcounter{equation}{0}
\setcounter{figure}{0}
\setcounter{table}{0}

\makeatletter

\renewcommand{\theequation}{A\arabic{equation}}
\renewcommand{\thefigure}{A\arabic{figure}}
\renewcommand{\theHfigure}{A\arabic{figure}}
\renewcommand{\bibnumfmt}[1]{[#1]}
\renewcommand{\citenumfont}[1]{#1}

\begin{figure*}[]
\includegraphics[width=1\linewidth]{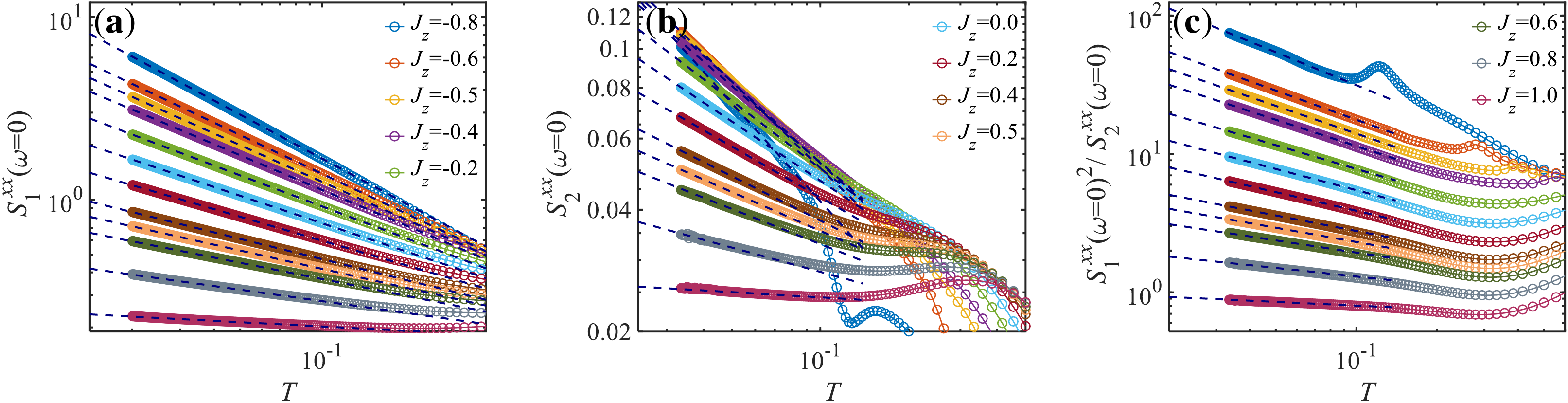}
\caption{\textbf{Supporting data of $\mathcal{S}_1(0)$ in the TLL phase 
of the spin-1/2 HAF chain.} We provide three different proxies to compute 
$1/T_1$ in the TLL phase, including (a) $\mathcal{S}_{1}^{xx}(0)$, 
(b) $\mathcal{S}_{2}^{xx}(0)$, and (c) $[\mathcal{S}_{1}^{xx}(0)]^2/\mathcal{S}_{2}^{xx}(0)$. 
The dashed lines are power-law fittings $1/T_1^{xx} \sim T^{\eta-1}$ 
at low temperature, from which the corresponding exponents $\eta$ 
obtained in three proxy simulations are found to be consistent. The fitted 
exponents $\eta$ are shown in the inset of Fig.~\ref{Fig2}(a) in the main text.
}
\label{FigS1}
\end{figure*}

In summary, our results highlight the importance of developing 
theoretical tools for low-energy spectroscopic properties for investigating 
quantum magnets. Above we use the MPO-based TTN approach to 
simulate infinite-size 1D chain and also large-scale 2D cylinder geometries, 
and showcase the power of our approach with representative model 
examples and realistic materials. Moreover, such a TTN approach can 
be generalized to infinite 2D systems using the projected entangled pair 
operator~\cite{Corboz2019}. Beyond the quantum spin systems, the TTN 
proxy can also be employed to study the many-electron problems, 
which are important for understanding the NMR measurements on 
correlated electron materials, which we leave for feature studies.

\section{Acknowledgments} The authors are indebted to Lei Wang, Yang Qi, 
Jiahao Yang, Yuchen Fan, and Weiqiang Yu for helpful discussions. This 
work was supported by the National Natural Science Foundation of China 
(Grant Nos.~12222412, 12347138, 12047503, 12334008 and 12174441), 
the National Key Projects for Research and Development of China 
(Grant Nos. 2023YFA1406500 and 2021YFA1400400), China Postdoctoral 
Science Foundation (Grant No. 2021TQ0355), and CAS Project for Young 
Scientists in Basic Research (Grant No.~YSBR-057). We thank HPC at 
ITP-CAS and RUC for the technical support and generous allocation of CPU time. 

\begin{appendix}

\section{Thermal Tensor Network Approach}

In this appendix, we review the thermal tensor network method that was 
employed to derive the equilibrium density operator and the imaginary time 
proxies associated with spin relaxation, specifically $1/T_1$ or spectral 
weight $\mathcal{S}_1(\omega=0)$, for correlated quantum spin systems.

\subsection{Linearized tensor renormalization group}
\label{App:LTRG}
We employ linearized tensor renormalization group (LTRG) method to 
simulate infinite-size and finite-size 1D quantum spin chains~\cite{Li2011}. 
The thermodynamics calculations resort to an efficient contraction of the 
(1+1)D TTN through the Trotter-Suzuki decomposition. Through (first-order) 
Trotter-Suzuki decomposition, the density matrix $\rho = [\exp{(-\tau H)}]^N$
can be expressed as
\begin{equation}\label{rho}
\rho \approx \sum_{\{\sigma_i^n\}} \prod_{i=1}^{L} \prod_{n=1}^{N} \nu_{\sigma_i^n,
\sigma_{i+1}^n, \sigma_i^{n+1}, \sigma_{i+1}^{n+1}},
\end{equation}
by inserting $N$ sets of orthonormal basis $\{ \sigma_i^n \}$ where $i$ is spatial 
index and $n$ the imaginary-time index. We arrive at a TTN consists of rank-four 
tensors,
\begin{equation}
\nu_{\sigma_i^n, \sigma_{i+1}^n, \sigma_i^{n+1}, \sigma_{i+1}^{n+1}} = \langle \sigma_i^{n},
\sigma_{i+1}^{n} | e^{-\tau h_i} | \sigma_{i}^{n+1}, \sigma_{i+1}^{n+1}\rangle,
\end{equation}
To contract the TTN, we start from an identity MPO which represents the density
matrix at infinite high temperature ($\beta=0$), and iteratively project $\nu$ tensors 
onto the MPO. The system cools down from $\beta=0$ to various lower temperatures,
following a linear inverse-temperature grid, i.e.,
$\beta=n \tau$ at $n$-th step.

In practical calculations, we regard the MPO as a super vector and 
follows the decimation technique developed for matrix product state
\cite{Zwolak2004,Li2011,Dong2017}. To further reduce the Trotter errors, 
we employ up to sixth-order Trotter-Suzuki decomposition so that the 
Trotter error is less than numerical precision and can be ignored in our 
calculations. Besides, the step length $\tau$ can also be adjusted dynamically 
during the cooling process, with the temperature grid also quite flexible.

\begin{figure}[]
\includegraphics[width=1\linewidth]{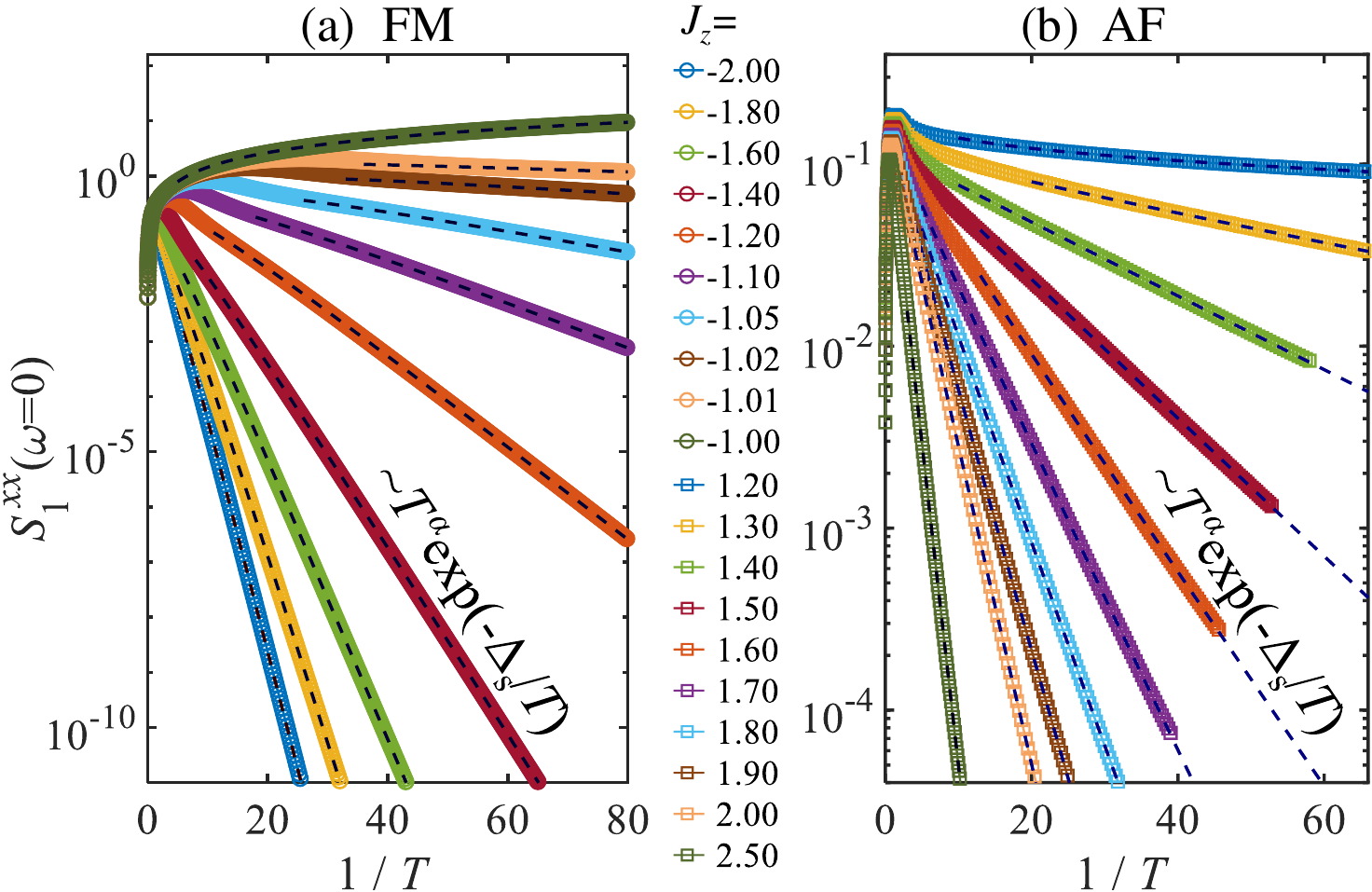}
\caption{\textbf{Supporting data of $\mathcal{S}_1(0)$ in the gapped phase 
of the spin-1/2 HAF chain.} The calculated $\mathcal{S}_1^{xx}(\omega=0)$ 
results and their exponential fittings in (a) FM and (b) AF phases of the 
easy-axis spin-1/2 HAF chain.}
\label{FigS2}
\end{figure}

\subsection{Tensor renormalization group for 2D lattice models}
We exploit exponential (XTRG)~\cite{Chen2018} and tangent-space (tanTRG)
\cite{tanTRG2023} tensor renormalization group to simulate 2D quantum spin 
models. In XTRG calculations, we start from the initial density matrix $\rho_0(\tau)$
at a very high temperature with $\tau = \mathcal{O}(10^{-4})$ ($T \equiv 1/ \tau$), 
represented in an MPO form via a series expansion
\cite{Chen2017}
\begin{equation}
\rho_0(\tau) = e^{-\tau H} = \sum_{n=0}^{\infty} \frac{(-\tau H)^n}{n!}
\simeq \sum_{n=0}^{N_{\rm cut}} \frac{(-\tau H)^n}{n!}.
\end{equation}
For a high temperature such as $\tau = 2.5\times 10^{-4}$, the 
expansion error is with machine precision for $N_{\rm cut}=6$-$7$.
Subsequently, by keeping squaring the density matrix repeatedly, i.e.,
$$\rho_n(2^n\tau) \cdot \rho_n(2^n\tau) \to \rho_{n+1}(2^{n+1}\tau),$$
we cool down the system along a logarithmic inverse-temperature
grid $\tau \to 2\tau \to 2^2\tau \to ... \to 2^n\tau$. Such a cooling process
reaches low temperature exponentially fast, with the projection and 
truncation steps significantly reduced. This renders higher accuracy 
than traditional linear evolution scheme, particularly for the challenging 
2D lattice models. Thus it constitutes a very powerful TTN method for 
ultra-low temperature simulations.

In tanTRG simulations, on the other hand, we consider the imaginary-time 
evolution of density matrix $\rho(\beta)$
\begin{equation}
\frac{d\rho}{d\beta} = -H\rho.
\label{EqS:Evo}
\end{equation}
In the tensor-network language, we prepare the density matrix
$\rho=e^{-\beta H}$ and the Hamiltonian $H$ in the MPO form,
and Eq.~\ref{EqS:Evo} can be solved by the tangent-space technique
\cite{TDVP2011,TDVP2016, MPSManifold2014} by introducing
a Choi transformation~\cite{Choi1972}
\begin{equation}
\frac{d}{d\beta} |\rho\rangle\rangle= -H \otimes I|\rho\rangle\rangle.
\label{EqS:TenEvo}
\end{equation}
Noticing that $e^{-\beta H\otimes I} = e^{-\beta H} \otimes I$, we
solve Eq.~\ref{EqS:TenEvo} within the MPO form~\cite{tanTRG2023}.

With tanTRG, we reduce the computational complexity from 
$\mathcal{O}(D^4)$ of XTRG to $\mathcal{O}(D^3)$, which 
allows us to retain much larger bond dimension in MPO.
 Thus, for challenging problems like the triangular 
lattice HAF model with width $W=6$, tanTRG is employed so 
we can retain up to $16,000$ U(1) states [$4000$ SU(2) multiplets] 
to obtain converged results.
\\


\section{Supplemental Results of Spin Lattice Models}
\label{App:B}

\begin{figure}[]
\includegraphics[width=1\linewidth]{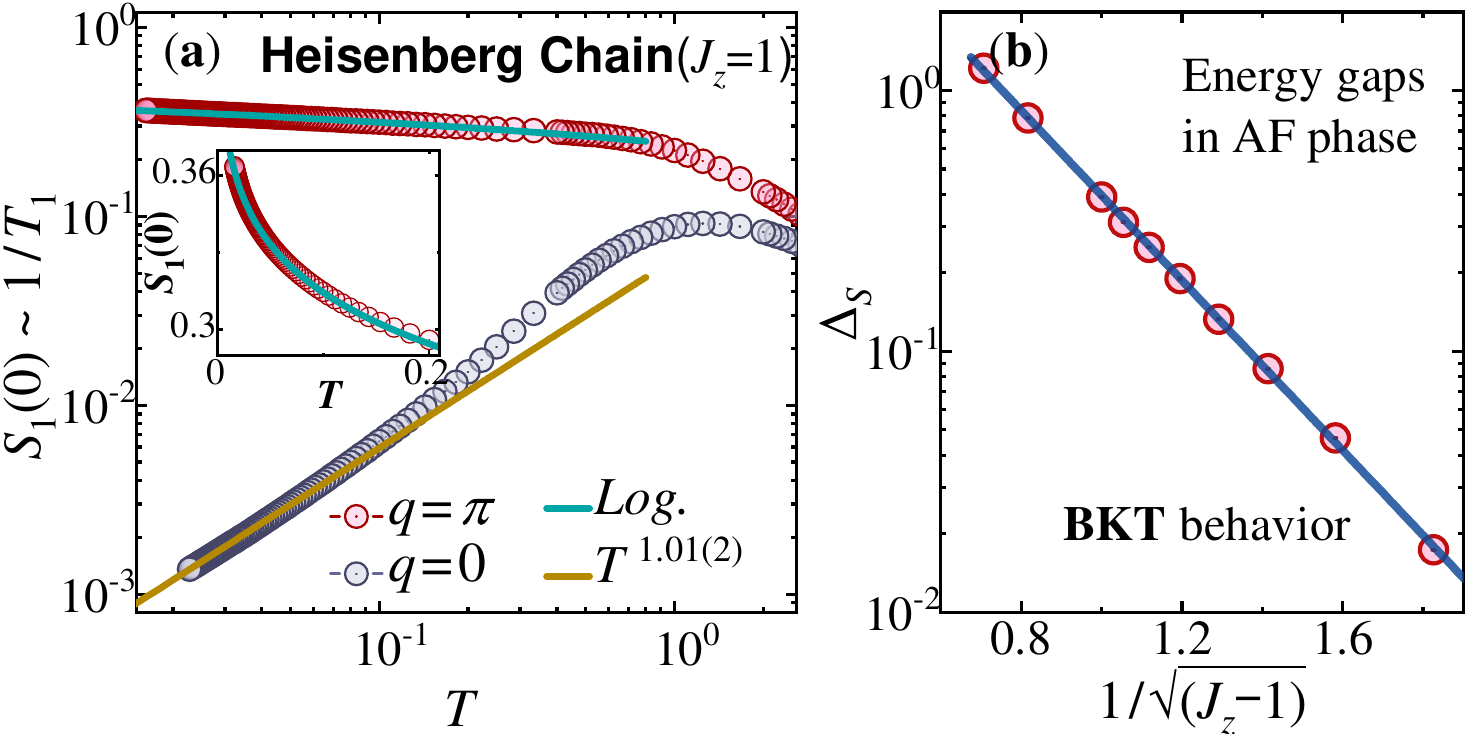}
\caption{\textbf{Calculated $\mathcal{S}_1(0)$ near the isotropic 
spin-1/2 HAF chain.} (a) The $\mathcal{S}_1(q, \omega=0)$ 
of the isotropic HAF chain ($J_{z} = 1$) for two different moments 
$q = 0$ and $\pi$, respectively, which show distinct behaviors as 
indicated by the two solid lines. The uniform $q=0$ contribution scales 
linearly with $T$, while the staggered $q=\pi$ part is nearly a constant 
with a logarithmic correction (see texts below). The inset demonstrates 
a zoomed-in linear plot of the data with $q=\pi$ at low temperatures. 
(b) The exponential gap opening that indicates a BKT transition 
from the gapless TLL to the gapped AF phase. 
}
\label{FigS3}
\end{figure}

\subsection{Spin-1/2 Heisenberg chain}
We start with more detailed $\mathcal{S}_{1,2}(\omega=0)$ results 
of the spin-1/2 HAF chains, with the results shown in Figs.~\ref{FigS1},
\ref{FigS2}, \ref{FigS3}. In the main text, we have illustrated that the 
critical exponent $\eta$ varies as $J_z$ is altered within the TLL phase 
[c.f., in Fig.~\ref{Fig2}(a)]. To support this conclusion, we show in 
Fig.~\ref{FigS1} the calculated results for $\mathcal{S}_{1}^{xx}(\omega=0)$, 
$\mathcal{S}_{2}^{xx}(\omega=0)$, and the ratio 
$[\mathcal{S}_{1}^{xx}(\omega=0)]^2/\mathcal{S}_{2}^{xx}(\omega=0)$ 
alongside their respective power-law fits in low temperatures.
We find the three approaches generate consistent results.
In Fig.~\ref{FigS2}, we show the computed $\mathcal{S}_{1}^{xx}(\omega=0)$ 
and their fittings at low temperatures, given by $T^{\alpha}\exp(\Delta_{S}/T)$, 
for the gapped FM and AF phases. From the fittings, we determine the 
values of spin gaps and present the results in Fig.~\ref{Fig2}(a) of the main text.

\begin{figure*}[]
\includegraphics[width=0.9\linewidth]{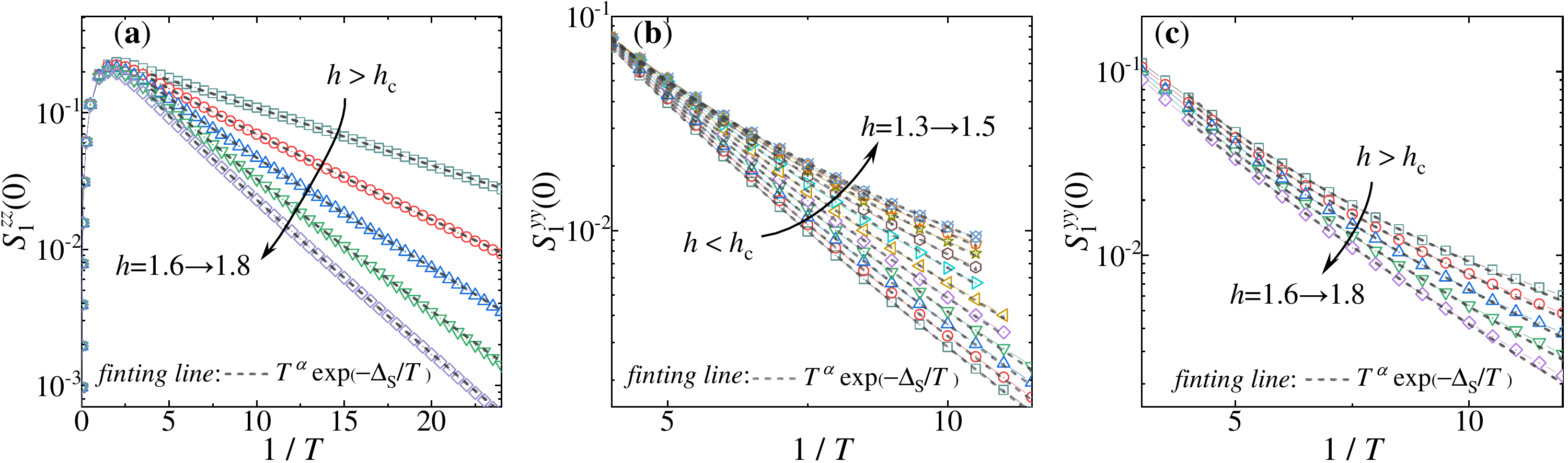}
\caption{\textbf{Spin-lattice relaxation rate of TFI model on the square lattice.}
The calculated results of (a) $\mathcal{S}_1^{zz}(\omega=0)$ and (b-c)
$\mathcal{S}_1^{yy}(\omega=0)$ for the square-lattice TFI model in the 
gapped phases. The results are found to show exponential decaying 
behaviors, and the fittings to the results are indicated by the dashed lines. 
The estimated gaps $\Delta_{\rm S}$ are collected and plotted in 
Fig.~\ref{Fig4}(b) of the main text.
}
\label{FigS4}
\end{figure*}

For the spin-1/2 XXZ chain, the Heisenberg point $J_{z}=1$ with 
SU(2) symmetry separates the gapped AF and gapless TLL phases. 
Figure~\ref{FigS3}(a) demonstrates distinct behaviors of $\mathcal{S}^{xx}_1(q, 
\omega=0)$ at $q=0$ (representing long-wave length fluctuations) and 
$q=\pi$ (staggered moment fluctuations) for the Heisenberg chain. Specifically, 
for $q=0$ it corresponding to $\eta=2$ and thus $\mathcal{S}^{xx}_1(\omega=0,
q=0)\propto T$. On the other hand, for $q=\pi$, the staggered fluctuations 
dominates and are nearly constant (with a multiplicative logarithmic correction) 
at low temperature~\cite{Sachdev1994,Sandvik1995}. The behavior of $\mathcal{S}^{xx}_1(\omega=0, q=\pi)$ is fitted with the formula,
$\sqrt{\ln(\Lambda/T)+\ln(\sqrt{\ln(\Lambda/T)})}$ with $\Lambda \simeq 
24.3$~\cite{Barzykin_2001}. The logarithmic correction can be ascribed to 
the presence of marginally irrelevant operator in critical HAF chain~\cite{Sachdev1994}. 
In Fig.~\ref{FigS3}(b) we show the spin gap $\Delta_{\rm S}$ vs. $1/\sqrt{(J_z-1)}$, 
from which we observe $\Delta_{\rm S} \sim e^{-\xi/\sqrt{(J_z-1)}}$ with $\xi$ a
constant. The exponential opening of spin gap reveals the existence of 
Berezinskii-Kosterlitz-Thouless transition at the isotropic Heisenberg point.

\subsection{Transverse-field Ising model on the square lattice}

In Fig.~\ref{FigS4}, we present supplementary data on $\mathcal{S}_{1}(\omega=0)$ 
for the TFI model on the square lattice, with the focus on the gapped 
paramagnetic ($h>h_c$) and ordered states ($h<h_c$) in the vicinity
of field-driven Ising QCP. In Fig.~\ref{FigS4}, we show the calculated 
$\mathcal{S}_1^{zz}(\omega=0)$ for $h>h_c$ (panel a) and 
$\mathcal{S}_1^{yy}(\omega=0)$ for both $h>h_c$ (panel b) and $h<h_c$ 
(panel c), along with the exponential fittings at low temperatures, following 
the expression $\mathcal{S}_1 \sim T^{\alpha}\exp(-\Delta_{S}/T)$,
to extract the spin gap $\Delta_{S}$. From the results in Fig.~\ref{FigS4}, 
we see the gap becomes increased as $h$ deviates $h_c$. By collecting 
the fitted values of $\Delta_{S}$, we show the gaps in Fig.~\ref{Fig4}(c) 
of the main text, which fall into algebraic scaling vs. $|h-h_c|$.

\end{appendix}

\bibliography{T1ref.bib}

\begin{thebibliography}{114}%
\makeatletter
\providecommand \@ifxundefined [1]{%
 \@ifx{#1\undefined}
}%
\providecommand \@ifnum [1]{%
 \ifnum #1\expandafter \@firstoftwo
 \else \expandafter \@secondoftwo
 \fi
}%
\providecommand \@ifx [1]{%
 \ifx #1\expandafter \@firstoftwo
 \else \expandafter \@secondoftwo
 \fi
}%
\providecommand \natexlab [1]{#1}%
\providecommand \enquote  [1]{``#1''}%
\providecommand \bibnamefont  [1]{#1}%
\providecommand \bibfnamefont [1]{#1}%
\providecommand \citenamefont [1]{#1}%
\providecommand \href@noop [0]{\@secondoftwo}%
\providecommand \href [0]{\begingroup \@sanitize@url \@href}%
\providecommand \@href[1]{\@@startlink{#1}\@@href}%
\providecommand \@@href[1]{\endgroup#1\@@endlink}%
\providecommand \@sanitize@url [0]{\catcode `\\12\catcode `\$12\catcode
  `\&12\catcode `\#12\catcode `\^12\catcode `\_12\catcode `\%12\relax}%
\providecommand \@@startlink[1]{}%
\providecommand \@@endlink[0]{}%
\providecommand \url  [0]{\begingroup\@sanitize@url \@url }%
\providecommand \@url [1]{\endgroup\@href {#1}{\urlprefix }}%
\providecommand \urlprefix  [0]{URL }%
\providecommand \Eprint [0]{\href }%
\providecommand \doibase [0]{https://doi.org/}%
\providecommand \selectlanguage [0]{\@gobble}%
\providecommand \bibinfo  [0]{\@secondoftwo}%
\providecommand \bibfield  [0]{\@secondoftwo}%
\providecommand \translation [1]{[#1]}%
\providecommand \BibitemOpen [0]{}%
\providecommand \bibitemStop [0]{}%
\providecommand \bibitemNoStop [0]{.\EOS\space}%
\providecommand \EOS [0]{\spacefactor3000\relax}%
\providecommand \BibitemShut  [1]{\csname bibitem#1\endcsname}%
\let\auto@bib@innerbib\@empty
\bibitem [{\citenamefont {Xiang}\ \emph {et~al.}(2024)\citenamefont {Xiang},
  \citenamefont {Zhang}, \citenamefont {Gao}, \citenamefont {Schmidt},
  \citenamefont {Schmalzl}, \citenamefont {Wang}, \citenamefont {Li},
  \citenamefont {Xi}, \citenamefont {Liu}, \citenamefont {Jin}, \citenamefont
  {Li}, \citenamefont {Shen}, \citenamefont {Chen}, \citenamefont {Qi},
  \citenamefont {Wan}, \citenamefont {Jin}, \citenamefont {Li}, \citenamefont
  {Sun},\ and\ \citenamefont {Su}}]{Xiang2024NBCP}%
  \BibitemOpen
  \bibfield  {author} {\bibinfo {author} {\bibfnamefont {J.}~\bibnamefont
  {Xiang}}, \bibinfo {author} {\bibfnamefont {C.}~\bibnamefont {Zhang}},
  \bibinfo {author} {\bibfnamefont {Y.}~\bibnamefont {Gao}}, \bibinfo {author}
  {\bibfnamefont {W.}~\bibnamefont {Schmidt}}, \bibinfo {author} {\bibfnamefont
  {K.}~\bibnamefont {Schmalzl}}, \bibinfo {author} {\bibfnamefont {C.-W.}\
  \bibnamefont {Wang}}, \bibinfo {author} {\bibfnamefont {B.}~\bibnamefont
  {Li}}, \bibinfo {author} {\bibfnamefont {N.}~\bibnamefont {Xi}}, \bibinfo
  {author} {\bibfnamefont {X.-Y.}\ \bibnamefont {Liu}}, \bibinfo {author}
  {\bibfnamefont {H.}~\bibnamefont {Jin}}, \bibinfo {author} {\bibfnamefont
  {G.}~\bibnamefont {Li}}, \bibinfo {author} {\bibfnamefont {J.}~\bibnamefont
  {Shen}}, \bibinfo {author} {\bibfnamefont {Z.}~\bibnamefont {Chen}}, \bibinfo
  {author} {\bibfnamefont {Y.}~\bibnamefont {Qi}}, \bibinfo {author}
  {\bibfnamefont {Y.}~\bibnamefont {Wan}}, \bibinfo {author} {\bibfnamefont
  {W.}~\bibnamefont {Jin}}, \bibinfo {author} {\bibfnamefont {W.}~\bibnamefont
  {Li}}, \bibinfo {author} {\bibfnamefont {P.}~\bibnamefont {Sun}},\ and\
  \bibinfo {author} {\bibfnamefont {G.}~\bibnamefont {Su}},\ }\bibfield
  {title} {\bibinfo {title} {Giant magnetocaloric effect in spin supersolid
  candidate {Na$_2$BaCo(PO$_4$)$_2$}},\ }\href
  {https://doi.org/10.1038/s41586-023-06885-w} {\bibfield  {journal} {\bibinfo
  {journal} {Nature}\ }\textbf {\bibinfo {volume} {625}},\ \bibinfo {pages}
  {270} (\bibinfo {year} {2024})}\BibitemShut {NoStop}%
\bibitem [{\citenamefont {Shangguan}\ \emph {et~al.}(2023)\citenamefont
  {Shangguan}, \citenamefont {Bao}, \citenamefont {Dong}, \citenamefont {Xi},
  \citenamefont {Gao}, \citenamefont {Ma}, \citenamefont {Wang}, \citenamefont
  {Qi}, \citenamefont {Zhang}, \citenamefont {Huang}, \citenamefont {Liao},
  \citenamefont {Zhao}, \citenamefont {Zhang}, \citenamefont {Cheng},
  \citenamefont {Xu}, \citenamefont {Yu}, \citenamefont {Mole}, \citenamefont
  {Murai}, \citenamefont {Ohira-Kawamura}, \citenamefont {He}, \citenamefont
  {Hao}, \citenamefont {Yan}, \citenamefont {Song}, \citenamefont {Li},
  \citenamefont {Yu}, \citenamefont {Li},\ and\ \citenamefont
  {Wen}}]{shangguan2023}%
  \BibitemOpen
  \bibfield  {author} {\bibinfo {author} {\bibfnamefont {Y.}~\bibnamefont
  {Shangguan}}, \bibinfo {author} {\bibfnamefont {S.}~\bibnamefont {Bao}},
  \bibinfo {author} {\bibfnamefont {Z.-Y.}\ \bibnamefont {Dong}}, \bibinfo
  {author} {\bibfnamefont {N.}~\bibnamefont {Xi}}, \bibinfo {author}
  {\bibfnamefont {Y.-P.}\ \bibnamefont {Gao}}, \bibinfo {author} {\bibfnamefont
  {Z.}~\bibnamefont {Ma}}, \bibinfo {author} {\bibfnamefont {W.}~\bibnamefont
  {Wang}}, \bibinfo {author} {\bibfnamefont {Z.}~\bibnamefont {Qi}}, \bibinfo
  {author} {\bibfnamefont {S.}~\bibnamefont {Zhang}}, \bibinfo {author}
  {\bibfnamefont {Z.}~\bibnamefont {Huang}}, \bibinfo {author} {\bibfnamefont
  {J.}~\bibnamefont {Liao}}, \bibinfo {author} {\bibfnamefont {X.}~\bibnamefont
  {Zhao}}, \bibinfo {author} {\bibfnamefont {B.}~\bibnamefont {Zhang}},
  \bibinfo {author} {\bibfnamefont {S.}~\bibnamefont {Cheng}}, \bibinfo
  {author} {\bibfnamefont {H.}~\bibnamefont {Xu}}, \bibinfo {author}
  {\bibfnamefont {D.}~\bibnamefont {Yu}}, \bibinfo {author} {\bibfnamefont
  {R.~A.}\ \bibnamefont {Mole}}, \bibinfo {author} {\bibfnamefont
  {N.}~\bibnamefont {Murai}}, \bibinfo {author} {\bibfnamefont
  {S.}~\bibnamefont {Ohira-Kawamura}}, \bibinfo {author} {\bibfnamefont
  {L.}~\bibnamefont {He}}, \bibinfo {author} {\bibfnamefont {J.}~\bibnamefont
  {Hao}}, \bibinfo {author} {\bibfnamefont {Q.-B.}\ \bibnamefont {Yan}},
  \bibinfo {author} {\bibfnamefont {F.}~\bibnamefont {Song}}, \bibinfo {author}
  {\bibfnamefont {W.}~\bibnamefont {Li}}, \bibinfo {author} {\bibfnamefont
  {S.-L.}\ \bibnamefont {Yu}}, \bibinfo {author} {\bibfnamefont {J.-X.}\
  \bibnamefont {Li}},\ and\ \bibinfo {author} {\bibfnamefont {J.}~\bibnamefont
  {Wen}},\ }\bibfield  {title} {\bibinfo {title} {A one-third magnetization
  plateau phase as evidence for the {Kitaev} interaction in a honeycomb-lattice
  antiferromagnet},\ }\href {https://doi.org/10.1038/s41567-023-02212-2}
  {\bibfield  {journal} {\bibinfo  {journal} {Nature Physics}\ }\textbf
  {\bibinfo {volume} {19}},\ \bibinfo {pages} {1883} (\bibinfo {year}
  {2023})}\BibitemShut {NoStop}%
\bibitem [{\citenamefont {Anderson}(1973)}]{Anderson1973}%
  \BibitemOpen
  \bibfield  {author} {\bibinfo {author} {\bibfnamefont {P.~W.}\ \bibnamefont
  {Anderson}},\ }\bibfield  {title} {\bibinfo {title} {Resonating valence
  bonds: A new kind of insulator?},\ }\href
  {https://doi.org/https://doi.org/10.1016/0025-5408(73)90167-0} {\bibfield
  {journal} {\bibinfo  {journal} {Mater. Res. Bull.}\ }\textbf {\bibinfo
  {volume} {8}},\ \bibinfo {pages} {153 } (\bibinfo {year} {1973})}\BibitemShut
  {NoStop}%
\bibitem [{\citenamefont {Balents}(2010)}]{Balents2010}%
  \BibitemOpen
  \bibfield  {author} {\bibinfo {author} {\bibfnamefont {L.}~\bibnamefont
  {Balents}},\ }\bibfield  {title} {\bibinfo {title} {Spin liquids in
  frustrated magnets},\ }\href {https://doi.org/10.1038/nature08917} {\bibfield
   {journal} {\bibinfo  {journal} {Nature}\ }\textbf {\bibinfo {volume}
  {464}},\ \bibinfo {pages} {199} (\bibinfo {year} {2010})}\BibitemShut
  {NoStop}%
\bibitem [{\citenamefont {Zhou}\ \emph {et~al.}(2017)\citenamefont {Zhou},
  \citenamefont {Kanoda},\ and\ \citenamefont {Ng}}]{Zhou2017}%
  \BibitemOpen
  \bibfield  {author} {\bibinfo {author} {\bibfnamefont {Y.}~\bibnamefont
  {Zhou}}, \bibinfo {author} {\bibfnamefont {K.}~\bibnamefont {Kanoda}},\ and\
  \bibinfo {author} {\bibfnamefont {T.-K.}\ \bibnamefont {Ng}},\ }\bibfield
  {title} {\bibinfo {title} {Quantum spin liquid states},\ }\href
  {https://doi.org/10.1103/RevModPhys.89.025003} {\bibfield  {journal}
  {\bibinfo  {journal} {Rev. Mod. Phys.}\ }\textbf {\bibinfo {volume} {89}},\
  \bibinfo {pages} {025003} (\bibinfo {year} {2017})}\BibitemShut {NoStop}%
\bibitem [{\citenamefont {Broholm}\ \emph {et~al.}(2020)\citenamefont
  {Broholm}, \citenamefont {Cava}, \citenamefont {Kivelson}, \citenamefont
  {Nocera}, \citenamefont {Norman},\ and\ \citenamefont
  {Senthil}}]{Broholm2020QSL}%
  \BibitemOpen
  \bibfield  {author} {\bibinfo {author} {\bibfnamefont {C.}~\bibnamefont
  {Broholm}}, \bibinfo {author} {\bibfnamefont {R.~J.}\ \bibnamefont {Cava}},
  \bibinfo {author} {\bibfnamefont {S.~A.}\ \bibnamefont {Kivelson}}, \bibinfo
  {author} {\bibfnamefont {D.~G.}\ \bibnamefont {Nocera}}, \bibinfo {author}
  {\bibfnamefont {M.~R.}\ \bibnamefont {Norman}},\ and\ \bibinfo {author}
  {\bibfnamefont {T.}~\bibnamefont {Senthil}},\ }\bibfield  {title} {\bibinfo
  {title} {Quantum spin liquids},\ }\href
  {https://doi.org/10.1126/science.aay0668} {\bibfield  {journal} {\bibinfo
  {journal} {Science}\ }\textbf {\bibinfo {volume} {367}},\ \bibinfo {pages}
  {eaay0668} (\bibinfo {year} {2020})}\BibitemShut {NoStop}%
\bibitem [{\citenamefont {Takigawa}\ \emph {et~al.}(1996)\citenamefont
  {Takigawa}, \citenamefont {Motoyama}, \citenamefont {Eisaki},\ and\
  \citenamefont {Uchida}}]{Takigawa1996SpinChain}%
  \BibitemOpen
  \bibfield  {author} {\bibinfo {author} {\bibfnamefont {M.}~\bibnamefont
  {Takigawa}}, \bibinfo {author} {\bibfnamefont {N.}~\bibnamefont {Motoyama}},
  \bibinfo {author} {\bibfnamefont {H.}~\bibnamefont {Eisaki}},\ and\ \bibinfo
  {author} {\bibfnamefont {S.}~\bibnamefont {Uchida}},\ }\bibfield  {title}
  {\bibinfo {title} {Dynamics in the
  {$\mathit{S}\phantom{\rule{0ex}{0ex}}=\phantom{\rule{0ex}{0ex}}1/2$}
  one-dimensional antiferromagnet {${\mathrm{Sr}}_{2}{\mathrm{CuO}}_{3}$ via
  ${}^{63}\mathrm{Cu}$ NMR}},\ }\href
  {https://doi.org/10.1103/PhysRevLett.76.4612} {\bibfield  {journal} {\bibinfo
   {journal} {Phys. Rev. Lett.}\ }\textbf {\bibinfo {volume} {76}},\ \bibinfo
  {pages} {4612} (\bibinfo {year} {1996})}\BibitemShut {NoStop}%
\bibitem [{\citenamefont {Sachdev}(1994)}]{Sachdev1994}%
  \BibitemOpen
  \bibfield  {author} {\bibinfo {author} {\bibfnamefont {S.}~\bibnamefont
  {Sachdev}},\ }\bibfield  {title} {\bibinfo {title} {{NMR} relaxation in
  half-integer antiferromagnetic spin chains},\ }\href
  {https://doi.org/10.1103/PhysRevB.50.13006} {\bibfield  {journal} {\bibinfo
  {journal} {Phys. Rev. B}\ }\textbf {\bibinfo {volume} {50}},\ \bibinfo
  {pages} {13006} (\bibinfo {year} {1994})}\BibitemShut {NoStop}%
\bibitem [{\citenamefont {Steinberg}\ \emph {et~al.}(2019)\citenamefont
  {Steinberg}, \citenamefont {Armitage}, \citenamefont {Essler},\ and\
  \citenamefont {Sachdev}}]{Sachdev2019}%
  \BibitemOpen
  \bibfield  {author} {\bibinfo {author} {\bibfnamefont {J.}~\bibnamefont
  {Steinberg}}, \bibinfo {author} {\bibfnamefont {N.~P.}\ \bibnamefont
  {Armitage}}, \bibinfo {author} {\bibfnamefont {F.~H.~L.}\ \bibnamefont
  {Essler}},\ and\ \bibinfo {author} {\bibfnamefont {S.}~\bibnamefont
  {Sachdev}},\ }\bibfield  {title} {\bibinfo {title} {{NMR} relaxation in
  {Ising} spin chains},\ }\href {https://doi.org/10.1103/PhysRevB.99.035156}
  {\bibfield  {journal} {\bibinfo  {journal} {Phys. Rev. B}\ }\textbf {\bibinfo
  {volume} {99}},\ \bibinfo {pages} {035156} (\bibinfo {year}
  {2019})}\BibitemShut {NoStop}%
\bibitem [{\citenamefont {Sandvik}(1995)}]{Sandvik1995}%
  \BibitemOpen
  \bibfield  {author} {\bibinfo {author} {\bibfnamefont {A.~W.}\ \bibnamefont
  {Sandvik}},\ }\bibfield  {title} {\bibinfo {title} {{NMR} relaxation rates
  for the spin-1/2 {Heisenberg} chain},\ }\href
  {https://doi.org/10.1103/PhysRevB.52.R9831} {\bibfield  {journal} {\bibinfo
  {journal} {Phys. Rev. B}\ }\textbf {\bibinfo {volume} {52}},\ \bibinfo
  {pages} {R9831} (\bibinfo {year} {1995})}\BibitemShut {NoStop}%
\bibitem [{\citenamefont {Sandvik}\ and\ \citenamefont
  {Scalapino}(1995)}]{Sandvik1995_2D}%
  \BibitemOpen
  \bibfield  {author} {\bibinfo {author} {\bibfnamefont {A.~W.}\ \bibnamefont
  {Sandvik}}\ and\ \bibinfo {author} {\bibfnamefont {D.~J.}\ \bibnamefont
  {Scalapino}},\ }\bibfield  {title} {\bibinfo {title} {Spin dynamics of
  {${\mathrm{La}}_{2}$${\mathrm{CuO}}_{4}$} and the two-dimensional
  {Heisenberg} model},\ }\href {https://doi.org/10.1103/PhysRevB.51.9403}
  {\bibfield  {journal} {\bibinfo  {journal} {Phys. Rev. B}\ }\textbf {\bibinfo
  {volume} {51}},\ \bibinfo {pages} {9403} (\bibinfo {year}
  {1995})}\BibitemShut {NoStop}%
\bibitem [{\citenamefont {Naef}\ and\ \citenamefont
  {Wang}(2000)}]{TMRGT1_2000}%
  \BibitemOpen
  \bibfield  {author} {\bibinfo {author} {\bibfnamefont {F.}~\bibnamefont
  {Naef}}\ and\ \bibinfo {author} {\bibfnamefont {X.}~\bibnamefont {Wang}},\
  }\bibfield  {title} {\bibinfo {title} {Nuclear spin relaxation rates in
  two-leg spin ladders},\ }\href {https://doi.org/10.1103/PhysRevLett.84.1320}
  {\bibfield  {journal} {\bibinfo  {journal} {Phys. Rev. Lett.}\ }\textbf
  {\bibinfo {volume} {84}},\ \bibinfo {pages} {1320} (\bibinfo {year}
  {2000})}\BibitemShut {NoStop}%
\bibitem [{\citenamefont {Kinross}\ \emph {et~al.}(2014)\citenamefont
  {Kinross}, \citenamefont {Fu}, \citenamefont {Munsie}, \citenamefont
  {Dabkowska}, \citenamefont {Luke}, \citenamefont {Sachdev},\ and\
  \citenamefont {Imai}}]{CoNbO2014PRX}%
  \BibitemOpen
  \bibfield  {author} {\bibinfo {author} {\bibfnamefont {A.~W.}\ \bibnamefont
  {Kinross}}, \bibinfo {author} {\bibfnamefont {M.}~\bibnamefont {Fu}},
  \bibinfo {author} {\bibfnamefont {T.~J.}\ \bibnamefont {Munsie}}, \bibinfo
  {author} {\bibfnamefont {H.~A.}\ \bibnamefont {Dabkowska}}, \bibinfo {author}
  {\bibfnamefont {G.~M.}\ \bibnamefont {Luke}}, \bibinfo {author}
  {\bibfnamefont {S.}~\bibnamefont {Sachdev}},\ and\ \bibinfo {author}
  {\bibfnamefont {T.}~\bibnamefont {Imai}},\ }\bibfield  {title} {\bibinfo
  {title} {Evolution of quantum fluctuations near the quantum critical point of
  the transverse field {Ising} chain system {CoNb$_{2}$O$_{6}$}},\ }\href
  {https://doi.org/10.1103/PhysRevX.4.031008} {\bibfield  {journal} {\bibinfo
  {journal} {Phys. Rev. X}\ }\textbf {\bibinfo {volume} {4}},\ \bibinfo {pages}
  {031008} (\bibinfo {year} {2014})}\BibitemShut {NoStop}%
\bibitem [{\citenamefont {Cui}\ \emph {et~al.}(2019)\citenamefont {Cui},
  \citenamefont {Zou}, \citenamefont {Xi}, \citenamefont {He}, \citenamefont
  {Yang}, \citenamefont {Shu}, \citenamefont {Zhang}, \citenamefont {Hu},
  \citenamefont {Chen}, \citenamefont {Yu}, \citenamefont {Wu},\ and\
  \citenamefont {Yu}}]{Cui2019SCVO}%
  \BibitemOpen
  \bibfield  {author} {\bibinfo {author} {\bibfnamefont {Y.}~\bibnamefont
  {Cui}}, \bibinfo {author} {\bibfnamefont {H.}~\bibnamefont {Zou}}, \bibinfo
  {author} {\bibfnamefont {N.}~\bibnamefont {Xi}}, \bibinfo {author}
  {\bibfnamefont {Z.}~\bibnamefont {He}}, \bibinfo {author} {\bibfnamefont
  {Y.~X.}\ \bibnamefont {Yang}}, \bibinfo {author} {\bibfnamefont
  {L.}~\bibnamefont {Shu}}, \bibinfo {author} {\bibfnamefont {G.~H.}\
  \bibnamefont {Zhang}}, \bibinfo {author} {\bibfnamefont {Z.}~\bibnamefont
  {Hu}}, \bibinfo {author} {\bibfnamefont {T.}~\bibnamefont {Chen}}, \bibinfo
  {author} {\bibfnamefont {R.}~\bibnamefont {Yu}}, \bibinfo {author}
  {\bibfnamefont {J.}~\bibnamefont {Wu}},\ and\ \bibinfo {author}
  {\bibfnamefont {W.}~\bibnamefont {Yu}},\ }\bibfield  {title} {\bibinfo
  {title} {Quantum criticality of the {Ising-like} screw chain antiferromagnet
  {${\mathrm{SrCo}}_{2}{\mathrm{V}}_{2}{\mathrm{O}}_{8}$} in a transverse
  magnetic field},\ }\href {https://doi.org/10.1103/PhysRevLett.123.067203}
  {\bibfield  {journal} {\bibinfo  {journal} {Phys. Rev. Lett.}\ }\textbf
  {\bibinfo {volume} {123}},\ \bibinfo {pages} {067203} (\bibinfo {year}
  {2019})}\BibitemShut {NoStop}%
\bibitem [{\citenamefont {Zou}\ \emph {et~al.}(2021)\citenamefont {Zou},
  \citenamefont {Cui}, \citenamefont {Wang}, \citenamefont {Zhang},
  \citenamefont {Yang}, \citenamefont {Xu}, \citenamefont {Okutani},
  \citenamefont {Hagiwara}, \citenamefont {Matsuda}, \citenamefont {Wang},
  \citenamefont {Mussardo}, \citenamefont {H\'ods\'agi}, \citenamefont
  {Kormos}, \citenamefont {He}, \citenamefont {Kimura}, \citenamefont {Yu},
  \citenamefont {Yu}, \citenamefont {Ma},\ and\ \citenamefont {Wu}}]{E8_2021}%
  \BibitemOpen
  \bibfield  {author} {\bibinfo {author} {\bibfnamefont {H.}~\bibnamefont
  {Zou}}, \bibinfo {author} {\bibfnamefont {Y.}~\bibnamefont {Cui}}, \bibinfo
  {author} {\bibfnamefont {X.}~\bibnamefont {Wang}}, \bibinfo {author}
  {\bibfnamefont {Z.}~\bibnamefont {Zhang}}, \bibinfo {author} {\bibfnamefont
  {J.}~\bibnamefont {Yang}}, \bibinfo {author} {\bibfnamefont {G.}~\bibnamefont
  {Xu}}, \bibinfo {author} {\bibfnamefont {A.}~\bibnamefont {Okutani}},
  \bibinfo {author} {\bibfnamefont {M.}~\bibnamefont {Hagiwara}}, \bibinfo
  {author} {\bibfnamefont {M.}~\bibnamefont {Matsuda}}, \bibinfo {author}
  {\bibfnamefont {G.}~\bibnamefont {Wang}}, \bibinfo {author} {\bibfnamefont
  {G.}~\bibnamefont {Mussardo}}, \bibinfo {author} {\bibfnamefont
  {K.}~\bibnamefont {H\'ods\'agi}}, \bibinfo {author} {\bibfnamefont
  {M.}~\bibnamefont {Kormos}}, \bibinfo {author} {\bibfnamefont
  {Z.}~\bibnamefont {He}}, \bibinfo {author} {\bibfnamefont {S.}~\bibnamefont
  {Kimura}}, \bibinfo {author} {\bibfnamefont {R.}~\bibnamefont {Yu}}, \bibinfo
  {author} {\bibfnamefont {W.}~\bibnamefont {Yu}}, \bibinfo {author}
  {\bibfnamefont {J.}~\bibnamefont {Ma}},\ and\ \bibinfo {author}
  {\bibfnamefont {J.}~\bibnamefont {Wu}},\ }\bibfield  {title} {\bibinfo
  {title} {${E}_{8}$ spectra of quasi-one-dimensional antiferromagnet
  {${\mathrm{BaCo}}_{2}{\mathrm{V}}_{2}{\mathrm{O}}_{8}$} under transverse
  field},\ }\href {https://doi.org/10.1103/PhysRevLett.127.077201} {\bibfield
  {journal} {\bibinfo  {journal} {Phys. Rev. Lett.}\ }\textbf {\bibinfo
  {volume} {127}},\ \bibinfo {pages} {077201} (\bibinfo {year}
  {2021})}\BibitemShut {NoStop}%
\bibitem [{\citenamefont {Hu}\ \emph {et~al.}(2020)\citenamefont {Hu},
  \citenamefont {Ma}, \citenamefont {Liao}, \citenamefont {Li}, \citenamefont
  {Ma}, \citenamefont {Cui}, \citenamefont {Shangguan}, \citenamefont {Huang},
  \citenamefont {Qi}, \citenamefont {Li}, \citenamefont {Meng}, \citenamefont
  {Wen},\ and\ \citenamefont {Yu}}]{TMGO_NMR}%
  \BibitemOpen
  \bibfield  {author} {\bibinfo {author} {\bibfnamefont {Z.}~\bibnamefont
  {Hu}}, \bibinfo {author} {\bibfnamefont {Z.}~\bibnamefont {Ma}}, \bibinfo
  {author} {\bibfnamefont {Y.-D.}\ \bibnamefont {Liao}}, \bibinfo {author}
  {\bibfnamefont {H.}~\bibnamefont {Li}}, \bibinfo {author} {\bibfnamefont
  {C.}~\bibnamefont {Ma}}, \bibinfo {author} {\bibfnamefont {Y.}~\bibnamefont
  {Cui}}, \bibinfo {author} {\bibfnamefont {Y.}~\bibnamefont {Shangguan}},
  \bibinfo {author} {\bibfnamefont {Z.}~\bibnamefont {Huang}}, \bibinfo
  {author} {\bibfnamefont {Y.}~\bibnamefont {Qi}}, \bibinfo {author}
  {\bibfnamefont {W.}~\bibnamefont {Li}}, \bibinfo {author} {\bibfnamefont
  {Z.~Y.}\ \bibnamefont {Meng}}, \bibinfo {author} {\bibfnamefont
  {J.}~\bibnamefont {Wen}},\ and\ \bibinfo {author} {\bibfnamefont
  {W.}~\bibnamefont {Yu}},\ }\bibfield  {title} {\bibinfo {title} {Evidence of
  the {Berezinskii-Kosterlitz-Thouless} phase in a frustrated magnet},\ }\href
  {https://doi.org/10.1038/s41467-020-19380-x} {\bibfield  {journal} {\bibinfo
  {journal} {Nat. Commun.}\ }\textbf {\bibinfo {volume} {11}},\ \bibinfo
  {pages} {5631} (\bibinfo {year} {2020})}\BibitemShut {NoStop}%
\bibitem [{\citenamefont {Zheng}\ \emph {et~al.}(2017)\citenamefont {Zheng},
  \citenamefont {Ran}, \citenamefont {Li}, \citenamefont {Wang}, \citenamefont
  {Wang}, \citenamefont {Liu}, \citenamefont {Liu}, \citenamefont {Normand},
  \citenamefont {Wen},\ and\ \citenamefont {Yu}}]{Zheng2017PRL}%
  \BibitemOpen
  \bibfield  {author} {\bibinfo {author} {\bibfnamefont {J.}~\bibnamefont
  {Zheng}}, \bibinfo {author} {\bibfnamefont {K.}~\bibnamefont {Ran}}, \bibinfo
  {author} {\bibfnamefont {T.}~\bibnamefont {Li}}, \bibinfo {author}
  {\bibfnamefont {J.}~\bibnamefont {Wang}}, \bibinfo {author} {\bibfnamefont
  {P.}~\bibnamefont {Wang}}, \bibinfo {author} {\bibfnamefont {B.}~\bibnamefont
  {Liu}}, \bibinfo {author} {\bibfnamefont {Z.-X.}\ \bibnamefont {Liu}},
  \bibinfo {author} {\bibfnamefont {B.}~\bibnamefont {Normand}}, \bibinfo
  {author} {\bibfnamefont {J.}~\bibnamefont {Wen}},\ and\ \bibinfo {author}
  {\bibfnamefont {W.}~\bibnamefont {Yu}},\ }\bibfield  {title} {\bibinfo
  {title} {Gapless spin excitations in the field-induced quantum spin liquid
  phase of {$\alpha$-RuCl$_3$}},\ }\href
  {https://doi.org/10.1103/PhysRevLett.119.227208} {\bibfield  {journal}
  {\bibinfo  {journal} {Phys. Rev. Lett.}\ }\textbf {\bibinfo {volume} {119}},\
  \bibinfo {pages} {227208} (\bibinfo {year} {2017})}\BibitemShut {NoStop}%
\bibitem [{\citenamefont {Cui}\ \emph {et~al.}(2023)\citenamefont {Cui},
  \citenamefont {Liu}, \citenamefont {Lin}, \citenamefont {Wu}, \citenamefont
  {Hong}, \citenamefont {Liu}, \citenamefont {Li}, \citenamefont {Hu},
  \citenamefont {Xi}, \citenamefont {Li}, \citenamefont {Yu}, \citenamefont
  {Sandvik},\ and\ \citenamefont {Yu}}]{Cui2023DQCP}%
  \BibitemOpen
  \bibfield  {author} {\bibinfo {author} {\bibfnamefont {Y.}~\bibnamefont
  {Cui}}, \bibinfo {author} {\bibfnamefont {L.}~\bibnamefont {Liu}}, \bibinfo
  {author} {\bibfnamefont {H.}~\bibnamefont {Lin}}, \bibinfo {author}
  {\bibfnamefont {K.-H.}\ \bibnamefont {Wu}}, \bibinfo {author} {\bibfnamefont
  {W.}~\bibnamefont {Hong}}, \bibinfo {author} {\bibfnamefont {X.}~\bibnamefont
  {Liu}}, \bibinfo {author} {\bibfnamefont {C.}~\bibnamefont {Li}}, \bibinfo
  {author} {\bibfnamefont {Z.}~\bibnamefont {Hu}}, \bibinfo {author}
  {\bibfnamefont {N.}~\bibnamefont {Xi}}, \bibinfo {author} {\bibfnamefont
  {S.}~\bibnamefont {Li}}, \bibinfo {author} {\bibfnamefont {R.}~\bibnamefont
  {Yu}}, \bibinfo {author} {\bibfnamefont {A.~W.}\ \bibnamefont {Sandvik}},\
  and\ \bibinfo {author} {\bibfnamefont {W.}~\bibnamefont {Yu}},\ }\bibfield
  {title} {\bibinfo {title} {Proximate deconfined quantum critical point in
  {SrCu$_2$(BO$_3$)$_2$}},\ }\href {https://doi.org/10.1126/science.adc9487}
  {\bibfield  {journal} {\bibinfo  {journal} {Science}\ }\textbf {\bibinfo
  {volume} {380}},\ \bibinfo {pages} {1179} (\bibinfo {year}
  {2023})}\BibitemShut {NoStop}%
\bibitem [{\citenamefont {Li}\ \emph {et~al.}(2020{\natexlab{a}})\citenamefont
  {Li}, \citenamefont {Liao}, \citenamefont {Chen}, \citenamefont {Zeng},
  \citenamefont {Sheng}, \citenamefont {Qi}, \citenamefont {Meng},\ and\
  \citenamefont {Li}}]{TMGO_Theo}%
  \BibitemOpen
  \bibfield  {author} {\bibinfo {author} {\bibfnamefont {H.}~\bibnamefont
  {Li}}, \bibinfo {author} {\bibfnamefont {Y.~D.}\ \bibnamefont {Liao}},
  \bibinfo {author} {\bibfnamefont {B.-B.}\ \bibnamefont {Chen}}, \bibinfo
  {author} {\bibfnamefont {X.-T.}\ \bibnamefont {Zeng}}, \bibinfo {author}
  {\bibfnamefont {X.-L.}\ \bibnamefont {Sheng}}, \bibinfo {author}
  {\bibfnamefont {Y.}~\bibnamefont {Qi}}, \bibinfo {author} {\bibfnamefont
  {Z.~Y.}\ \bibnamefont {Meng}},\ and\ \bibinfo {author} {\bibfnamefont
  {W.}~\bibnamefont {Li}},\ }\bibfield  {title} {\bibinfo {title}
  {{Kosterlitz-Thouless} melting of magnetic order in the triangular quantum
  {Ising} material {TmMgGaO$_4$}},\ }\href
  {https://doi.org/10.1038/s41467-020-14907-8} {\bibfield  {journal} {\bibinfo
  {journal} {Nat. Commun.}\ }\textbf {\bibinfo {volume} {11}},\ \bibinfo
  {pages} {1111} (\bibinfo {year} {2020}{\natexlab{a}})}\BibitemShut {NoStop}%
\bibitem [{\citenamefont {Wang}\ \emph {et~al.}(2023)\citenamefont {Wang},
  \citenamefont {Li}, \citenamefont {Xi}, \citenamefont {Gao}, \citenamefont
  {Yan}, \citenamefont {Li},\ and\ \citenamefont {Su}}]{Wang2023SSM}%
  \BibitemOpen
  \bibfield  {author} {\bibinfo {author} {\bibfnamefont {J.}~\bibnamefont
  {Wang}}, \bibinfo {author} {\bibfnamefont {H.}~\bibnamefont {Li}}, \bibinfo
  {author} {\bibfnamefont {N.}~\bibnamefont {Xi}}, \bibinfo {author}
  {\bibfnamefont {Y.}~\bibnamefont {Gao}}, \bibinfo {author} {\bibfnamefont
  {Q.-B.}\ \bibnamefont {Yan}}, \bibinfo {author} {\bibfnamefont
  {W.}~\bibnamefont {Li}},\ and\ \bibinfo {author} {\bibfnamefont
  {G.}~\bibnamefont {Su}},\ }\bibfield  {title} {\bibinfo {title} {Plaquette
  singlet transition, magnetic barocaloric effect, and spin supersolidity in
  the {Shastry-Sutherland} model},\ }\href
  {https://doi.org/10.1103/PhysRevLett.131.116702} {\bibfield  {journal}
  {\bibinfo  {journal} {Phys. Rev. Lett.}\ }\textbf {\bibinfo {volume} {131}},\
  \bibinfo {pages} {116702} (\bibinfo {year} {2023})}\BibitemShut {NoStop}%
\bibitem [{\citenamefont {Yang}\ \emph {et~al.}(2022)\citenamefont {Yang},
  \citenamefont {Yuan}, \citenamefont {Imai}, \citenamefont {Si}, \citenamefont
  {Wu},\ and\ \citenamefont {Kormos}}]{Yang2022}%
  \BibitemOpen
  \bibfield  {author} {\bibinfo {author} {\bibfnamefont {J.}~\bibnamefont
  {Yang}}, \bibinfo {author} {\bibfnamefont {W.}~\bibnamefont {Yuan}}, \bibinfo
  {author} {\bibfnamefont {T.}~\bibnamefont {Imai}}, \bibinfo {author}
  {\bibfnamefont {Q.}~\bibnamefont {Si}}, \bibinfo {author} {\bibfnamefont
  {J.}~\bibnamefont {Wu}},\ and\ \bibinfo {author} {\bibfnamefont
  {M.}~\bibnamefont {Kormos}},\ }\bibfield  {title} {\bibinfo {title} {Local
  dynamics and thermal activation in the transverse-field {Ising} chain},\
  }\href {https://doi.org/10.1103/PhysRevB.106.125149} {\bibfield  {journal}
  {\bibinfo  {journal} {Phys. Rev. B}\ }\textbf {\bibinfo {volume} {106}},\
  \bibinfo {pages} {125149} (\bibinfo {year} {2022})}\BibitemShut {NoStop}%
\bibitem [{\citenamefont {Barzykin}(2001)}]{Barzykin_2001}%
  \BibitemOpen
  \bibfield  {author} {\bibinfo {author} {\bibfnamefont {V.}~\bibnamefont
  {Barzykin}},\ }\bibfield  {title} {\bibinfo {title} {{NMR} relaxation rates
  in a spin-$\frac{1}{2}$ antiferromagnetic chain},\ }\href
  {https://doi.org/10.1103/PhysRevB.63.140412} {\bibfield  {journal} {\bibinfo
  {journal} {Phys. Rev. B}\ }\textbf {\bibinfo {volume} {63}},\ \bibinfo
  {pages} {140412} (\bibinfo {year} {2001})}\BibitemShut {NoStop}%
\bibitem [{\citenamefont {Dupont}\ \emph {et~al.}(2016)\citenamefont {Dupont},
  \citenamefont {Capponi},\ and\ \citenamefont {Laflorencie}}]{Realtime_2016}%
  \BibitemOpen
  \bibfield  {author} {\bibinfo {author} {\bibfnamefont {M.}~\bibnamefont
  {Dupont}}, \bibinfo {author} {\bibfnamefont {S.}~\bibnamefont {Capponi}},\
  and\ \bibinfo {author} {\bibfnamefont {N.}~\bibnamefont {Laflorencie}},\
  }\bibfield  {title} {\bibinfo {title} {Temperature dependence of the {NMR}
  relaxation rate $1/{T}_{1}$ for quantum spin chains},\ }\href
  {https://doi.org/10.1103/PhysRevB.94.144409} {\bibfield  {journal} {\bibinfo
  {journal} {Phys. Rev. B}\ }\textbf {\bibinfo {volume} {94}},\ \bibinfo
  {pages} {144409} (\bibinfo {year} {2016})}\BibitemShut {NoStop}%
\bibitem [{\citenamefont {Coira}\ \emph {et~al.}(2016)\citenamefont {Coira},
  \citenamefont {Barmettler}, \citenamefont {Giamarchi},\ and\ \citenamefont
  {Kollath}}]{Realtime2_2016}%
  \BibitemOpen
  \bibfield  {author} {\bibinfo {author} {\bibfnamefont {E.}~\bibnamefont
  {Coira}}, \bibinfo {author} {\bibfnamefont {P.}~\bibnamefont {Barmettler}},
  \bibinfo {author} {\bibfnamefont {T.}~\bibnamefont {Giamarchi}},\ and\
  \bibinfo {author} {\bibfnamefont {C.}~\bibnamefont {Kollath}},\ }\bibfield
  {title} {\bibinfo {title} {Temperature dependence of the {NMR} spin-lattice
  relaxation rate for spin-$\frac{1}{2}$ chains},\ }\href
  {https://doi.org/10.1103/PhysRevB.94.144408} {\bibfield  {journal} {\bibinfo
  {journal} {Phys. Rev. B}\ }\textbf {\bibinfo {volume} {94}},\ \bibinfo
  {pages} {144408} (\bibinfo {year} {2016})}\BibitemShut {NoStop}%
\bibitem [{\citenamefont {Capponi}\ \emph {et~al.}(2019)\citenamefont
  {Capponi}, \citenamefont {Dupont}, \citenamefont {Sandvik},\ and\
  \citenamefont {Sengupta}}]{Capponi_2019}%
  \BibitemOpen
  \bibfield  {author} {\bibinfo {author} {\bibfnamefont {S.}~\bibnamefont
  {Capponi}}, \bibinfo {author} {\bibfnamefont {M.}~\bibnamefont {Dupont}},
  \bibinfo {author} {\bibfnamefont {A.~W.}\ \bibnamefont {Sandvik}},\ and\
  \bibinfo {author} {\bibfnamefont {P.}~\bibnamefont {Sengupta}},\ }\bibfield
  {title} {\bibinfo {title} {{NMR} relaxation in the spin-1 heisenberg chain},\
  }\href {https://doi.org/10.1103/PhysRevB.100.094411} {\bibfield  {journal}
  {\bibinfo  {journal} {Phys. Rev. B}\ }\textbf {\bibinfo {volume} {100}},\
  \bibinfo {pages} {094411} (\bibinfo {year} {2019})}\BibitemShut {NoStop}%
\bibitem [{\citenamefont {Bertini}\ \emph {et~al.}(2021)\citenamefont
  {Bertini}, \citenamefont {Heidrich-Meisner}, \citenamefont {Karrasch},
  \citenamefont {Prosen}, \citenamefont {Steinigeweg},\ and\ \citenamefont
  {\ifmmode \check{Z}\else \v{Z}\fi{}nidari\ifmmode~\check{c}\else
  \v{c}\fi{}}}]{RMP_2021}%
  \BibitemOpen
  \bibfield  {author} {\bibinfo {author} {\bibfnamefont {B.}~\bibnamefont
  {Bertini}}, \bibinfo {author} {\bibfnamefont {F.}~\bibnamefont
  {Heidrich-Meisner}}, \bibinfo {author} {\bibfnamefont {C.}~\bibnamefont
  {Karrasch}}, \bibinfo {author} {\bibfnamefont {T.}~\bibnamefont {Prosen}},
  \bibinfo {author} {\bibfnamefont {R.}~\bibnamefont {Steinigeweg}},\ and\
  \bibinfo {author} {\bibfnamefont {M.}~\bibnamefont {\ifmmode \check{Z}\else
  \v{Z}\fi{}nidari\ifmmode~\check{c}\else \v{c}\fi{}}},\ }\bibfield  {title}
  {\bibinfo {title} {Finite-temperature transport in one-dimensional quantum
  lattice models},\ }\href {https://doi.org/10.1103/RevModPhys.93.025003}
  {\bibfield  {journal} {\bibinfo  {journal} {Rev. Mod. Phys.}\ }\textbf
  {\bibinfo {volume} {93}},\ \bibinfo {pages} {025003} (\bibinfo {year}
  {2021})}\BibitemShut {NoStop}%
\bibitem [{\citenamefont {Shu}\ \emph {et~al.}(2018)\citenamefont {Shu},
  \citenamefont {Dupont}, \citenamefont {Yao}, \citenamefont {Capponi},\ and\
  \citenamefont {Sandvik}}]{MCT1_2018}%
  \BibitemOpen
  \bibfield  {author} {\bibinfo {author} {\bibfnamefont {Y.-R.}\ \bibnamefont
  {Shu}}, \bibinfo {author} {\bibfnamefont {M.}~\bibnamefont {Dupont}},
  \bibinfo {author} {\bibfnamefont {D.-X.}\ \bibnamefont {Yao}}, \bibinfo
  {author} {\bibfnamefont {S.}~\bibnamefont {Capponi}},\ and\ \bibinfo {author}
  {\bibfnamefont {A.~W.}\ \bibnamefont {Sandvik}},\ }\bibfield  {title}
  {\bibinfo {title} {Dynamical properties of the {$S=\frac{1}{2}$ random
  Heisenberg} chain},\ }\href {https://doi.org/10.1103/PhysRevB.97.104424}
  {\bibfield  {journal} {\bibinfo  {journal} {Phys. Rev. B}\ }\textbf {\bibinfo
  {volume} {97}},\ \bibinfo {pages} {104424} (\bibinfo {year}
  {2018})}\BibitemShut {NoStop}%
\bibitem [{\citenamefont {Tang}\ \emph {et~al.}(2020)\citenamefont {Tang},
  \citenamefont {Tu},\ and\ \citenamefont {Wang}}]{Tang2020}%
  \BibitemOpen
  \bibfield  {author} {\bibinfo {author} {\bibfnamefont {W.}~\bibnamefont
  {Tang}}, \bibinfo {author} {\bibfnamefont {H.-H.}\ \bibnamefont {Tu}},\ and\
  \bibinfo {author} {\bibfnamefont {L.}~\bibnamefont {Wang}},\ }\bibfield
  {title} {\bibinfo {title} {Continuous matrix product operator approach to
  finite temperature quantum states},\ }\href
  {https://doi.org/10.1103/PhysRevLett.125.170604} {\bibfield  {journal}
  {\bibinfo  {journal} {Phys. Rev. Lett.}\ }\textbf {\bibinfo {volume} {125}},\
  \bibinfo {pages} {170604} (\bibinfo {year} {2020})}\BibitemShut {NoStop}%
\bibitem [{\citenamefont {Huscroft}\ \emph {et~al.}(2000)\citenamefont
  {Huscroft}, \citenamefont {Gass},\ and\ \citenamefont
  {Jarrell}}]{Huscroft_2000}%
  \BibitemOpen
  \bibfield  {author} {\bibinfo {author} {\bibfnamefont {C.}~\bibnamefont
  {Huscroft}}, \bibinfo {author} {\bibfnamefont {R.}~\bibnamefont {Gass}},\
  and\ \bibinfo {author} {\bibfnamefont {M.}~\bibnamefont {Jarrell}},\
  }\bibfield  {title} {\bibinfo {title} {Maximum entropy method of obtaining
  thermodynamic properties from quantum {Monte Carlo} simulations},\ }\href
  {https://doi.org/10.1103/PhysRevB.61.9300} {\bibfield  {journal} {\bibinfo
  {journal} {Phys. Rev. B}\ }\textbf {\bibinfo {volume} {61}},\ \bibinfo
  {pages} {9300} (\bibinfo {year} {2000})}\BibitemShut {NoStop}%
\bibitem [{\citenamefont {Shao}\ \emph {et~al.}(2017)\citenamefont {Shao},
  \citenamefont {Qin}, \citenamefont {Capponi}, \citenamefont {Chesi},
  \citenamefont {Meng},\ and\ \citenamefont {Sandvik}}]{ShaoH2017}%
  \BibitemOpen
  \bibfield  {author} {\bibinfo {author} {\bibfnamefont {H.}~\bibnamefont
  {Shao}}, \bibinfo {author} {\bibfnamefont {Y.~Q.}\ \bibnamefont {Qin}},
  \bibinfo {author} {\bibfnamefont {S.}~\bibnamefont {Capponi}}, \bibinfo
  {author} {\bibfnamefont {S.}~\bibnamefont {Chesi}}, \bibinfo {author}
  {\bibfnamefont {Z.~Y.}\ \bibnamefont {Meng}},\ and\ \bibinfo {author}
  {\bibfnamefont {A.~W.}\ \bibnamefont {Sandvik}},\ }\bibfield  {title}
  {\bibinfo {title} {Nearly deconfined spinon excitations in the square-lattice
  spin-$1/2$ {Heisenberg} antiferromagnet},\ }\href
  {https://doi.org/10.1103/PhysRevX.7.041072} {\bibfield  {journal} {\bibinfo
  {journal} {Phys. Rev. X}\ }\textbf {\bibinfo {volume} {7}},\ \bibinfo {pages}
  {041072} (\bibinfo {year} {2017})}\BibitemShut {NoStop}%
\bibitem [{\citenamefont {Huang}\ and\ \citenamefont
  {Liang}(2023{\natexlab{a}})}]{huang2023stochastic}%
  \BibitemOpen
  \bibfield  {author} {\bibinfo {author} {\bibfnamefont {L.}~\bibnamefont
  {Huang}}\ and\ \bibinfo {author} {\bibfnamefont {S.}~\bibnamefont {Liang}},\
  }\href@noop {} {\bibinfo {title} {Stochastic pole expansion method}}
  (\bibinfo {year} {2023}{\natexlab{a}}),\ \Eprint
  {https://arxiv.org/abs/2307.11324} {arXiv:2307.11324} \BibitemShut {NoStop}%
\bibitem [{\citenamefont {Huang}\ and\ \citenamefont
  {Liang}(2023{\natexlab{b}})}]{huang2023reconstructing}%
  \BibitemOpen
  \bibfield  {author} {\bibinfo {author} {\bibfnamefont {L.}~\bibnamefont
  {Huang}}\ and\ \bibinfo {author} {\bibfnamefont {S.}~\bibnamefont {Liang}},\
  }\href@noop {} {\bibinfo {title} {Reconstructing lattice qcd spectral
  functions with stochastic pole expansion and nevanlinna analytic
  continuation}} (\bibinfo {year} {2023}{\natexlab{b}}),\ \Eprint
  {https://arxiv.org/abs/2309.11114} {arXiv:2309.11114} \BibitemShut {NoStop}%
\bibitem [{\citenamefont {Fei}\ \emph {et~al.}(2021)\citenamefont {Fei},
  \citenamefont {Yeh},\ and\ \citenamefont {Gull}}]{Nevanlinna_2021}%
  \BibitemOpen
  \bibfield  {author} {\bibinfo {author} {\bibfnamefont {J.}~\bibnamefont
  {Fei}}, \bibinfo {author} {\bibfnamefont {C.-N.}\ \bibnamefont {Yeh}},\ and\
  \bibinfo {author} {\bibfnamefont {E.}~\bibnamefont {Gull}},\ }\bibfield
  {title} {\bibinfo {title} {Nevanlinna analytical continuation},\ }\href
  {https://doi.org/10.1103/PhysRevLett.126.056402} {\bibfield  {journal}
  {\bibinfo  {journal} {Phys. Rev. Lett.}\ }\textbf {\bibinfo {volume} {126}},\
  \bibinfo {pages} {056402} (\bibinfo {year} {2021})}\BibitemShut {NoStop}%
\bibitem [{\citenamefont {Randeria}\ \emph {et~al.}(1992)\citenamefont
  {Randeria}, \citenamefont {Trivedi}, \citenamefont {Moreo},\ and\
  \citenamefont {Scalettar}}]{Scalettar1992}%
  \BibitemOpen
  \bibfield  {author} {\bibinfo {author} {\bibfnamefont {M.}~\bibnamefont
  {Randeria}}, \bibinfo {author} {\bibfnamefont {N.}~\bibnamefont {Trivedi}},
  \bibinfo {author} {\bibfnamefont {A.}~\bibnamefont {Moreo}},\ and\ \bibinfo
  {author} {\bibfnamefont {R.~T.}\ \bibnamefont {Scalettar}},\ }\bibfield
  {title} {\bibinfo {title} {Pairing and spin gap in the normal state of short
  coherence length superconductors},\ }\href
  {https://doi.org/10.1103/PhysRevLett.69.2001} {\bibfield  {journal} {\bibinfo
   {journal} {Phys. Rev. Lett.}\ }\textbf {\bibinfo {volume} {69}},\ \bibinfo
  {pages} {2001} (\bibinfo {year} {1992})}\BibitemShut {NoStop}%
\bibitem [{\citenamefont {Fan}\ \emph {et~al.}(2020)\citenamefont {Fan},
  \citenamefont {Yang}, \citenamefont {Yu}, \citenamefont {Wu},\ and\
  \citenamefont {Yu}}]{FYC_2020}%
  \BibitemOpen
  \bibfield  {author} {\bibinfo {author} {\bibfnamefont {Y.}~\bibnamefont
  {Fan}}, \bibinfo {author} {\bibfnamefont {J.}~\bibnamefont {Yang}}, \bibinfo
  {author} {\bibfnamefont {W.}~\bibnamefont {Yu}}, \bibinfo {author}
  {\bibfnamefont {J.}~\bibnamefont {Wu}},\ and\ \bibinfo {author}
  {\bibfnamefont {R.}~\bibnamefont {Yu}},\ }\bibfield  {title} {\bibinfo
  {title} {Phase diagram and quantum criticality of {Heisenberg} spin chains
  with {Ising} anisotropic interchain couplings},\ }\href
  {https://doi.org/10.1103/PhysRevResearch.2.013345} {\bibfield  {journal}
  {\bibinfo  {journal} {Phys. Rev. Res.}\ }\textbf {\bibinfo {volume} {2}},\
  \bibinfo {pages} {013345} (\bibinfo {year} {2020})}\BibitemShut {NoStop}%
\bibitem [{\citenamefont {Bursill}\ \emph {et~al.}(1996)\citenamefont
  {Bursill}, \citenamefont {Xiang},\ and\ \citenamefont
  {Gehring}}]{Bursill1996DMRG}%
  \BibitemOpen
  \bibfield  {author} {\bibinfo {author} {\bibfnamefont {R.~J.}\ \bibnamefont
  {Bursill}}, \bibinfo {author} {\bibfnamefont {T.}~\bibnamefont {Xiang}},\
  and\ \bibinfo {author} {\bibfnamefont {G.~A.}\ \bibnamefont {Gehring}},\
  }\bibfield  {title} {\bibinfo {title} {The density matrix renormalization
  group for a quantum spin chain at non-zero temperature},\ }\href
  {http://stacks.iop.org/0953-8984/8/i=40/a=003} {\bibfield  {journal}
  {\bibinfo  {journal} {J. Phys.: Condens. Matter}\ }\textbf {\bibinfo {volume}
  {8}},\ \bibinfo {pages} {L583} (\bibinfo {year} {1996})}\BibitemShut
  {NoStop}%
\bibitem [{\citenamefont {Wang}\ and\ \citenamefont
  {Xiang}(1997)}]{Wang1997TMRG}%
  \BibitemOpen
  \bibfield  {author} {\bibinfo {author} {\bibfnamefont {X.}~\bibnamefont
  {Wang}}\ and\ \bibinfo {author} {\bibfnamefont {T.}~\bibnamefont {Xiang}},\
  }\bibfield  {title} {\bibinfo {title} {Transfer-matrix density-matrix
  renormalization-group theory for thermodynamics of one-dimensional quantum
  systems},\ }\href {https://doi.org/10.1103/PhysRevB.56.5061} {\bibfield
  {journal} {\bibinfo  {journal} {Phys. Rev. B}\ }\textbf {\bibinfo {volume}
  {56}},\ \bibinfo {pages} {5061} (\bibinfo {year} {1997})}\BibitemShut
  {NoStop}%
\bibitem [{\citenamefont {Zwolak}\ and\ \citenamefont
  {Vidal}(2004)}]{Zwolak2004}%
  \BibitemOpen
  \bibfield  {author} {\bibinfo {author} {\bibfnamefont {M.}~\bibnamefont
  {Zwolak}}\ and\ \bibinfo {author} {\bibfnamefont {G.}~\bibnamefont {Vidal}},\
  }\bibfield  {title} {\bibinfo {title} {Mixed-state dynamics in
  one-dimensional quantum lattice systems: {A} time-dependent superoperator
  renormalization algorithm},\ }\href
  {https://doi.org/10.1103/PhysRevLett.93.207205} {\bibfield  {journal}
  {\bibinfo  {journal} {Phys. Rev. Lett.}\ }\textbf {\bibinfo {volume} {93}},\
  \bibinfo {pages} {207205} (\bibinfo {year} {2004})}\BibitemShut {NoStop}%
\bibitem [{\citenamefont {Feiguin}\ and\ \citenamefont
  {White}(2005)}]{Feiguin2005}%
  \BibitemOpen
  \bibfield  {author} {\bibinfo {author} {\bibfnamefont {A.~E.}\ \bibnamefont
  {Feiguin}}\ and\ \bibinfo {author} {\bibfnamefont {S.~R.}\ \bibnamefont
  {White}},\ }\bibfield  {title} {\bibinfo {title} {Finite-temperature density
  matrix renormalization using an enlarged hilbert space},\ }\href
  {https://doi.org/10.1103/PhysRevB.72.220401} {\bibfield  {journal} {\bibinfo
  {journal} {Phys. Rev. B}\ }\textbf {\bibinfo {volume} {72}},\ \bibinfo
  {pages} {220401(R)} (\bibinfo {year} {2005})}\BibitemShut {NoStop}%
\bibitem [{\citenamefont {White}(2009)}]{White2009METTS}%
  \BibitemOpen
  \bibfield  {author} {\bibinfo {author} {\bibfnamefont {S.~R.}\ \bibnamefont
  {White}},\ }\bibfield  {title} {\bibinfo {title} {Minimally entangled typical
  quantum states at finite temperature},\ }\href
  {https://doi.org/10.1103/PhysRevLett.102.190601} {\bibfield  {journal}
  {\bibinfo  {journal} {Phys. Rev. Lett.}\ }\textbf {\bibinfo {volume} {102}},\
  \bibinfo {pages} {190601} (\bibinfo {year} {2009})}\BibitemShut {NoStop}%
\bibitem [{\citenamefont {Li}\ \emph {et~al.}(2011)\citenamefont {Li},
  \citenamefont {Ran}, \citenamefont {Gong}, \citenamefont {Zhao},
  \citenamefont {Xi}, \citenamefont {Ye},\ and\ \citenamefont {Su}}]{Li2011}%
  \BibitemOpen
  \bibfield  {author} {\bibinfo {author} {\bibfnamefont {W.}~\bibnamefont
  {Li}}, \bibinfo {author} {\bibfnamefont {S.-J.}\ \bibnamefont {Ran}},
  \bibinfo {author} {\bibfnamefont {S.-S.}\ \bibnamefont {Gong}}, \bibinfo
  {author} {\bibfnamefont {Y.}~\bibnamefont {Zhao}}, \bibinfo {author}
  {\bibfnamefont {B.}~\bibnamefont {Xi}}, \bibinfo {author} {\bibfnamefont
  {F.}~\bibnamefont {Ye}},\ and\ \bibinfo {author} {\bibfnamefont
  {G.}~\bibnamefont {Su}},\ }\bibfield  {title} {\bibinfo {title} {Linearized
  tensor renormalization group algorithm for the calculation of thermodynamic
  properties of quantum lattice models},\ }\href
  {https://doi.org/10.1103/PhysRevLett.106.127202} {\bibfield  {journal}
  {\bibinfo  {journal} {Phys. Rev. Lett.}\ }\textbf {\bibinfo {volume} {106}},\
  \bibinfo {pages} {127202} (\bibinfo {year} {2011})}\BibitemShut {NoStop}%
\bibitem [{\citenamefont {Czarnik}\ \emph {et~al.}(2012)\citenamefont
  {Czarnik}, \citenamefont {Cincio},\ and\ \citenamefont
  {Dziarmaga}}]{Czarnik2012PEPS}%
  \BibitemOpen
  \bibfield  {author} {\bibinfo {author} {\bibfnamefont {P.}~\bibnamefont
  {Czarnik}}, \bibinfo {author} {\bibfnamefont {L.}~\bibnamefont {Cincio}},\
  and\ \bibinfo {author} {\bibfnamefont {J.}~\bibnamefont {Dziarmaga}},\
  }\bibfield  {title} {\bibinfo {title} {Projected entangled pair states at
  finite temperature: Imaginary time evolution with ancillas},\ }\href
  {https://doi.org/10.1103/PhysRevB.86.245101} {\bibfield  {journal} {\bibinfo
  {journal} {Phys. Rev. B}\ }\textbf {\bibinfo {volume} {86}},\ \bibinfo
  {pages} {245101} (\bibinfo {year} {2012})}\BibitemShut {NoStop}%
\bibitem [{\citenamefont {Czarnik}\ and\ \citenamefont
  {Dziarmaga}(2015)}]{Czarnik2015PEPS}%
  \BibitemOpen
  \bibfield  {author} {\bibinfo {author} {\bibfnamefont {P.}~\bibnamefont
  {Czarnik}}\ and\ \bibinfo {author} {\bibfnamefont {J.}~\bibnamefont
  {Dziarmaga}},\ }\bibfield  {title} {\bibinfo {title} {Variational approach to
  projected entangled pair states at finite temperature},\ }\href
  {https://doi.org/10.1103/PhysRevB.92.035152} {\bibfield  {journal} {\bibinfo
  {journal} {Phys. Rev. B}\ }\textbf {\bibinfo {volume} {92}},\ \bibinfo
  {pages} {035152} (\bibinfo {year} {2015})}\BibitemShut {NoStop}%
\bibitem [{\citenamefont {Chen}\ \emph {et~al.}(2017)\citenamefont {Chen},
  \citenamefont {Liu}, \citenamefont {Chen},\ and\ \citenamefont
  {Li}}]{Chen2017}%
  \BibitemOpen
  \bibfield  {author} {\bibinfo {author} {\bibfnamefont {B.-B.}\ \bibnamefont
  {Chen}}, \bibinfo {author} {\bibfnamefont {Y.-J.}\ \bibnamefont {Liu}},
  \bibinfo {author} {\bibfnamefont {Z.}~\bibnamefont {Chen}},\ and\ \bibinfo
  {author} {\bibfnamefont {W.}~\bibnamefont {Li}},\ }\bibfield  {title}
  {\bibinfo {title} {Series-expansion thermal tensor network approach for
  quantum lattice models},\ }\href {https://doi.org/10.1103/PhysRevB.95.161104}
  {\bibfield  {journal} {\bibinfo  {journal} {Phys. Rev. B}\ }\textbf {\bibinfo
  {volume} {95}},\ \bibinfo {pages} {161104(R)} (\bibinfo {year}
  {2017})}\BibitemShut {NoStop}%
\bibitem [{\citenamefont {Dong}\ \emph {et~al.}(2017)\citenamefont {Dong},
  \citenamefont {Chen}, \citenamefont {Liu},\ and\ \citenamefont
  {Li}}]{Dong2017}%
  \BibitemOpen
  \bibfield  {author} {\bibinfo {author} {\bibfnamefont {Y.-L.}\ \bibnamefont
  {Dong}}, \bibinfo {author} {\bibfnamefont {L.}~\bibnamefont {Chen}}, \bibinfo
  {author} {\bibfnamefont {Y.-J.}\ \bibnamefont {Liu}},\ and\ \bibinfo {author}
  {\bibfnamefont {W.}~\bibnamefont {Li}},\ }\bibfield  {title} {\bibinfo
  {title} {Bilayer linearized tensor renormalization group approach for thermal
  tensor networks},\ }\href {https://doi.org/10.1103/PhysRevB.95.144428}
  {\bibfield  {journal} {\bibinfo  {journal} {Phys. Rev. B}\ }\textbf {\bibinfo
  {volume} {95}},\ \bibinfo {pages} {144428} (\bibinfo {year}
  {2017})}\BibitemShut {NoStop}%
\bibitem [{\citenamefont {Chen}\ \emph {et~al.}(2018)\citenamefont {Chen},
  \citenamefont {Chen}, \citenamefont {Chen}, \citenamefont {Li},\ and\
  \citenamefont {Weichselbaum}}]{Chen2018}%
  \BibitemOpen
  \bibfield  {author} {\bibinfo {author} {\bibfnamefont {B.-B.}\ \bibnamefont
  {Chen}}, \bibinfo {author} {\bibfnamefont {L.}~\bibnamefont {Chen}}, \bibinfo
  {author} {\bibfnamefont {Z.}~\bibnamefont {Chen}}, \bibinfo {author}
  {\bibfnamefont {W.}~\bibnamefont {Li}},\ and\ \bibinfo {author}
  {\bibfnamefont {A.}~\bibnamefont {Weichselbaum}},\ }\bibfield  {title}
  {\bibinfo {title} {Exponential {{Thermal Tensor Network Approach}} for
  {{Quantum Lattice Models}}},\ }\href
  {https://doi.org/10.1103/PhysRevX.8.031082} {\bibfield  {journal} {\bibinfo
  {journal} {Phys. Rev. X}\ }\textbf {\bibinfo {volume} {8}},\ \bibinfo {pages}
  {031082} (\bibinfo {year} {2018})}\BibitemShut {NoStop}%
\bibitem [{\citenamefont {Kshetrimayum}\ \emph {et~al.}(2019)\citenamefont
  {Kshetrimayum}, \citenamefont {Rizzi}, \citenamefont {Eisert},\ and\
  \citenamefont {Or\'us}}]{Kshetrimayum2019annealing}%
  \BibitemOpen
  \bibfield  {author} {\bibinfo {author} {\bibfnamefont {A.}~\bibnamefont
  {Kshetrimayum}}, \bibinfo {author} {\bibfnamefont {M.}~\bibnamefont {Rizzi}},
  \bibinfo {author} {\bibfnamefont {J.}~\bibnamefont {Eisert}},\ and\ \bibinfo
  {author} {\bibfnamefont {R.}~\bibnamefont {Or\'us}},\ }\bibfield  {title}
  {\bibinfo {title} {Tensor network annealing algorithm for two-dimensional
  thermal states},\ }\href {https://doi.org/10.1103/PhysRevLett.122.070502}
  {\bibfield  {journal} {\bibinfo  {journal} {Phys. Rev. Lett.}\ }\textbf
  {\bibinfo {volume} {122}},\ \bibinfo {pages} {070502} (\bibinfo {year}
  {2019})}\BibitemShut {NoStop}%
\bibitem [{\citenamefont {Li}\ \emph {et~al.}(2023{\natexlab{a}})\citenamefont
  {Li}, \citenamefont {Gao}, \citenamefont {He}, \citenamefont {Qi},
  \citenamefont {Chen},\ and\ \citenamefont {Li}}]{tanTRG2023}%
  \BibitemOpen
  \bibfield  {author} {\bibinfo {author} {\bibfnamefont {Q.}~\bibnamefont
  {Li}}, \bibinfo {author} {\bibfnamefont {Y.}~\bibnamefont {Gao}}, \bibinfo
  {author} {\bibfnamefont {Y.-Y.}\ \bibnamefont {He}}, \bibinfo {author}
  {\bibfnamefont {Y.}~\bibnamefont {Qi}}, \bibinfo {author} {\bibfnamefont
  {B.-B.}\ \bibnamefont {Chen}},\ and\ \bibinfo {author} {\bibfnamefont
  {W.}~\bibnamefont {Li}},\ }\bibfield  {title} {\bibinfo {title} {Tangent
  space approach for thermal tensor network simulations of the {2D Hubbard}
  model},\ }\href {https://doi.org/10.1103/PhysRevLett.130.226502} {\bibfield
  {journal} {\bibinfo  {journal} {Phys. Rev. Lett.}\ }\textbf {\bibinfo
  {volume} {130}},\ \bibinfo {pages} {226502} (\bibinfo {year}
  {2023}{\natexlab{a}})}\BibitemShut {NoStop}%
\bibitem [{\citenamefont {Li}\ \emph {et~al.}(2019)\citenamefont {Li},
  \citenamefont {Chen}, \citenamefont {Chen}, \citenamefont {{von Delft}},
  \citenamefont {Weichselbaum},\ and\ \citenamefont {Li}}]{Li2019}%
  \BibitemOpen
  \bibfield  {author} {\bibinfo {author} {\bibfnamefont {H.}~\bibnamefont
  {Li}}, \bibinfo {author} {\bibfnamefont {B.-B.}\ \bibnamefont {Chen}},
  \bibinfo {author} {\bibfnamefont {Z.}~\bibnamefont {Chen}}, \bibinfo {author}
  {\bibfnamefont {J.}~\bibnamefont {{von Delft}}}, \bibinfo {author}
  {\bibfnamefont {A.}~\bibnamefont {Weichselbaum}},\ and\ \bibinfo {author}
  {\bibfnamefont {W.}~\bibnamefont {Li}},\ }\bibfield  {title} {\bibinfo
  {title} {Thermal tensor renormalization group simulations of square-lattice
  quantum spin models},\ }\href {https://doi.org/10.1103/PhysRevB.100.045110}
  {\bibfield  {journal} {\bibinfo  {journal} {Phys. Rev. B}\ }\textbf {\bibinfo
  {volume} {100}},\ \bibinfo {pages} {045110} (\bibinfo {year}
  {2019})}\BibitemShut {NoStop}%
\bibitem [{\citenamefont {Chen}\ \emph {et~al.}(2019)\citenamefont {Chen},
  \citenamefont {Qu}, \citenamefont {Li}, \citenamefont {Chen}, \citenamefont
  {Gong}, \citenamefont {von Delft}, \citenamefont {Weichselbaum},\ and\
  \citenamefont {Li}}]{Chen2019}%
  \BibitemOpen
  \bibfield  {author} {\bibinfo {author} {\bibfnamefont {L.}~\bibnamefont
  {Chen}}, \bibinfo {author} {\bibfnamefont {D.-W.}\ \bibnamefont {Qu}},
  \bibinfo {author} {\bibfnamefont {H.}~\bibnamefont {Li}}, \bibinfo {author}
  {\bibfnamefont {B.-B.}\ \bibnamefont {Chen}}, \bibinfo {author}
  {\bibfnamefont {S.-S.}\ \bibnamefont {Gong}}, \bibinfo {author}
  {\bibfnamefont {J.}~\bibnamefont {von Delft}}, \bibinfo {author}
  {\bibfnamefont {A.}~\bibnamefont {Weichselbaum}},\ and\ \bibinfo {author}
  {\bibfnamefont {W.}~\bibnamefont {Li}},\ }\bibfield  {title} {\bibinfo
  {title} {Two-temperature scales in the triangular-lattice {Heisenberg}
  antiferromagnet},\ }\href {https://doi.org/10.1103/PhysRevB.99.140404}
  {\bibfield  {journal} {\bibinfo  {journal} {Phys. Rev. B}\ }\textbf {\bibinfo
  {volume} {99}},\ \bibinfo {pages} {140404(R)} (\bibinfo {year}
  {2019})}\BibitemShut {NoStop}%
\bibitem [{\citenamefont {Chen}\ \emph {et~al.}(2021)\citenamefont {Chen},
  \citenamefont {Chen}, \citenamefont {Chen}, \citenamefont {Cui},
  \citenamefont {Zhai}, \citenamefont {Weichselbaum}, \citenamefont {von
  Delft}, \citenamefont {Meng},\ and\ \citenamefont {Li}}]{Chen2021SLU}%
  \BibitemOpen
  \bibfield  {author} {\bibinfo {author} {\bibfnamefont {B.-B.}\ \bibnamefont
  {Chen}}, \bibinfo {author} {\bibfnamefont {C.}~\bibnamefont {Chen}}, \bibinfo
  {author} {\bibfnamefont {Z.}~\bibnamefont {Chen}}, \bibinfo {author}
  {\bibfnamefont {J.}~\bibnamefont {Cui}}, \bibinfo {author} {\bibfnamefont
  {Y.}~\bibnamefont {Zhai}}, \bibinfo {author} {\bibfnamefont {A.}~\bibnamefont
  {Weichselbaum}}, \bibinfo {author} {\bibfnamefont {J.}~\bibnamefont {von
  Delft}}, \bibinfo {author} {\bibfnamefont {Z.~Y.}\ \bibnamefont {Meng}},\
  and\ \bibinfo {author} {\bibfnamefont {W.}~\bibnamefont {Li}},\ }\bibfield
  {title} {\bibinfo {title} {{Quantum Many-Body Simulations of the
  Two-Dimensional Fermi-Hubbard Model in Ultracold Optical Lattices}},\ }\href
  {https://doi.org/10.1103/PhysRevB.103.L041107} {\bibfield  {journal}
  {\bibinfo  {journal} {Phys. Rev. B}\ }\textbf {\bibinfo {volume} {103}},\
  \bibinfo {pages} {L041107} (\bibinfo {year} {2021})}\BibitemShut {NoStop}%
\bibitem [{\citenamefont {Lin}\ \emph {et~al.}(2022)\citenamefont {Lin},
  \citenamefont {Chen}, \citenamefont {Li}, \citenamefont {Meng},\ and\
  \citenamefont {Shi}}]{Chen2022tbg}%
  \BibitemOpen
  \bibfield  {author} {\bibinfo {author} {\bibfnamefont {X.}~\bibnamefont
  {Lin}}, \bibinfo {author} {\bibfnamefont {B.-B.}\ \bibnamefont {Chen}},
  \bibinfo {author} {\bibfnamefont {W.}~\bibnamefont {Li}}, \bibinfo {author}
  {\bibfnamefont {Z.~Y.}\ \bibnamefont {Meng}},\ and\ \bibinfo {author}
  {\bibfnamefont {T.}~\bibnamefont {Shi}},\ }\bibfield  {title} {\bibinfo
  {title} {Exciton proliferation and fate of the topological mott insulator in
  a twisted bilayer graphene lattice model},\ }\href
  {https://doi.org/10.1103/PhysRevLett.128.157201} {\bibfield  {journal}
  {\bibinfo  {journal} {Phys. Rev. Lett.}\ }\textbf {\bibinfo {volume} {128}},\
  \bibinfo {pages} {157201} (\bibinfo {year} {2022})}\BibitemShut {NoStop}%
\bibitem [{\citenamefont {Liu}\ \emph {et~al.}(2021)\citenamefont {Liu},
  \citenamefont {Liu}, \citenamefont {Gao}, \citenamefont {Jin}, \citenamefont
  {He}, \citenamefont {Sheng}, \citenamefont {Jin}, \citenamefont {Chen},\ and\
  \citenamefont {Li}}]{LiuSpin1Chain}%
  \BibitemOpen
  \bibfield  {author} {\bibinfo {author} {\bibfnamefont {T.}~\bibnamefont
  {Liu}}, \bibinfo {author} {\bibfnamefont {X.-Y.}\ \bibnamefont {Liu}},
  \bibinfo {author} {\bibfnamefont {Y.}~\bibnamefont {Gao}}, \bibinfo {author}
  {\bibfnamefont {H.}~\bibnamefont {Jin}}, \bibinfo {author} {\bibfnamefont
  {J.}~\bibnamefont {He}}, \bibinfo {author} {\bibfnamefont {X.-L.}\
  \bibnamefont {Sheng}}, \bibinfo {author} {\bibfnamefont {W.}~\bibnamefont
  {Jin}}, \bibinfo {author} {\bibfnamefont {Z.}~\bibnamefont {Chen}},\ and\
  \bibinfo {author} {\bibfnamefont {W.}~\bibnamefont {Li}},\ }\bibfield
  {title} {\bibinfo {title} {Significant inverse magnetocaloric effect induced
  by quantum criticality},\ }\href
  {https://doi.org/10.1103/PhysRevResearch.3.033094} {\bibfield  {journal}
  {\bibinfo  {journal} {Phys. Rev. Research}\ }\textbf {\bibinfo {volume}
  {3}},\ \bibinfo {pages} {033094} (\bibinfo {year} {2021})}\BibitemShut
  {NoStop}%
\bibitem [{\citenamefont {Yu}\ \emph {et~al.}(2021)\citenamefont {Yu},
  \citenamefont {Gao}, \citenamefont {Chen},\ and\ \citenamefont
  {Li}}]{YuCPL2021}%
  \BibitemOpen
  \bibfield  {author} {\bibinfo {author} {\bibfnamefont {S.}~\bibnamefont
  {Yu}}, \bibinfo {author} {\bibfnamefont {Y.}~\bibnamefont {Gao}}, \bibinfo
  {author} {\bibfnamefont {B.-B.}\ \bibnamefont {Chen}},\ and\ \bibinfo
  {author} {\bibfnamefont {W.}~\bibnamefont {Li}},\ }\bibfield  {title}
  {\bibinfo {title} {Learning the effective spin hamiltonian of a quantum
  magnet},\ }\href {https://doi.org/10.1088/0256-307X/38/9/097502} {\bibfield
  {journal} {\bibinfo  {journal} {Chin. Phys. Lett.}\ }\textbf {\bibinfo
  {volume} {38}},\ \bibinfo {eid} {097502} (\bibinfo {year}
  {2021})}\BibitemShut {NoStop}%
\bibitem [{\citenamefont {Li}\ \emph {et~al.}(2020{\natexlab{b}})\citenamefont
  {Li}, \citenamefont {Liao}, \citenamefont {Chen}, \citenamefont {Zeng},
  \citenamefont {Sheng}, \citenamefont {Qi}, \citenamefont {Meng},\ and\
  \citenamefont {Li}}]{Li2020TMGO}%
  \BibitemOpen
  \bibfield  {author} {\bibinfo {author} {\bibfnamefont {H.}~\bibnamefont
  {Li}}, \bibinfo {author} {\bibfnamefont {Y.~D.}\ \bibnamefont {Liao}},
  \bibinfo {author} {\bibfnamefont {B.-B.}\ \bibnamefont {Chen}}, \bibinfo
  {author} {\bibfnamefont {X.-T.}\ \bibnamefont {Zeng}}, \bibinfo {author}
  {\bibfnamefont {X.-L.}\ \bibnamefont {Sheng}}, \bibinfo {author}
  {\bibfnamefont {Y.}~\bibnamefont {Qi}}, \bibinfo {author} {\bibfnamefont
  {Z.~Y.}\ \bibnamefont {Meng}},\ and\ \bibinfo {author} {\bibfnamefont
  {W.}~\bibnamefont {Li}},\ }\bibfield  {title} {\bibinfo {title}
  {{Kosterlitz-Thouless} melting of magnetic order in the triangular quantum
  {Ising} material {TmMgGaO$_4$}},\ }\href
  {https://doi.org/10.1038/s41467-020-14907-8} {\bibfield  {journal} {\bibinfo
  {journal} {Nat. Commun.}\ }\textbf {\bibinfo {volume} {11}},\ \bibinfo
  {pages} {1111} (\bibinfo {year} {2020}{\natexlab{b}})}\BibitemShut {NoStop}%
\bibitem [{\citenamefont {Liu}\ \emph {et~al.}()\citenamefont {Liu},
  \citenamefont {Gao}, \citenamefont {Li}, \citenamefont {Jin}, \citenamefont
  {Xiang}, \citenamefont {Jin}, \citenamefont {Chen}, \citenamefont {Li},\ and\
  \citenamefont {Su}}]{Liu2022}%
  \BibitemOpen
  \bibfield  {author} {\bibinfo {author} {\bibfnamefont {X.-Y.}\ \bibnamefont
  {Liu}}, \bibinfo {author} {\bibfnamefont {Y.}~\bibnamefont {Gao}}, \bibinfo
  {author} {\bibfnamefont {H.}~\bibnamefont {Li}}, \bibinfo {author}
  {\bibfnamefont {W.}~\bibnamefont {Jin}}, \bibinfo {author} {\bibfnamefont
  {J.}~\bibnamefont {Xiang}}, \bibinfo {author} {\bibfnamefont
  {H.}~\bibnamefont {Jin}}, \bibinfo {author} {\bibfnamefont {Z.}~\bibnamefont
  {Chen}}, \bibinfo {author} {\bibfnamefont {W.}~\bibnamefont {Li}},\ and\
  \bibinfo {author} {\bibfnamefont {G.}~\bibnamefont {Su}},\ }\bibfield
  {title} {\bibinfo {title} {Quantum spin liquid candidate as superior
  refrigerant in cascade demagnetization cooling},\ }\href
  {https://doi.org/10.1038/s42005-022-01010-1} {\bibfield  {journal} {\bibinfo
  {journal} {Communications Physics}\ }\textbf {\bibinfo {volume} {5}},\
  \bibinfo {pages} {233}}\BibitemShut {NoStop}%
\bibitem [{\citenamefont {Gao}\ \emph {et~al.}()\citenamefont {Gao},
  \citenamefont {Fan}, \citenamefont {Li}, \citenamefont {Yang}, \citenamefont
  {Zeng}, \citenamefont {Sheng}, \citenamefont {Zhong}, \citenamefont {Qi},
  \citenamefont {Wan},\ and\ \citenamefont {Li}}]{Gao2022NBCP}%
  \BibitemOpen
  \bibfield  {author} {\bibinfo {author} {\bibfnamefont {Y.}~\bibnamefont
  {Gao}}, \bibinfo {author} {\bibfnamefont {Y.-C.}\ \bibnamefont {Fan}},
  \bibinfo {author} {\bibfnamefont {H.}~\bibnamefont {Li}}, \bibinfo {author}
  {\bibfnamefont {F.}~\bibnamefont {Yang}}, \bibinfo {author} {\bibfnamefont
  {X.-T.}\ \bibnamefont {Zeng}}, \bibinfo {author} {\bibfnamefont {X.-L.}\
  \bibnamefont {Sheng}}, \bibinfo {author} {\bibfnamefont {R.}~\bibnamefont
  {Zhong}}, \bibinfo {author} {\bibfnamefont {Y.}~\bibnamefont {Qi}}, \bibinfo
  {author} {\bibfnamefont {Y.}~\bibnamefont {Wan}},\ and\ \bibinfo {author}
  {\bibfnamefont {W.}~\bibnamefont {Li}},\ }\bibfield  {title} {\bibinfo
  {title} {Spin supersolidity in nearly ideal easy-axis triangular quantum
  antiferromagnet {Na$_2$BaCo(PO$_4$)$_2$}},\ }\href
  {https://doi.org/10.1038/s41535-022-00500-3} {\bibfield  {journal} {\bibinfo
  {journal} {npj Quantum Materials}\ }\textbf {\bibinfo {volume} {7}},\
  \bibinfo {pages} {89}}\BibitemShut {NoStop}%
\bibitem [{\citenamefont {Li}\ \emph {et~al.}(2020{\natexlab{c}})\citenamefont
  {Li}, \citenamefont {Qu}, \citenamefont {Zhang}, \citenamefont {Jia},
  \citenamefont {Gong}, \citenamefont {Qi},\ and\ \citenamefont
  {Li}}]{HLi2020PRR}%
  \BibitemOpen
  \bibfield  {author} {\bibinfo {author} {\bibfnamefont {H.}~\bibnamefont
  {Li}}, \bibinfo {author} {\bibfnamefont {D.-W.}\ \bibnamefont {Qu}}, \bibinfo
  {author} {\bibfnamefont {H.-K.}\ \bibnamefont {Zhang}}, \bibinfo {author}
  {\bibfnamefont {Y.-Z.}\ \bibnamefont {Jia}}, \bibinfo {author} {\bibfnamefont
  {S.-S.}\ \bibnamefont {Gong}}, \bibinfo {author} {\bibfnamefont
  {Y.}~\bibnamefont {Qi}},\ and\ \bibinfo {author} {\bibfnamefont
  {W.}~\bibnamefont {Li}},\ }\bibfield  {title} {\bibinfo {title} {Universal
  thermodynamics in the {Kitaev} fractional liquid},\ }\href
  {https://doi.org/10.1103/PhysRevResearch.2.043015} {\bibfield  {journal}
  {\bibinfo  {journal} {Phys. Rev. Research}\ }\textbf {\bibinfo {volume}
  {2}},\ \bibinfo {pages} {043015} (\bibinfo {year}
  {2020}{\natexlab{c}})}\BibitemShut {NoStop}%
\bibitem [{\citenamefont {Li}\ \emph {et~al.}(2021)\citenamefont {Li},
  \citenamefont {Zhang}, \citenamefont {Wang}, \citenamefont {Wu},
  \citenamefont {Gao}, \citenamefont {Qu}, \citenamefont {Liu}, \citenamefont
  {Gong},\ and\ \citenamefont {Li}}]{Li2021NC}%
  \BibitemOpen
  \bibfield  {author} {\bibinfo {author} {\bibfnamefont {H.}~\bibnamefont
  {Li}}, \bibinfo {author} {\bibfnamefont {H.-K.}\ \bibnamefont {Zhang}},
  \bibinfo {author} {\bibfnamefont {J.}~\bibnamefont {Wang}}, \bibinfo {author}
  {\bibfnamefont {H.-Q.}\ \bibnamefont {Wu}}, \bibinfo {author} {\bibfnamefont
  {Y.}~\bibnamefont {Gao}}, \bibinfo {author} {\bibfnamefont {D.-W.}\
  \bibnamefont {Qu}}, \bibinfo {author} {\bibfnamefont {Z.-X.}\ \bibnamefont
  {Liu}}, \bibinfo {author} {\bibfnamefont {S.-S.}\ \bibnamefont {Gong}},\ and\
  \bibinfo {author} {\bibfnamefont {W.}~\bibnamefont {Li}},\ }\bibfield
  {title} {\bibinfo {title} {Identification of magnetic interactions and
  high-field quantum spin liquid in {$\alpha$-RuCl$_3$}},\ }\href
  {https://doi.org/10.1038/s41467-021-24257-8} {\bibfield  {journal} {\bibinfo
  {journal} {Nat. Commun.}\ }\textbf {\bibinfo {volume} {12}},\ \bibinfo
  {pages} {4007} (\bibinfo {year} {2021})}\BibitemShut {NoStop}%
\bibitem [{\citenamefont {Li}\ \emph {et~al.}(2023{\natexlab{b}})\citenamefont
  {Li}, \citenamefont {Li},\ and\ \citenamefont {Su}}]{Li2023KitaevPRB}%
  \BibitemOpen
  \bibfield  {author} {\bibinfo {author} {\bibfnamefont {H.}~\bibnamefont
  {Li}}, \bibinfo {author} {\bibfnamefont {W.}~\bibnamefont {Li}},\ and\
  \bibinfo {author} {\bibfnamefont {G.}~\bibnamefont {Su}},\ }\bibfield
  {title} {\bibinfo {title} {High-field quantum spin liquid transitions and
  angle-field phase diagram of the {Kitaev} magnet {$\alpha$-RuCl$_3$}},\
  }\href {https://doi.org/10.1103/PhysRevB.107.115124} {\bibfield  {journal}
  {\bibinfo  {journal} {Phys. Rev. B}\ }\textbf {\bibinfo {volume} {107}},\
  \bibinfo {pages} {115124} (\bibinfo {year} {2023}{\natexlab{b}})}\BibitemShut
  {NoStop}%
\bibitem [{\citenamefont {Zhou}\ \emph {et~al.}(2023)\citenamefont {Zhou},
  \citenamefont {Li}, \citenamefont {Matsuda}, \citenamefont {Matsuo},
  \citenamefont {Li}, \citenamefont {Kurita}, \citenamefont {Su}, \citenamefont
  {Kindo},\ and\ \citenamefont {Tanaka}}]{Zhou2023NC}%
  \BibitemOpen
  \bibfield  {author} {\bibinfo {author} {\bibfnamefont {X.-G.}\ \bibnamefont
  {Zhou}}, \bibinfo {author} {\bibfnamefont {H.}~\bibnamefont {Li}}, \bibinfo
  {author} {\bibfnamefont {Y.~H.}\ \bibnamefont {Matsuda}}, \bibinfo {author}
  {\bibfnamefont {A.}~\bibnamefont {Matsuo}}, \bibinfo {author} {\bibfnamefont
  {W.}~\bibnamefont {Li}}, \bibinfo {author} {\bibfnamefont {N.}~\bibnamefont
  {Kurita}}, \bibinfo {author} {\bibfnamefont {G.}~\bibnamefont {Su}}, \bibinfo
  {author} {\bibfnamefont {K.}~\bibnamefont {Kindo}},\ and\ \bibinfo {author}
  {\bibfnamefont {H.}~\bibnamefont {Tanaka}},\ }\bibfield  {title} {\bibinfo
  {title} {Possible intermediate quantum spin liquid phase in
  {$\alpha$-RuCl$_3$} under high magnetic fields up to {100 T}},\ }\href
  {https://doi.org/10.1038/s41467-023-41232-7} {\bibfield  {journal} {\bibinfo
  {journal} {Nat. Commun.}\ }\textbf {\bibinfo {volume} {14}},\ \bibinfo
  {pages} {5630} (\bibinfo {year} {2023})}\BibitemShut {NoStop}%
\bibitem [{\citenamefont {Liu}\ \emph {et~al.}(2018)\citenamefont {Liu},
  \citenamefont {Zhang}, \citenamefont {Ji}, \citenamefont {Liu}, \citenamefont
  {Li}, \citenamefont {Wang}, \citenamefont {Lei}, \citenamefont {Chen},\ and\
  \citenamefont {Zhang}}]{Liu2018ARX}%
  \BibitemOpen
  \bibfield  {author} {\bibinfo {author} {\bibfnamefont {W.}~\bibnamefont
  {Liu}}, \bibinfo {author} {\bibfnamefont {Z.}~\bibnamefont {Zhang}}, \bibinfo
  {author} {\bibfnamefont {J.}~\bibnamefont {Ji}}, \bibinfo {author}
  {\bibfnamefont {Y.}~\bibnamefont {Liu}}, \bibinfo {author} {\bibfnamefont
  {J.}~\bibnamefont {Li}}, \bibinfo {author} {\bibfnamefont {X.}~\bibnamefont
  {Wang}}, \bibinfo {author} {\bibfnamefont {H.}~\bibnamefont {Lei}}, \bibinfo
  {author} {\bibfnamefont {G.}~\bibnamefont {Chen}},\ and\ \bibinfo {author}
  {\bibfnamefont {Q.}~\bibnamefont {Zhang}},\ }\bibfield  {title} {\bibinfo
  {title} {Rare-earth {Chalcogenides}: A large family of triangular lattice
  spin liquid candidates},\ }\href
  {https://doi.org/10.1088/0256-307X/35/11/117501} {\bibfield  {journal}
  {\bibinfo  {journal} {Chinese Physics Letters}\ }\textbf {\bibinfo {volume}
  {35}},\ \bibinfo {pages} {117501} (\bibinfo {year} {2018})}\BibitemShut
  {NoStop}%
\bibitem [{\citenamefont {Ranjith}\ \emph
  {et~al.}(2019{\natexlab{a}})\citenamefont {Ranjith}, \citenamefont {Luther},
  \citenamefont {Reimann}, \citenamefont {Schmidt}, \citenamefont {Schlender},
  \citenamefont {Sichelschmidt}, \citenamefont {Yasuoka}, \citenamefont
  {Strydom}, \citenamefont {Skourski}, \citenamefont {Wosnitza}, \citenamefont
  {K\"uhne}, \citenamefont {Doert},\ and\ \citenamefont
  {Baenitz}}]{Ranjith2019NYS}%
  \BibitemOpen
  \bibfield  {author} {\bibinfo {author} {\bibfnamefont {K.~M.}\ \bibnamefont
  {Ranjith}}, \bibinfo {author} {\bibfnamefont {S.}~\bibnamefont {Luther}},
  \bibinfo {author} {\bibfnamefont {T.}~\bibnamefont {Reimann}}, \bibinfo
  {author} {\bibfnamefont {B.}~\bibnamefont {Schmidt}}, \bibinfo {author}
  {\bibfnamefont {P.}~\bibnamefont {Schlender}}, \bibinfo {author}
  {\bibfnamefont {J.}~\bibnamefont {Sichelschmidt}}, \bibinfo {author}
  {\bibfnamefont {H.}~\bibnamefont {Yasuoka}}, \bibinfo {author} {\bibfnamefont
  {A.~M.}\ \bibnamefont {Strydom}}, \bibinfo {author} {\bibfnamefont
  {Y.}~\bibnamefont {Skourski}}, \bibinfo {author} {\bibfnamefont
  {J.}~\bibnamefont {Wosnitza}}, \bibinfo {author} {\bibfnamefont
  {H.}~\bibnamefont {K\"uhne}}, \bibinfo {author} {\bibfnamefont
  {T.}~\bibnamefont {Doert}},\ and\ \bibinfo {author} {\bibfnamefont
  {M.}~\bibnamefont {Baenitz}},\ }\bibfield  {title} {\bibinfo {title}
  {Anisotropic field-induced ordering in the triangular-lattice quantum spin
  liquid {${\mathrm{NaYbSe}}_{2}$}},\ }\href
  {https://doi.org/10.1103/PhysRevB.100.224417} {\bibfield  {journal} {\bibinfo
   {journal} {Phys. Rev. B}\ }\textbf {\bibinfo {volume} {100}},\ \bibinfo
  {pages} {224417} (\bibinfo {year} {2019}{\natexlab{a}})}\BibitemShut
  {NoStop}%
\bibitem [{\citenamefont {Dai}\ \emph {et~al.}(2021)\citenamefont {Dai},
  \citenamefont {Zhang}, \citenamefont {Xie}, \citenamefont {Duan},
  \citenamefont {Gao}, \citenamefont {Zhu}, \citenamefont {Feng}, \citenamefont
  {Tao}, \citenamefont {Huang}, \citenamefont {Cao}, \citenamefont
  {Podlesnyak}, \citenamefont {Granroth}, \citenamefont {Everett},
  \citenamefont {Neuefeind}, \citenamefont {Voneshen}, \citenamefont {Wang},
  \citenamefont {Tan}, \citenamefont {Morosan}, \citenamefont {Wang},
  \citenamefont {Lin}, \citenamefont {Shu}, \citenamefont {Chen}, \citenamefont
  {Guo}, \citenamefont {Lu},\ and\ \citenamefont {Dai}}]{Dai2021NYS}%
  \BibitemOpen
  \bibfield  {author} {\bibinfo {author} {\bibfnamefont {P.-L.}\ \bibnamefont
  {Dai}}, \bibinfo {author} {\bibfnamefont {G.}~\bibnamefont {Zhang}}, \bibinfo
  {author} {\bibfnamefont {Y.}~\bibnamefont {Xie}}, \bibinfo {author}
  {\bibfnamefont {C.}~\bibnamefont {Duan}}, \bibinfo {author} {\bibfnamefont
  {Y.}~\bibnamefont {Gao}}, \bibinfo {author} {\bibfnamefont {Z.}~\bibnamefont
  {Zhu}}, \bibinfo {author} {\bibfnamefont {E.}~\bibnamefont {Feng}}, \bibinfo
  {author} {\bibfnamefont {Z.}~\bibnamefont {Tao}}, \bibinfo {author}
  {\bibfnamefont {C.-L.}\ \bibnamefont {Huang}}, \bibinfo {author}
  {\bibfnamefont {H.}~\bibnamefont {Cao}}, \bibinfo {author} {\bibfnamefont
  {A.}~\bibnamefont {Podlesnyak}}, \bibinfo {author} {\bibfnamefont {G.~E.}\
  \bibnamefont {Granroth}}, \bibinfo {author} {\bibfnamefont {M.~S.}\
  \bibnamefont {Everett}}, \bibinfo {author} {\bibfnamefont {J.~C.}\
  \bibnamefont {Neuefeind}}, \bibinfo {author} {\bibfnamefont {D.}~\bibnamefont
  {Voneshen}}, \bibinfo {author} {\bibfnamefont {S.}~\bibnamefont {Wang}},
  \bibinfo {author} {\bibfnamefont {G.}~\bibnamefont {Tan}}, \bibinfo {author}
  {\bibfnamefont {E.}~\bibnamefont {Morosan}}, \bibinfo {author} {\bibfnamefont
  {X.}~\bibnamefont {Wang}}, \bibinfo {author} {\bibfnamefont {H.-Q.}\
  \bibnamefont {Lin}}, \bibinfo {author} {\bibfnamefont {L.}~\bibnamefont
  {Shu}}, \bibinfo {author} {\bibfnamefont {G.}~\bibnamefont {Chen}}, \bibinfo
  {author} {\bibfnamefont {Y.}~\bibnamefont {Guo}}, \bibinfo {author}
  {\bibfnamefont {X.}~\bibnamefont {Lu}},\ and\ \bibinfo {author}
  {\bibfnamefont {P.}~\bibnamefont {Dai}},\ }\bibfield  {title} {\bibinfo
  {title} {Spinon fermi surface spin liquid in a triangular lattice
  antiferromagnet {${\mathrm{NaYbSe}}_{2}$}},\ }\href
  {https://doi.org/10.1103/PhysRevX.11.021044} {\bibfield  {journal} {\bibinfo
  {journal} {Phys. Rev. X}\ }\textbf {\bibinfo {volume} {11}},\ \bibinfo
  {pages} {021044} (\bibinfo {year} {2021})}\BibitemShut {NoStop}%
\bibitem [{\citenamefont {Scheie}\ \emph {et~al.}(2024)\citenamefont {Scheie},
  \citenamefont {Ghioldi}, \citenamefont {Xing}, \citenamefont {Paddison},
  \citenamefont {Sherman}, \citenamefont {Dupont}, \citenamefont {Sanjeewa},
  \citenamefont {Lee}, \citenamefont {Woods}, \citenamefont {Abernathy},
  \citenamefont {Pajerowski}, \citenamefont {Williams}, \citenamefont {Zhang},
  \citenamefont {Manuel}, \citenamefont {Trumper}, \citenamefont {Pemmaraju},
  \citenamefont {Sefat}, \citenamefont {Parker}, \citenamefont {Devereaux},
  \citenamefont {Movshovich}, \citenamefont {Moore}, \citenamefont {Batista},\
  and\ \citenamefont {Tennant}}]{Scheie2024KYS}%
  \BibitemOpen
  \bibfield  {author} {\bibinfo {author} {\bibfnamefont {A.~O.}\ \bibnamefont
  {Scheie}}, \bibinfo {author} {\bibfnamefont {E.~A.}\ \bibnamefont {Ghioldi}},
  \bibinfo {author} {\bibfnamefont {J.}~\bibnamefont {Xing}}, \bibinfo {author}
  {\bibfnamefont {J.~A.~M.}\ \bibnamefont {Paddison}}, \bibinfo {author}
  {\bibfnamefont {N.~E.}\ \bibnamefont {Sherman}}, \bibinfo {author}
  {\bibfnamefont {M.}~\bibnamefont {Dupont}}, \bibinfo {author} {\bibfnamefont
  {L.~D.}\ \bibnamefont {Sanjeewa}}, \bibinfo {author} {\bibfnamefont
  {S.}~\bibnamefont {Lee}}, \bibinfo {author} {\bibfnamefont {A.~J.}\
  \bibnamefont {Woods}}, \bibinfo {author} {\bibfnamefont {D.}~\bibnamefont
  {Abernathy}}, \bibinfo {author} {\bibfnamefont {D.~M.}\ \bibnamefont
  {Pajerowski}}, \bibinfo {author} {\bibfnamefont {T.~J.}\ \bibnamefont
  {Williams}}, \bibinfo {author} {\bibfnamefont {S.-S.}\ \bibnamefont {Zhang}},
  \bibinfo {author} {\bibfnamefont {L.~O.}\ \bibnamefont {Manuel}}, \bibinfo
  {author} {\bibfnamefont {A.~E.}\ \bibnamefont {Trumper}}, \bibinfo {author}
  {\bibfnamefont {C.~D.}\ \bibnamefont {Pemmaraju}}, \bibinfo {author}
  {\bibfnamefont {A.~S.}\ \bibnamefont {Sefat}}, \bibinfo {author}
  {\bibfnamefont {D.~S.}\ \bibnamefont {Parker}}, \bibinfo {author}
  {\bibfnamefont {T.~P.}\ \bibnamefont {Devereaux}}, \bibinfo {author}
  {\bibfnamefont {R.}~\bibnamefont {Movshovich}}, \bibinfo {author}
  {\bibfnamefont {J.~E.}\ \bibnamefont {Moore}}, \bibinfo {author}
  {\bibfnamefont {C.~D.}\ \bibnamefont {Batista}},\ and\ \bibinfo {author}
  {\bibfnamefont {D.~A.}\ \bibnamefont {Tennant}},\ }\bibfield  {title}
  {\bibinfo {title} {Proximate spin liquid and fractionalization in the
  triangular antiferromagnet {KYbSe$_2$}},\ }\href
  {https://doi.org/10.1038/s41567-023-02259-1} {\bibfield  {journal} {\bibinfo
  {journal} {Nature Physics}\ }\textbf {\bibinfo {volume} {20}},\ \bibinfo
  {pages} {74} (\bibinfo {year} {2024})}\BibitemShut {NoStop}%
\bibitem [{\citenamefont {Bl\"ote}\ and\ \citenamefont
  {Deng}(2002)}]{Deng2002}%
  \BibitemOpen
  \bibfield  {author} {\bibinfo {author} {\bibfnamefont {H.~W.~J.}\
  \bibnamefont {Bl\"ote}}\ and\ \bibinfo {author} {\bibfnamefont
  {Y.}~\bibnamefont {Deng}},\ }\bibfield  {title} {\bibinfo {title} {Cluster
  {Monte Carlo} simulation of the transverse {Ising} model},\ }\href
  {https://doi.org/10.1103/PhysRevE.66.066110} {\bibfield  {journal} {\bibinfo
  {journal} {Phys. Rev. E}\ }\textbf {\bibinfo {volume} {66}},\ \bibinfo
  {pages} {066110} (\bibinfo {year} {2002})}\BibitemShut {NoStop}%
\bibitem [{\citenamefont {Wiebe}\ \emph {et~al.}(2010)\citenamefont {Wiebe},
  \citenamefont {Berry}, \citenamefont {Høyer},\ and\ \citenamefont
  {Sanders}}]{trotter2010}%
  \BibitemOpen
  \bibfield  {author} {\bibinfo {author} {\bibfnamefont {N.}~\bibnamefont
  {Wiebe}}, \bibinfo {author} {\bibfnamefont {D.}~\bibnamefont {Berry}},
  \bibinfo {author} {\bibfnamefont {P.}~\bibnamefont {Høyer}},\ and\ \bibinfo
  {author} {\bibfnamefont {B.~C.}\ \bibnamefont {Sanders}},\ }\bibfield
  {title} {\bibinfo {title} {Higher order decompositions of ordered operator
  exponentials},\ }\href {https://doi.org/10.1088/1751-8113/43/6/065203}
  {\bibfield  {journal} {\bibinfo  {journal} {J. Phys. A: Math. Theor.}\
  }\textbf {\bibinfo {volume} {43}},\ \bibinfo {pages} {065203} (\bibinfo
  {year} {2010})}\BibitemShut {NoStop}%
\bibitem [{\citenamefont {Qu}\ \emph {et~al.}(2023)\citenamefont {Qu},
  \citenamefont {Chen}, \citenamefont {Lu}, \citenamefont {Li}, \citenamefont
  {Qi}, \citenamefont {Gong}, \citenamefont {Li},\ and\ \citenamefont
  {Su}}]{Qu2023dwave}%
  \BibitemOpen
  \bibfield  {author} {\bibinfo {author} {\bibfnamefont {D.-W.}\ \bibnamefont
  {Qu}}, \bibinfo {author} {\bibfnamefont {B.-B.}\ \bibnamefont {Chen}},
  \bibinfo {author} {\bibfnamefont {X.}~\bibnamefont {Lu}}, \bibinfo {author}
  {\bibfnamefont {Q.}~\bibnamefont {Li}}, \bibinfo {author} {\bibfnamefont
  {Y.}~\bibnamefont {Qi}}, \bibinfo {author} {\bibfnamefont {S.-S.}\
  \bibnamefont {Gong}}, \bibinfo {author} {\bibfnamefont {W.}~\bibnamefont
  {Li}},\ and\ \bibinfo {author} {\bibfnamefont {G.}~\bibnamefont {Su}},\
  }\href@noop {} {\bibinfo {title} {$d$-wave superconductivity, pseudogap, and
  the phase diagram of {$t$-$t'$-$J$} model at finite temperature}} (\bibinfo
  {year} {2023}),\ \Eprint {https://arxiv.org/abs/2211.06322}
  {arXiv:2211.06322} \BibitemShut {NoStop}%
\bibitem [{\citenamefont {Haegeman}\ \emph {et~al.}(2011)\citenamefont
  {Haegeman}, \citenamefont {Cirac}, \citenamefont {Osborne}, \citenamefont
  {Pi\ifmmode~\check{z}\else \v{z}\fi{}orn}, \citenamefont {Verschelde},\ and\
  \citenamefont {Verstraete}}]{TDVP2011}%
  \BibitemOpen
  \bibfield  {author} {\bibinfo {author} {\bibfnamefont {J.}~\bibnamefont
  {Haegeman}}, \bibinfo {author} {\bibfnamefont {J.~I.}\ \bibnamefont {Cirac}},
  \bibinfo {author} {\bibfnamefont {T.~J.}\ \bibnamefont {Osborne}}, \bibinfo
  {author} {\bibfnamefont {I.}~\bibnamefont {Pi\ifmmode~\check{z}\else
  \v{z}\fi{}orn}}, \bibinfo {author} {\bibfnamefont {H.}~\bibnamefont
  {Verschelde}},\ and\ \bibinfo {author} {\bibfnamefont {F.}~\bibnamefont
  {Verstraete}},\ }\bibfield  {title} {\bibinfo {title} {Time-dependent
  variational principle for quantum lattices},\ }\href
  {https://doi.org/10.1103/PhysRevLett.107.070601} {\bibfield  {journal}
  {\bibinfo  {journal} {Phys. Rev. Lett.}\ }\textbf {\bibinfo {volume} {107}},\
  \bibinfo {pages} {070601} (\bibinfo {year} {2011})}\BibitemShut {NoStop}%
\bibitem [{\citenamefont {Hackl}\ \emph {et~al.}(2020)\citenamefont {Hackl},
  \citenamefont {Guaita}, \citenamefont {Shi}, \citenamefont {Haegeman},
  \citenamefont {Demler},\ and\ \citenamefont {Cirac}}]{GeometryTDVP2020}%
  \BibitemOpen
  \bibfield  {author} {\bibinfo {author} {\bibfnamefont {L.}~\bibnamefont
  {Hackl}}, \bibinfo {author} {\bibfnamefont {T.}~\bibnamefont {Guaita}},
  \bibinfo {author} {\bibfnamefont {T.}~\bibnamefont {Shi}}, \bibinfo {author}
  {\bibfnamefont {J.}~\bibnamefont {Haegeman}}, \bibinfo {author}
  {\bibfnamefont {E.}~\bibnamefont {Demler}},\ and\ \bibinfo {author}
  {\bibfnamefont {J.~I.}\ \bibnamefont {Cirac}},\ }\bibfield  {title} {\bibinfo
  {title} {Geometry of variational methods: dynamics of closed quantum
  systems},\ }\href {https://doi.org/10.21468/SciPostPhys.9.4.048} {\bibfield
  {journal} {\bibinfo  {journal} {SciPost Phys.}\ }\textbf {\bibinfo {volume}
  {9}},\ \bibinfo {pages} {048} (\bibinfo {year} {2020})}\BibitemShut {NoStop}%
\bibitem [{\citenamefont {Fradkin}(2013)}]{fradkin_2013}%
  \BibitemOpen
  \bibfield  {author} {\bibinfo {author} {\bibfnamefont {E.}~\bibnamefont
  {Fradkin}},\ }\href {https://doi.org/10.1017/CBO9781139015509} {\emph
  {\bibinfo {title} {Field Theories of Condensed Matter Physics}}},\ \bibinfo
  {edition} {2nd}\ ed.\ (\bibinfo  {publisher} {Cambridge University Press},\
  \bibinfo {year} {2013})\BibitemShut {NoStop}%
\bibitem [{\citenamefont {Hong}\ \emph {et~al.}(2021)\citenamefont {Hong},
  \citenamefont {Liu}, \citenamefont {Liu}, \citenamefont {Ma}, \citenamefont
  {Koda}, \citenamefont {Li}, \citenamefont {Song}, \citenamefont {Yang},
  \citenamefont {Yang}, \citenamefont {Cheng}, \citenamefont {Zhang},
  \citenamefont {Bao}, \citenamefont {Ma}, \citenamefont {Chen}, \citenamefont
  {Sun}, \citenamefont {Guo}, \citenamefont {Luo}, \citenamefont {Sandvik},\
  and\ \citenamefont {Li}}]{Hong2021}%
  \BibitemOpen
  \bibfield  {author} {\bibinfo {author} {\bibfnamefont {W.}~\bibnamefont
  {Hong}}, \bibinfo {author} {\bibfnamefont {L.}~\bibnamefont {Liu}}, \bibinfo
  {author} {\bibfnamefont {C.}~\bibnamefont {Liu}}, \bibinfo {author}
  {\bibfnamefont {X.}~\bibnamefont {Ma}}, \bibinfo {author} {\bibfnamefont
  {A.}~\bibnamefont {Koda}}, \bibinfo {author} {\bibfnamefont {X.}~\bibnamefont
  {Li}}, \bibinfo {author} {\bibfnamefont {J.}~\bibnamefont {Song}}, \bibinfo
  {author} {\bibfnamefont {W.}~\bibnamefont {Yang}}, \bibinfo {author}
  {\bibfnamefont {J.}~\bibnamefont {Yang}}, \bibinfo {author} {\bibfnamefont
  {P.}~\bibnamefont {Cheng}}, \bibinfo {author} {\bibfnamefont
  {H.}~\bibnamefont {Zhang}}, \bibinfo {author} {\bibfnamefont
  {W.}~\bibnamefont {Bao}}, \bibinfo {author} {\bibfnamefont {X.}~\bibnamefont
  {Ma}}, \bibinfo {author} {\bibfnamefont {D.}~\bibnamefont {Chen}}, \bibinfo
  {author} {\bibfnamefont {K.}~\bibnamefont {Sun}}, \bibinfo {author}
  {\bibfnamefont {W.}~\bibnamefont {Guo}}, \bibinfo {author} {\bibfnamefont
  {H.}~\bibnamefont {Luo}}, \bibinfo {author} {\bibfnamefont {A.~W.}\
  \bibnamefont {Sandvik}},\ and\ \bibinfo {author} {\bibfnamefont
  {S.}~\bibnamefont {Li}},\ }\bibfield  {title} {\bibinfo {title} {Extreme
  suppression of antiferromagnetic order and critical scaling in a
  two-dimensional random quantum magnet},\ }\href
  {https://doi.org/10.1103/PhysRevLett.126.037201} {\bibfield  {journal}
  {\bibinfo  {journal} {Phys. Rev. Lett.}\ }\textbf {\bibinfo {volume} {126}},\
  \bibinfo {pages} {037201} (\bibinfo {year} {2021})}\BibitemShut {NoStop}%
\bibitem [{\citenamefont {Luther}\ and\ \citenamefont
  {Peschel}(1975)}]{Luther1975}%
  \BibitemOpen
  \bibfield  {author} {\bibinfo {author} {\bibfnamefont {A.}~\bibnamefont
  {Luther}}\ and\ \bibinfo {author} {\bibfnamefont {I.}~\bibnamefont
  {Peschel}},\ }\bibfield  {title} {\bibinfo {title} {Calculation of critical
  exponents in two dimensions from quantum field theory in one dimension},\
  }\href {https://doi.org/10.1103/PhysRevB.12.3908} {\bibfield  {journal}
  {\bibinfo  {journal} {Phys. Rev. B}\ }\textbf {\bibinfo {volume} {12}},\
  \bibinfo {pages} {3908} (\bibinfo {year} {1975})}\BibitemShut {NoStop}%
\bibitem [{\citenamefont {White}\ and\ \citenamefont
  {Huse}(1993)}]{White1993Spin1}%
  \BibitemOpen
  \bibfield  {author} {\bibinfo {author} {\bibfnamefont {S.~R.}\ \bibnamefont
  {White}}\ and\ \bibinfo {author} {\bibfnamefont {D.~A.}\ \bibnamefont
  {Huse}},\ }\bibfield  {title} {\bibinfo {title} {Numerical
  renormalization-group study of low-lying eigenstates of the antiferromagnetic
  {{\it S}=1 Heisenberg} chain},\ }\href
  {https://doi.org/10.1103/PhysRevB.48.3844} {\bibfield  {journal} {\bibinfo
  {journal} {Phys. Rev. B}\ }\textbf {\bibinfo {volume} {48}},\ \bibinfo
  {pages} {3844} (\bibinfo {year} {1993})}\BibitemShut {NoStop}%
\bibitem [{\citenamefont {Haldane}(1983)}]{Haldane1983}%
  \BibitemOpen
  \bibfield  {author} {\bibinfo {author} {\bibfnamefont {F.~D.~M.}\
  \bibnamefont {Haldane}},\ }\bibfield  {title} {\bibinfo {title} {Nonlinear
  field theory of large-spin {Heisenberg} antiferromagnets: Semiclassically
  quantized solitons of the one-dimensional easy-axis {N\'eel} state},\ }\href
  {https://doi.org/10.1103/PhysRevLett.50.1153} {\bibfield  {journal} {\bibinfo
   {journal} {Phys. Rev. Lett.}\ }\textbf {\bibinfo {volume} {50}},\ \bibinfo
  {pages} {1153} (\bibinfo {year} {1983})}\BibitemShut {NoStop}%
\bibitem [{\citenamefont {Affleck}\ \emph {et~al.}(1987)\citenamefont
  {Affleck}, \citenamefont {Kennedy}, \citenamefont {Lieb},\ and\ \citenamefont
  {Tasaki}}]{AKLT1987}%
  \BibitemOpen
  \bibfield  {author} {\bibinfo {author} {\bibfnamefont {I.}~\bibnamefont
  {Affleck}}, \bibinfo {author} {\bibfnamefont {T.}~\bibnamefont {Kennedy}},
  \bibinfo {author} {\bibfnamefont {E.~H.}\ \bibnamefont {Lieb}},\ and\
  \bibinfo {author} {\bibfnamefont {H.}~\bibnamefont {Tasaki}},\ }\bibfield
  {title} {\bibinfo {title} {Rigorous results on valence-bond ground states in
  antiferromagnets},\ }\href {https://doi.org/10.1103/PhysRevLett.59.799}
  {\bibfield  {journal} {\bibinfo  {journal} {Phys. Rev. Lett.}\ }\textbf
  {\bibinfo {volume} {59}},\ \bibinfo {pages} {799} (\bibinfo {year}
  {1987})}\BibitemShut {NoStop}%
\bibitem [{\citenamefont {Peschel}\ \emph {et~al.}(1999)\citenamefont
  {Peschel}, \citenamefont {Wang}, \citenamefont {Kaulke},\ and\ \citenamefont
  {Hallberg}}]{DMRGLec}%
  \BibitemOpen
  \bibfield  {author} {\bibinfo {author} {\bibfnamefont {I.}~\bibnamefont
  {Peschel}}, \bibinfo {author} {\bibfnamefont {X.}~\bibnamefont {Wang}},
  \bibinfo {author} {\bibfnamefont {M.}~\bibnamefont {Kaulke}},\ and\ \bibinfo
  {author} {\bibfnamefont {K.}~\bibnamefont {Hallberg}},\ }\href
  {https://doi.org/10.1007/BFb0106062} {\emph {\bibinfo {title} {Density-Matrix
  Renormalization - A New Numerical Method in Physics}}},\ \bibinfo {edition}
  {2nd}\ ed.\ (\bibinfo  {publisher} {Springer Berlin, Heidelberg},\ \bibinfo
  {year} {1999})\BibitemShut {NoStop}%
\bibitem [{\citenamefont {Gu}\ and\ \citenamefont {Wen}(2009)}]{Gu2009tensor}%
  \BibitemOpen
  \bibfield  {author} {\bibinfo {author} {\bibfnamefont {Z.-C.}\ \bibnamefont
  {Gu}}\ and\ \bibinfo {author} {\bibfnamefont {X.-G.}\ \bibnamefont {Wen}},\
  }\bibfield  {title} {\bibinfo {title} {Tensor-entanglement-filtering
  renormalization approach and symmetry-protected topological order},\ }\href
  {https://doi.org/10.1103/PhysRevB.80.155131} {\bibfield  {journal} {\bibinfo
  {journal} {Phys. Rev. B}\ }\textbf {\bibinfo {volume} {80}},\ \bibinfo
  {pages} {155131} (\bibinfo {year} {2009})}\BibitemShut {NoStop}%
\bibitem [{\citenamefont {Chubukov}\ \emph
  {et~al.}(1994{\natexlab{a}})\citenamefont {Chubukov}, \citenamefont
  {Sachdev},\ and\ \citenamefont {Ye}}]{Chubukov1994SL}%
  \BibitemOpen
  \bibfield  {author} {\bibinfo {author} {\bibfnamefont {A.~V.}\ \bibnamefont
  {Chubukov}}, \bibinfo {author} {\bibfnamefont {S.}~\bibnamefont {Sachdev}},\
  and\ \bibinfo {author} {\bibfnamefont {J.}~\bibnamefont {Ye}},\ }\bibfield
  {title} {\bibinfo {title} {Theory of two-dimensional quantum {Heisenberg}
  antiferromagnets with a nearly critical ground state},\ }\href
  {https://doi.org/10.1103/PhysRevB.49.11919} {\bibfield  {journal} {\bibinfo
  {journal} {Phys. Rev. B}\ }\textbf {\bibinfo {volume} {49}},\ \bibinfo
  {pages} {11919} (\bibinfo {year} {1994}{\natexlab{a}})}\BibitemShut {NoStop}%
\bibitem [{\citenamefont {Coldea}\ \emph {et~al.}(2010)\citenamefont {Coldea},
  \citenamefont {Tennant}, \citenamefont {Wheeler}, \citenamefont {Wawrzynska},
  \citenamefont {Prabhakaran}, \citenamefont {Telling}, \citenamefont
  {Habicht}, \citenamefont {Smeibidl},\ and\ \citenamefont
  {Kiefer}}]{Coldea2010}%
  \BibitemOpen
  \bibfield  {author} {\bibinfo {author} {\bibfnamefont {R.}~\bibnamefont
  {Coldea}}, \bibinfo {author} {\bibfnamefont {D.~A.}\ \bibnamefont {Tennant}},
  \bibinfo {author} {\bibfnamefont {E.~M.}\ \bibnamefont {Wheeler}}, \bibinfo
  {author} {\bibfnamefont {E.}~\bibnamefont {Wawrzynska}}, \bibinfo {author}
  {\bibfnamefont {D.}~\bibnamefont {Prabhakaran}}, \bibinfo {author}
  {\bibfnamefont {M.}~\bibnamefont {Telling}}, \bibinfo {author} {\bibfnamefont
  {K.}~\bibnamefont {Habicht}}, \bibinfo {author} {\bibfnamefont
  {P.}~\bibnamefont {Smeibidl}},\ and\ \bibinfo {author} {\bibfnamefont
  {K.}~\bibnamefont {Kiefer}},\ }\bibfield  {title} {\bibinfo {title} {Quantum
  criticality in an {Ising} chain: Experimental evidence for emergent {{\it
  E}$_8$} symmetry},\ }\href {https://doi.org/10.1126/science.1180085}
  {\bibfield  {journal} {\bibinfo  {journal} {Science}\ }\textbf {\bibinfo
  {volume} {327}},\ \bibinfo {pages} {177} (\bibinfo {year}
  {2010})}\BibitemShut {NoStop}%
\bibitem [{\citenamefont {Wang}\ \emph {et~al.}(2018)\citenamefont {Wang},
  \citenamefont {Lorenz}, \citenamefont {Gorbunov}, \citenamefont {Cong},
  \citenamefont {Kohama}, \citenamefont {Niesen}, \citenamefont {Breunig},
  \citenamefont {Engelmayer}, \citenamefont {Herman}, \citenamefont {Wu},
  \citenamefont {Kindo}, \citenamefont {Wosnitza}, \citenamefont {Zherlitsyn},\
  and\ \citenamefont {Loidl}}]{WangZ2018}%
  \BibitemOpen
  \bibfield  {author} {\bibinfo {author} {\bibfnamefont {Z.}~\bibnamefont
  {Wang}}, \bibinfo {author} {\bibfnamefont {T.}~\bibnamefont {Lorenz}},
  \bibinfo {author} {\bibfnamefont {D.~I.}\ \bibnamefont {Gorbunov}}, \bibinfo
  {author} {\bibfnamefont {P.~T.}\ \bibnamefont {Cong}}, \bibinfo {author}
  {\bibfnamefont {Y.}~\bibnamefont {Kohama}}, \bibinfo {author} {\bibfnamefont
  {S.}~\bibnamefont {Niesen}}, \bibinfo {author} {\bibfnamefont
  {O.}~\bibnamefont {Breunig}}, \bibinfo {author} {\bibfnamefont
  {J.}~\bibnamefont {Engelmayer}}, \bibinfo {author} {\bibfnamefont
  {A.}~\bibnamefont {Herman}}, \bibinfo {author} {\bibfnamefont
  {J.}~\bibnamefont {Wu}}, \bibinfo {author} {\bibfnamefont {K.}~\bibnamefont
  {Kindo}}, \bibinfo {author} {\bibfnamefont {J.}~\bibnamefont {Wosnitza}},
  \bibinfo {author} {\bibfnamefont {S.}~\bibnamefont {Zherlitsyn}},\ and\
  \bibinfo {author} {\bibfnamefont {A.}~\bibnamefont {Loidl}},\ }\bibfield
  {title} {\bibinfo {title} {Quantum criticality of an {Ising-like} spin-$1/2$
  antiferromagnetic chain in a transverse magnetic field},\ }\href
  {https://doi.org/10.1103/PhysRevLett.120.207205} {\bibfield  {journal}
  {\bibinfo  {journal} {Phys. Rev. Lett.}\ }\textbf {\bibinfo {volume} {120}},\
  \bibinfo {pages} {207205} (\bibinfo {year} {2018})}\BibitemShut {NoStop}%
\bibitem [{\citenamefont {Fava}\ \emph {et~al.}(2020)\citenamefont {Fava},
  \citenamefont {Coldea},\ and\ \citenamefont {Parameswaran}}]{Fava2020}%
  \BibitemOpen
  \bibfield  {author} {\bibinfo {author} {\bibfnamefont {M.}~\bibnamefont
  {Fava}}, \bibinfo {author} {\bibfnamefont {R.}~\bibnamefont {Coldea}},\ and\
  \bibinfo {author} {\bibfnamefont {S.~A.}\ \bibnamefont {Parameswaran}},\
  }\bibfield  {title} {\bibinfo {title} {Glide symmetry breaking and ising
  criticality in the quasi-1d magnet {CoNb$_2$O$_6$}},\ }\href
  {https://doi.org/10.1073/pnas.2007986117} {\bibfield  {journal} {\bibinfo
  {journal} {PNAS}\ }\textbf {\bibinfo {volume} {117}},\ \bibinfo {pages}
  {25219} (\bibinfo {year} {2020})}\BibitemShut {NoStop}%
\bibitem [{\citenamefont {Morris}\ \emph {et~al.}(2021)\citenamefont {Morris},
  \citenamefont {Desai}, \citenamefont {Viirok}, \citenamefont {Hüvonen},
  \citenamefont {Nagel}, \citenamefont {R{\~{o}}{\~{o}}m}, \citenamefont
  {Krizan}, \citenamefont {Cava}, \citenamefont {McQueen}, \citenamefont
  {Koohpayeh}, \citenamefont {Kaul},\ and\ \citenamefont
  {Armitage}}]{Morris2021}%
  \BibitemOpen
  \bibfield  {author} {\bibinfo {author} {\bibfnamefont {C.~M.}\ \bibnamefont
  {Morris}}, \bibinfo {author} {\bibfnamefont {N.}~\bibnamefont {Desai}},
  \bibinfo {author} {\bibfnamefont {J.}~\bibnamefont {Viirok}}, \bibinfo
  {author} {\bibfnamefont {D.}~\bibnamefont {Hüvonen}}, \bibinfo {author}
  {\bibfnamefont {U.}~\bibnamefont {Nagel}}, \bibinfo {author} {\bibfnamefont
  {T.}~\bibnamefont {R{\~{o}}{\~{o}}m}}, \bibinfo {author} {\bibfnamefont
  {J.~W.}\ \bibnamefont {Krizan}}, \bibinfo {author} {\bibfnamefont {R.~J.}\
  \bibnamefont {Cava}}, \bibinfo {author} {\bibfnamefont {T.~M.}\ \bibnamefont
  {McQueen}}, \bibinfo {author} {\bibfnamefont {S.~M.}\ \bibnamefont
  {Koohpayeh}}, \bibinfo {author} {\bibfnamefont {R.~K.}\ \bibnamefont
  {Kaul}},\ and\ \bibinfo {author} {\bibfnamefont {N.~P.}\ \bibnamefont
  {Armitage}},\ }\bibfield  {title} {\bibinfo {title} {Duality and domain wall
  dynamics in a twisted {Kitaev} chain},\ }\href
  {https://doi.org/10.1038/s41567-021-01208-0} {\bibfield  {journal} {\bibinfo
  {journal} {Nature Physics}\ }\textbf {\bibinfo {volume} {17}},\ \bibinfo
  {pages} {832} (\bibinfo {year} {2021})}\BibitemShut {NoStop}%
\bibitem [{\citenamefont {Kj\"all}\ \emph {et~al.}(2011)\citenamefont
  {Kj\"all}, \citenamefont {Pollmann},\ and\ \citenamefont
  {Moore}}]{Moore2011}%
  \BibitemOpen
  \bibfield  {author} {\bibinfo {author} {\bibfnamefont {J.~A.}\ \bibnamefont
  {Kj\"all}}, \bibinfo {author} {\bibfnamefont {F.}~\bibnamefont {Pollmann}},\
  and\ \bibinfo {author} {\bibfnamefont {J.~E.}\ \bibnamefont {Moore}},\
  }\bibfield  {title} {\bibinfo {title} {Bound states and ${E}_{8}$ symmetry
  effects in perturbed quantum {Ising} chains},\ }\href
  {https://doi.org/10.1103/PhysRevB.83.020407} {\bibfield  {journal} {\bibinfo
  {journal} {Phys. Rev. B}\ }\textbf {\bibinfo {volume} {83}},\ \bibinfo
  {pages} {020407} (\bibinfo {year} {2011})}\BibitemShut {NoStop}%
\bibitem [{\citenamefont {Liang}\ \emph {et~al.}(2015)\citenamefont {Liang},
  \citenamefont {Koohpayeh}, \citenamefont {Krizan}, \citenamefont {McQueen},
  \citenamefont {Cava},\ and\ \citenamefont {Ong}}]{Liang2015}%
  \BibitemOpen
  \bibfield  {author} {\bibinfo {author} {\bibfnamefont {T.}~\bibnamefont
  {Liang}}, \bibinfo {author} {\bibfnamefont {S.~M.}\ \bibnamefont
  {Koohpayeh}}, \bibinfo {author} {\bibfnamefont {J.~W.}\ \bibnamefont
  {Krizan}}, \bibinfo {author} {\bibfnamefont {T.~M.}\ \bibnamefont {McQueen}},
  \bibinfo {author} {\bibfnamefont {R.~J.}\ \bibnamefont {Cava}},\ and\
  \bibinfo {author} {\bibfnamefont {N.~P.}\ \bibnamefont {Ong}},\ }\bibfield
  {title} {\bibinfo {title} {Heat capacity peak at the quantum critical point
  of the transverse {Ising} magnet {CoNb$_2$O$_6$}},\ }\href
  {https://doi.org/10.1038/ncomms8611} {\bibfield  {journal} {\bibinfo
  {journal} {Nat. Commun.}\ }\textbf {\bibinfo {volume} {6}},\ \bibinfo {pages}
  {7611} (\bibinfo {year} {2015})}\BibitemShut {NoStop}%
\bibitem [{\citenamefont {Xu}\ \emph {et~al.}(2022)\citenamefont {Xu},
  \citenamefont {Wang}, \citenamefont {Huang}, \citenamefont {Ni},
  \citenamefont {Zhao}, \citenamefont {Dai}, \citenamefont {Pan}, \citenamefont
  {Hong}, \citenamefont {Chauhan}, \citenamefont {Koohpayeh}, \citenamefont
  {Armitage},\ and\ \citenamefont {Li}}]{Xu2022}%
  \BibitemOpen
  \bibfield  {author} {\bibinfo {author} {\bibfnamefont {Y.}~\bibnamefont
  {Xu}}, \bibinfo {author} {\bibfnamefont {L.~S.}\ \bibnamefont {Wang}},
  \bibinfo {author} {\bibfnamefont {Y.~Y.}\ \bibnamefont {Huang}}, \bibinfo
  {author} {\bibfnamefont {J.~M.}\ \bibnamefont {Ni}}, \bibinfo {author}
  {\bibfnamefont {C.~C.}\ \bibnamefont {Zhao}}, \bibinfo {author}
  {\bibfnamefont {Y.~F.}\ \bibnamefont {Dai}}, \bibinfo {author} {\bibfnamefont
  {B.~Y.}\ \bibnamefont {Pan}}, \bibinfo {author} {\bibfnamefont {X.~C.}\
  \bibnamefont {Hong}}, \bibinfo {author} {\bibfnamefont {P.}~\bibnamefont
  {Chauhan}}, \bibinfo {author} {\bibfnamefont {S.~M.}\ \bibnamefont
  {Koohpayeh}}, \bibinfo {author} {\bibfnamefont {N.~P.}\ \bibnamefont
  {Armitage}},\ and\ \bibinfo {author} {\bibfnamefont {S.~Y.}\ \bibnamefont
  {Li}},\ }\bibfield  {title} {\bibinfo {title} {Quantum critical magnetic
  excitations in spin-$1/2$ and spin-1 chain systems},\ }\href
  {https://doi.org/10.1103/PhysRevX.12.021020} {\bibfield  {journal} {\bibinfo
  {journal} {Phys. Rev. X}\ }\textbf {\bibinfo {volume} {12}},\ \bibinfo
  {pages} {021020} (\bibinfo {year} {2022})}\BibitemShut {NoStop}%
\bibitem [{\citenamefont {Woodland}\ \emph {et~al.}(2023)\citenamefont
  {Woodland}, \citenamefont {Macdougal}, \citenamefont {Cabrera}, \citenamefont
  {Thompson}, \citenamefont {Prabhakaran}, \citenamefont {Bewley},\ and\
  \citenamefont {Coldea}}]{woodland2023tuning}%
  \BibitemOpen
  \bibfield  {author} {\bibinfo {author} {\bibfnamefont {L.}~\bibnamefont
  {Woodland}}, \bibinfo {author} {\bibfnamefont {D.}~\bibnamefont {Macdougal}},
  \bibinfo {author} {\bibfnamefont {I.~M.}\ \bibnamefont {Cabrera}}, \bibinfo
  {author} {\bibfnamefont {J.~D.}\ \bibnamefont {Thompson}}, \bibinfo {author}
  {\bibfnamefont {D.}~\bibnamefont {Prabhakaran}}, \bibinfo {author}
  {\bibfnamefont {R.~I.}\ \bibnamefont {Bewley}},\ and\ \bibinfo {author}
  {\bibfnamefont {R.}~\bibnamefont {Coldea}},\ }\bibfield  {title} {\bibinfo
  {title} {Tuning the confinement potential between spinons in the ising chain
  compound {CoNb$_{2}$O$_{6}$} using longitudinal fields and quantitative
  determination of the microscopic hamiltonian},\ }\href
  {https://doi.org/10.1103/PhysRevB.108.184416} {\bibfield  {journal} {\bibinfo
   {journal} {Phys. Rev. B}\ }\textbf {\bibinfo {volume} {108}},\ \bibinfo
  {pages} {184416} (\bibinfo {year} {2023})}\BibitemShut {NoStop}%
\bibitem [{\citenamefont {Cui}\ \emph {et~al.}(2018)\citenamefont {Cui},
  \citenamefont {Dai}, \citenamefont {Zhou}, \citenamefont {Wang},
  \citenamefont {Li}, \citenamefont {Song}, \citenamefont {Wang}, \citenamefont
  {Ma}, \citenamefont {Zhang}, \citenamefont {Li}, \citenamefont {Luke},
  \citenamefont {Normand}, \citenamefont {Xiang},\ and\ \citenamefont
  {Yu}}]{BCNO2018}%
  \BibitemOpen
  \bibfield  {author} {\bibinfo {author} {\bibfnamefont {Y.}~\bibnamefont
  {Cui}}, \bibinfo {author} {\bibfnamefont {J.}~\bibnamefont {Dai}}, \bibinfo
  {author} {\bibfnamefont {P.}~\bibnamefont {Zhou}}, \bibinfo {author}
  {\bibfnamefont {P.~S.}\ \bibnamefont {Wang}}, \bibinfo {author}
  {\bibfnamefont {T.~R.}\ \bibnamefont {Li}}, \bibinfo {author} {\bibfnamefont
  {W.~H.}\ \bibnamefont {Song}}, \bibinfo {author} {\bibfnamefont {J.~C.}\
  \bibnamefont {Wang}}, \bibinfo {author} {\bibfnamefont {L.}~\bibnamefont
  {Ma}}, \bibinfo {author} {\bibfnamefont {Z.}~\bibnamefont {Zhang}}, \bibinfo
  {author} {\bibfnamefont {S.~Y.}\ \bibnamefont {Li}}, \bibinfo {author}
  {\bibfnamefont {G.~M.}\ \bibnamefont {Luke}}, \bibinfo {author}
  {\bibfnamefont {B.}~\bibnamefont {Normand}}, \bibinfo {author} {\bibfnamefont
  {T.}~\bibnamefont {Xiang}},\ and\ \bibinfo {author} {\bibfnamefont
  {W.}~\bibnamefont {Yu}},\ }\bibfield  {title} {\bibinfo {title}
  {{Mermin-Wagner} physics, {$(H,T)$} phase diagram, and candidate quantum
  spin-liquid phase in the spin-$\frac{1}{2}$ triangular-lattice
  antiferromagnet {${\mathrm{Ba}}_{8}{\mathrm{CoNb}}_{6}{\mathrm{O}}_{24}$}},\
  }\href {https://doi.org/10.1103/PhysRevMaterials.2.044403} {\bibfield
  {journal} {\bibinfo  {journal} {Phys. Rev. Mater.}\ }\textbf {\bibinfo
  {volume} {2}},\ \bibinfo {pages} {044403} (\bibinfo {year}
  {2018})}\BibitemShut {NoStop}%
\bibitem [{\citenamefont {Chubukov}\ \emph
  {et~al.}(1994{\natexlab{b}})\citenamefont {Chubukov}, \citenamefont
  {Senthil},\ and\ \citenamefont {Sachdev}}]{Chubukov1994TL}%
  \BibitemOpen
  \bibfield  {author} {\bibinfo {author} {\bibfnamefont {A.~V.}\ \bibnamefont
  {Chubukov}}, \bibinfo {author} {\bibfnamefont {T.}~\bibnamefont {Senthil}},\
  and\ \bibinfo {author} {\bibfnamefont {S.}~\bibnamefont {Sachdev}},\
  }\bibfield  {title} {\bibinfo {title} {Universal magnetic properties of
  frustrated quantum antiferromagnets in two dimensions},\ }\href
  {https://doi.org/10.1103/physrevlett.72.2089} {\bibfield  {journal} {\bibinfo
   {journal} {Phys. Rev. Lett.}\ }\textbf {\bibinfo {volume} {72}},\ \bibinfo
  {pages} {2089} (\bibinfo {year} {1994}{\natexlab{b}})}\BibitemShut {NoStop}%
\bibitem [{\citenamefont {Chubukov}\ and\ \citenamefont
  {Golosov}(1991)}]{Chubukov1991}%
  \BibitemOpen
  \bibfield  {author} {\bibinfo {author} {\bibfnamefont {A.~V.}\ \bibnamefont
  {Chubukov}}\ and\ \bibinfo {author} {\bibfnamefont {D.~I.}\ \bibnamefont
  {Golosov}},\ }\bibfield  {title} {\bibinfo {title} {Quantum theory of an
  antiferromagnet on a triangular lattice in a magnetic field},\ }\href
  {https://doi.org/10.1088/0953-8984/3/1/005} {\bibfield  {journal} {\bibinfo
  {journal} {J. Phys.: Condens. Matter}\ }\textbf {\bibinfo {volume} {3}},\
  \bibinfo {pages} {69} (\bibinfo {year} {1991})}\BibitemShut {NoStop}%
\bibitem [{\citenamefont {Starykh}(2015)}]{Starykh2015}%
  \BibitemOpen
  \bibfield  {author} {\bibinfo {author} {\bibfnamefont {O.~A.}\ \bibnamefont
  {Starykh}},\ }\bibfield  {title} {\bibinfo {title} {Unusual ordered phases of
  highly frustrated magnets: a review},\ }\href
  {http://stacks.iop.org/0034-4885/78/i=5/a=052502} {\bibfield  {journal}
  {\bibinfo  {journal} {Rep. Prog. Phys.}\ }\textbf {\bibinfo {volume} {78}},\
  \bibinfo {pages} {052502} (\bibinfo {year} {2015})}\BibitemShut {NoStop}%
\bibitem [{\citenamefont {Bernu}\ \emph {et~al.}(1992)\citenamefont {Bernu},
  \citenamefont {Lhuillier},\ and\ \citenamefont {Pierre}}]{Bernu1992}%
  \BibitemOpen
  \bibfield  {author} {\bibinfo {author} {\bibfnamefont {B.}~\bibnamefont
  {Bernu}}, \bibinfo {author} {\bibfnamefont {C.}~\bibnamefont {Lhuillier}},\
  and\ \bibinfo {author} {\bibfnamefont {L.}~\bibnamefont {Pierre}},\
  }\bibfield  {title} {\bibinfo {title} {Signature of {N\'eel} order in exact
  spectra of quantum antiferromagnets on finite lattices},\ }\href
  {https://doi.org/10.1103/PhysRevLett.69.2590} {\bibfield  {journal} {\bibinfo
   {journal} {Phys. Rev. Lett.}\ }\textbf {\bibinfo {volume} {69}},\ \bibinfo
  {pages} {2590} (\bibinfo {year} {1992})}\BibitemShut {NoStop}%
\bibitem [{\citenamefont {Capriotti}\ \emph {et~al.}(1999)\citenamefont
  {Capriotti}, \citenamefont {Trumper},\ and\ \citenamefont
  {Sorella}}]{Capriotti1999}%
  \BibitemOpen
  \bibfield  {author} {\bibinfo {author} {\bibfnamefont {L.}~\bibnamefont
  {Capriotti}}, \bibinfo {author} {\bibfnamefont {A.~E.}\ \bibnamefont
  {Trumper}},\ and\ \bibinfo {author} {\bibfnamefont {S.}~\bibnamefont
  {Sorella}},\ }\bibfield  {title} {\bibinfo {title} {Long-range {N\'eel} order
  in the triangular {Heisenberg} model},\ }\href
  {https://doi.org/10.1103/PhysRevLett.82.3899} {\bibfield  {journal} {\bibinfo
   {journal} {Phys. Rev. Lett.}\ }\textbf {\bibinfo {volume} {82}},\ \bibinfo
  {pages} {3899} (\bibinfo {year} {1999})}\BibitemShut {NoStop}%
\bibitem [{\citenamefont {White}\ and\ \citenamefont
  {Chernyshev}(2007)}]{White2007}%
  \BibitemOpen
  \bibfield  {author} {\bibinfo {author} {\bibfnamefont {S.~R.}\ \bibnamefont
  {White}}\ and\ \bibinfo {author} {\bibfnamefont {A.~L.}\ \bibnamefont
  {Chernyshev}},\ }\bibfield  {title} {\bibinfo {title} {N\'eel order in square
  and triangular lattice {Heisenberg} models},\ }\href
  {https://doi.org/10.1103/PhysRevLett.99.127004} {\bibfield  {journal}
  {\bibinfo  {journal} {Phys. Rev. Lett.}\ }\textbf {\bibinfo {volume} {99}},\
  \bibinfo {pages} {127004} (\bibinfo {year} {2007})}\BibitemShut {NoStop}%
\bibitem [{\citenamefont {Elstner}\ \emph {et~al.}(1993)\citenamefont
  {Elstner}, \citenamefont {Singh},\ and\ \citenamefont {Young}}]{Elstner1993}%
  \BibitemOpen
  \bibfield  {author} {\bibinfo {author} {\bibfnamefont {N.}~\bibnamefont
  {Elstner}}, \bibinfo {author} {\bibfnamefont {R.~R.~P.}\ \bibnamefont
  {Singh}},\ and\ \bibinfo {author} {\bibfnamefont {A.~P.}\ \bibnamefont
  {Young}},\ }\bibfield  {title} {\bibinfo {title} {Finite temperature
  properties of the spin-1/2 {{Heisenberg}} antiferromagnet on the triangular
  lattice},\ }\href {https://doi.org/10.1103/PhysRevLett.71.1629} {\bibfield
  {journal} {\bibinfo  {journal} {Phys. Rev. Lett.}\ }\textbf {\bibinfo
  {volume} {71}},\ \bibinfo {pages} {1629} (\bibinfo {year}
  {1993})}\BibitemShut {NoStop}%
\bibitem [{\citenamefont {Kulagin}\ \emph {et~al.}(2013)\citenamefont
  {Kulagin}, \citenamefont {Prokof'ev}, \citenamefont {Starykh}, \citenamefont
  {Svistunov},\ and\ \citenamefont {Varney}}]{Kulagin2013}%
  \BibitemOpen
  \bibfield  {author} {\bibinfo {author} {\bibfnamefont {S.~A.}\ \bibnamefont
  {Kulagin}}, \bibinfo {author} {\bibfnamefont {N.}~\bibnamefont {Prokof'ev}},
  \bibinfo {author} {\bibfnamefont {O.~A.}\ \bibnamefont {Starykh}}, \bibinfo
  {author} {\bibfnamefont {B.}~\bibnamefont {Svistunov}},\ and\ \bibinfo
  {author} {\bibfnamefont {C.~N.}\ \bibnamefont {Varney}},\ }\bibfield  {title}
  {\bibinfo {title} {Bold diagrammatic monte carlo method applied to
  fermionized frustrated spins},\ }\href
  {https://doi.org/10.1103/PhysRevLett.110.070601} {\bibfield  {journal}
  {\bibinfo  {journal} {Phys. Rev. Lett.}\ }\textbf {\bibinfo {volume} {110}},\
  \bibinfo {pages} {070601} (\bibinfo {year} {2013})}\BibitemShut {NoStop}%
\bibitem [{\citenamefont {Doi}\ \emph {et~al.}(2004)\citenamefont {Doi},
  \citenamefont {Hinatsu},\ and\ \citenamefont {Ohoyama}}]{Doi2004}%
  \BibitemOpen
  \bibfield  {author} {\bibinfo {author} {\bibfnamefont {Y.}~\bibnamefont
  {Doi}}, \bibinfo {author} {\bibfnamefont {Y.}~\bibnamefont {Hinatsu}},\ and\
  \bibinfo {author} {\bibfnamefont {K.}~\bibnamefont {Ohoyama}},\ }\bibfield
  {title} {\bibinfo {title} {{Structural and magnetic properties of
  pseudo-two-dimensional triangular antiferromagnets Ba$_3$MSb$_2$O$_9$ (M =
  Mn, Co, and Ni)}},\ }\href {http://stacks.iop.org/0953-8984/16/i=49/a=009}
  {\bibfield  {journal} {\bibinfo  {journal} {J. Phys.: Condens. Matter}\
  }\textbf {\bibinfo {volume} {16}},\ \bibinfo {pages} {8923} (\bibinfo {year}
  {2004})}\BibitemShut {NoStop}%
\bibitem [{\citenamefont {Zhou}\ \emph {et~al.}(2012)\citenamefont {Zhou},
  \citenamefont {Xu}, \citenamefont {Hallas}, \citenamefont {Silverstein},
  \citenamefont {Wiebe}, \citenamefont {Umegaki}, \citenamefont {Yan},
  \citenamefont {Murphy}, \citenamefont {Park}, \citenamefont {Qiu},
  \citenamefont {Copley}, \citenamefont {Gardner},\ and\ \citenamefont
  {Takano}}]{Zhou2012}%
  \BibitemOpen
  \bibfield  {author} {\bibinfo {author} {\bibfnamefont {H.~D.}\ \bibnamefont
  {Zhou}}, \bibinfo {author} {\bibfnamefont {C.}~\bibnamefont {Xu}}, \bibinfo
  {author} {\bibfnamefont {A.~M.}\ \bibnamefont {Hallas}}, \bibinfo {author}
  {\bibfnamefont {H.~J.}\ \bibnamefont {Silverstein}}, \bibinfo {author}
  {\bibfnamefont {C.~R.}\ \bibnamefont {Wiebe}}, \bibinfo {author}
  {\bibfnamefont {I.}~\bibnamefont {Umegaki}}, \bibinfo {author} {\bibfnamefont
  {J.~Q.}\ \bibnamefont {Yan}}, \bibinfo {author} {\bibfnamefont {T.~P.}\
  \bibnamefont {Murphy}}, \bibinfo {author} {\bibfnamefont {J.-H.}\
  \bibnamefont {Park}}, \bibinfo {author} {\bibfnamefont {Y.}~\bibnamefont
  {Qiu}}, \bibinfo {author} {\bibfnamefont {J.~R.~D.}\ \bibnamefont {Copley}},
  \bibinfo {author} {\bibfnamefont {J.~S.}\ \bibnamefont {Gardner}},\ and\
  \bibinfo {author} {\bibfnamefont {Y.}~\bibnamefont {Takano}},\ }\bibfield
  {title} {\bibinfo {title} {Successive phase transitions and extended
  spin-excitation continuum in the ${S}\mathbf{=}\frac{1}{2}$
  triangular-lattice antiferromagnet {Ba$_3$CoSb$_2$O$_9$}},\ }\href
  {https://doi.org/10.1103/PhysRevLett.109.267206} {\bibfield  {journal}
  {\bibinfo  {journal} {Phys. Rev. Lett.}\ }\textbf {\bibinfo {volume} {109}},\
  \bibinfo {pages} {267206} (\bibinfo {year} {2012})}\BibitemShut {NoStop}%
\bibitem [{\citenamefont {Susuki}\ \emph {et~al.}(2013)\citenamefont {Susuki},
  \citenamefont {Kurita}, \citenamefont {Tanaka}, \citenamefont {Nojiri},
  \citenamefont {Matsuo}, \citenamefont {Kindo},\ and\ \citenamefont
  {Tanaka}}]{Susuki2013}%
  \BibitemOpen
  \bibfield  {author} {\bibinfo {author} {\bibfnamefont {T.}~\bibnamefont
  {Susuki}}, \bibinfo {author} {\bibfnamefont {N.}~\bibnamefont {Kurita}},
  \bibinfo {author} {\bibfnamefont {T.}~\bibnamefont {Tanaka}}, \bibinfo
  {author} {\bibfnamefont {H.}~\bibnamefont {Nojiri}}, \bibinfo {author}
  {\bibfnamefont {A.}~\bibnamefont {Matsuo}}, \bibinfo {author} {\bibfnamefont
  {K.}~\bibnamefont {Kindo}},\ and\ \bibinfo {author} {\bibfnamefont
  {H.}~\bibnamefont {Tanaka}},\ }\bibfield  {title} {\bibinfo {title}
  {Magnetization process and collective excitations in the {$S\mathbf{=}1/2$}
  triangular-lattice {Heisenberg} antiferromagnet {Ba$_3$CoSb$_2$O$_9$}},\
  }\href {https://doi.org/10.1103/PhysRevLett.110.267201} {\bibfield  {journal}
  {\bibinfo  {journal} {Phys. Rev. Lett.}\ }\textbf {\bibinfo {volume} {110}},\
  \bibinfo {pages} {267201} (\bibinfo {year} {2013})}\BibitemShut {NoStop}%
\bibitem [{\citenamefont {Ma}\ \emph {et~al.}(2016)\citenamefont {Ma},
  \citenamefont {Kamiya}, \citenamefont {Hong}, \citenamefont {Cao},
  \citenamefont {Ehlers}, \citenamefont {Tian}, \citenamefont {Batista},
  \citenamefont {Dun}, \citenamefont {Zhou},\ and\ \citenamefont
  {Matsuda}}]{Ma2016}%
  \BibitemOpen
  \bibfield  {author} {\bibinfo {author} {\bibfnamefont {J.}~\bibnamefont
  {Ma}}, \bibinfo {author} {\bibfnamefont {Y.}~\bibnamefont {Kamiya}}, \bibinfo
  {author} {\bibfnamefont {T.}~\bibnamefont {Hong}}, \bibinfo {author}
  {\bibfnamefont {H.~B.}\ \bibnamefont {Cao}}, \bibinfo {author} {\bibfnamefont
  {G.}~\bibnamefont {Ehlers}}, \bibinfo {author} {\bibfnamefont
  {W.}~\bibnamefont {Tian}}, \bibinfo {author} {\bibfnamefont {C.~D.}\
  \bibnamefont {Batista}}, \bibinfo {author} {\bibfnamefont {Z.~L.}\
  \bibnamefont {Dun}}, \bibinfo {author} {\bibfnamefont {H.~D.}\ \bibnamefont
  {Zhou}},\ and\ \bibinfo {author} {\bibfnamefont {M.}~\bibnamefont
  {Matsuda}},\ }\bibfield  {title} {\bibinfo {title} {Static and dynamical
  properties of the spin-$1/2$ equilateral triangular-lattice antiferromagnet
  {Ba$_3$CoSb$_2$O$_9$}},\ }\href
  {https://doi.org/10.1103/PhysRevLett.116.087201} {\bibfield  {journal}
  {\bibinfo  {journal} {Phys. Rev. Lett.}\ }\textbf {\bibinfo {volume} {116}},\
  \bibinfo {pages} {087201} (\bibinfo {year} {2016})}\BibitemShut {NoStop}%
\bibitem [{\citenamefont {Ito}\ \emph {et~al.}(2017)\citenamefont {Ito},
  \citenamefont {Kurita}, \citenamefont {Tanaka}, \citenamefont
  {Ohira-Kawamura}, \citenamefont {Nakajima}, \citenamefont {Itoh},
  \citenamefont {Kuwahara},\ and\ \citenamefont {Kakurai}}]{Ito2017}%
  \BibitemOpen
  \bibfield  {author} {\bibinfo {author} {\bibfnamefont {S.}~\bibnamefont
  {Ito}}, \bibinfo {author} {\bibfnamefont {N.}~\bibnamefont {Kurita}},
  \bibinfo {author} {\bibfnamefont {H.}~\bibnamefont {Tanaka}}, \bibinfo
  {author} {\bibfnamefont {S.}~\bibnamefont {Ohira-Kawamura}}, \bibinfo
  {author} {\bibfnamefont {K.}~\bibnamefont {Nakajima}}, \bibinfo {author}
  {\bibfnamefont {S.}~\bibnamefont {Itoh}}, \bibinfo {author} {\bibfnamefont
  {K.}~\bibnamefont {Kuwahara}},\ and\ \bibinfo {author} {\bibfnamefont
  {K.}~\bibnamefont {Kakurai}},\ }\bibfield  {title} {\bibinfo {title}
  {Structure of the magnetic excitations in the spin-1/2 triangular-lattice
  {{Heisenberg}} antiferromagnet {Ba$_3$CoSb$_2$O$_9$}},\ }\href
  {https://www.nature.com/articles/s41467-017-00316-x} {\bibfield  {journal}
  {\bibinfo  {journal} {Nature Commun.}\ }\textbf {\bibinfo {volume} {8}},\
  \bibinfo {pages} {235} (\bibinfo {year} {2017})}\BibitemShut {NoStop}%
\bibitem [{\citenamefont {Rawl}\ \emph {et~al.}(2017)\citenamefont {Rawl},
  \citenamefont {Ge}, \citenamefont {Agrawal}, \citenamefont {Kamiya},
  \citenamefont {Dela~Cruz}, \citenamefont {Butch}, \citenamefont {Sun},
  \citenamefont {Lee}, \citenamefont {Choi}, \citenamefont {Oitmaa},
  \citenamefont {Batista}, \citenamefont {Mourigal}, \citenamefont {Zhou},\
  and\ \citenamefont {Ma}}]{BCNO2017}%
  \BibitemOpen
  \bibfield  {author} {\bibinfo {author} {\bibfnamefont {R.}~\bibnamefont
  {Rawl}}, \bibinfo {author} {\bibfnamefont {L.}~\bibnamefont {Ge}}, \bibinfo
  {author} {\bibfnamefont {H.}~\bibnamefont {Agrawal}}, \bibinfo {author}
  {\bibfnamefont {Y.}~\bibnamefont {Kamiya}}, \bibinfo {author} {\bibfnamefont
  {C.~R.}\ \bibnamefont {Dela~Cruz}}, \bibinfo {author} {\bibfnamefont {N.~P.}\
  \bibnamefont {Butch}}, \bibinfo {author} {\bibfnamefont {X.~F.}\ \bibnamefont
  {Sun}}, \bibinfo {author} {\bibfnamefont {M.}~\bibnamefont {Lee}}, \bibinfo
  {author} {\bibfnamefont {E.~S.}\ \bibnamefont {Choi}}, \bibinfo {author}
  {\bibfnamefont {J.}~\bibnamefont {Oitmaa}}, \bibinfo {author} {\bibfnamefont
  {C.~D.}\ \bibnamefont {Batista}}, \bibinfo {author} {\bibfnamefont
  {M.}~\bibnamefont {Mourigal}}, \bibinfo {author} {\bibfnamefont {H.~D.}\
  \bibnamefont {Zhou}},\ and\ \bibinfo {author} {\bibfnamefont
  {J.}~\bibnamefont {Ma}},\ }\bibfield  {title} {\bibinfo {title}
  {{Ba$_{8}$CoNb$_{6}$O$_{24}$}: A spin-$\frac{1}{2}$ triangular-lattice
  {Heisenberg} antiferromagnet in the two-dimensional limit},\ }\href
  {https://doi.org/10.1103/PhysRevB.95.060412} {\bibfield  {journal} {\bibinfo
  {journal} {Phys. Rev. B}\ }\textbf {\bibinfo {volume} {95}},\ \bibinfo
  {pages} {060412} (\bibinfo {year} {2017})}\BibitemShut {NoStop}%
\bibitem [{\citenamefont {Bordelon}\ \emph {et~al.}(2019)\citenamefont
  {Bordelon}, \citenamefont {Kenney}, \citenamefont {Liu}, \citenamefont
  {Hogan}, \citenamefont {Posthuma}, \citenamefont {Kavand}, \citenamefont
  {Lyu}, \citenamefont {Sherwin}, \citenamefont {Butch}, \citenamefont {Brown},
  \citenamefont {Graf}, \citenamefont {Balents},\ and\ \citenamefont
  {Wilson}}]{Wilson2019NYO}%
  \BibitemOpen
  \bibfield  {author} {\bibinfo {author} {\bibfnamefont {M.~M.}\ \bibnamefont
  {Bordelon}}, \bibinfo {author} {\bibfnamefont {E.}~\bibnamefont {Kenney}},
  \bibinfo {author} {\bibfnamefont {C.}~\bibnamefont {Liu}}, \bibinfo {author}
  {\bibfnamefont {T.}~\bibnamefont {Hogan}}, \bibinfo {author} {\bibfnamefont
  {L.}~\bibnamefont {Posthuma}}, \bibinfo {author} {\bibfnamefont
  {M.}~\bibnamefont {Kavand}}, \bibinfo {author} {\bibfnamefont
  {Y.}~\bibnamefont {Lyu}}, \bibinfo {author} {\bibfnamefont {M.}~\bibnamefont
  {Sherwin}}, \bibinfo {author} {\bibfnamefont {N.~P.}\ \bibnamefont {Butch}},
  \bibinfo {author} {\bibfnamefont {C.}~\bibnamefont {Brown}}, \bibinfo
  {author} {\bibfnamefont {M.~J.}\ \bibnamefont {Graf}}, \bibinfo {author}
  {\bibfnamefont {L.}~\bibnamefont {Balents}},\ and\ \bibinfo {author}
  {\bibfnamefont {S.~D.}\ \bibnamefont {Wilson}},\ }\bibfield  {title}
  {\bibinfo {title} {Field-tunable quantum disordered ground state in the
  triangular-lattice antiferromagnet {NaYbO$_2$}},\ }\href
  {https://doi.org/10.1038/s41567-019-0594-5} {\bibfield  {journal} {\bibinfo
  {journal} {Nature Physics}\ }\textbf {\bibinfo {volume} {15}},\ \bibinfo
  {pages} {1058} (\bibinfo {year} {2019})}\BibitemShut {NoStop}%
\bibitem [{\citenamefont {Ranjith}\ \emph
  {et~al.}(2019{\natexlab{b}})\citenamefont {Ranjith}, \citenamefont
  {Dmytriieva}, \citenamefont {Khim}, \citenamefont {Sichelschmidt},
  \citenamefont {Luther}, \citenamefont {Ehlers}, \citenamefont {Yasuoka},
  \citenamefont {Wosnitza}, \citenamefont {Tsirlin}, \citenamefont {K\"uhne},\
  and\ \citenamefont {Baenitz}}]{Ranjith2019NYO}%
  \BibitemOpen
  \bibfield  {author} {\bibinfo {author} {\bibfnamefont {K.~M.}\ \bibnamefont
  {Ranjith}}, \bibinfo {author} {\bibfnamefont {D.}~\bibnamefont {Dmytriieva}},
  \bibinfo {author} {\bibfnamefont {S.}~\bibnamefont {Khim}}, \bibinfo {author}
  {\bibfnamefont {J.}~\bibnamefont {Sichelschmidt}}, \bibinfo {author}
  {\bibfnamefont {S.}~\bibnamefont {Luther}}, \bibinfo {author} {\bibfnamefont
  {D.}~\bibnamefont {Ehlers}}, \bibinfo {author} {\bibfnamefont
  {H.}~\bibnamefont {Yasuoka}}, \bibinfo {author} {\bibfnamefont
  {J.}~\bibnamefont {Wosnitza}}, \bibinfo {author} {\bibfnamefont {A.~A.}\
  \bibnamefont {Tsirlin}}, \bibinfo {author} {\bibfnamefont {H.}~\bibnamefont
  {K\"uhne}},\ and\ \bibinfo {author} {\bibfnamefont {M.}~\bibnamefont
  {Baenitz}},\ }\bibfield  {title} {\bibinfo {title} {Field-induced instability
  of the quantum spin liquid ground state in the
  ${J}_{\mathrm{eff}}=\frac{1}{2}$ triangular-lattice compound
  {${\mathrm{NaYbO}}_{2}$}},\ }\href
  {https://doi.org/10.1103/PhysRevB.99.180401} {\bibfield  {journal} {\bibinfo
  {journal} {Phys. Rev. B}\ }\textbf {\bibinfo {volume} {99}},\ \bibinfo
  {pages} {180401} (\bibinfo {year} {2019}{\natexlab{b}})}\BibitemShut
  {NoStop}%
\bibitem [{\citenamefont {Zhu}\ and\ \citenamefont {White}(2015)}]{Zhu2015}%
  \BibitemOpen
  \bibfield  {author} {\bibinfo {author} {\bibfnamefont {Z.}~\bibnamefont
  {Zhu}}\ and\ \bibinfo {author} {\bibfnamefont {S.~R.}\ \bibnamefont
  {White}},\ }\bibfield  {title} {\bibinfo {title} {Spin liquid phase of the
  $s=\frac{1}{2}\phantom{\rule{4.pt}{0ex}}{J}_{1}\ensuremath{-}{J}_{2}$
  heisenberg model on the triangular lattice},\ }\href
  {https://doi.org/10.1103/PhysRevB.92.041105} {\bibfield  {journal} {\bibinfo
  {journal} {Phys. Rev. B}\ }\textbf {\bibinfo {volume} {92}},\ \bibinfo
  {pages} {041105} (\bibinfo {year} {2015})}\BibitemShut {NoStop}%
\bibitem [{\citenamefont {Hu}\ \emph {et~al.}(2015)\citenamefont {Hu},
  \citenamefont {Gong}, \citenamefont {Zhu},\ and\ \citenamefont
  {Sheng}}]{Hu2015}%
  \BibitemOpen
  \bibfield  {author} {\bibinfo {author} {\bibfnamefont {W.-J.}\ \bibnamefont
  {Hu}}, \bibinfo {author} {\bibfnamefont {S.-S.}\ \bibnamefont {Gong}},
  \bibinfo {author} {\bibfnamefont {W.}~\bibnamefont {Zhu}},\ and\ \bibinfo
  {author} {\bibfnamefont {D.~N.}\ \bibnamefont {Sheng}},\ }\bibfield  {title}
  {\bibinfo {title} {Competing spin-liquid states in the spin-$\frac{1}{2}$
  heisenberg model on the triangular lattice},\ }\href
  {https://doi.org/10.1103/PhysRevB.92.140403} {\bibfield  {journal} {\bibinfo
  {journal} {Phys. Rev. B}\ }\textbf {\bibinfo {volume} {92}},\ \bibinfo
  {pages} {140403} (\bibinfo {year} {2015})}\BibitemShut {NoStop}%
\bibitem [{\citenamefont {Iqbal}\ \emph {et~al.}(2016)\citenamefont {Iqbal},
  \citenamefont {Hu}, \citenamefont {Thomale}, \citenamefont {Poilblanc},\ and\
  \citenamefont {Becca}}]{Iqbal2016}%
  \BibitemOpen
  \bibfield  {author} {\bibinfo {author} {\bibfnamefont {Y.}~\bibnamefont
  {Iqbal}}, \bibinfo {author} {\bibfnamefont {W.-J.}\ \bibnamefont {Hu}},
  \bibinfo {author} {\bibfnamefont {R.}~\bibnamefont {Thomale}}, \bibinfo
  {author} {\bibfnamefont {D.}~\bibnamefont {Poilblanc}},\ and\ \bibinfo
  {author} {\bibfnamefont {F.}~\bibnamefont {Becca}},\ }\bibfield  {title}
  {\bibinfo {title} {Spin liquid nature in the heisenberg
  ${J}_{1}\ensuremath{-}{J}_{2}$ triangular antiferromagnet},\ }\href
  {https://doi.org/10.1103/PhysRevB.93.144411} {\bibfield  {journal} {\bibinfo
  {journal} {Phys. Rev. B}\ }\textbf {\bibinfo {volume} {93}},\ \bibinfo
  {pages} {144411} (\bibinfo {year} {2016})}\BibitemShut {NoStop}%
\bibitem [{\citenamefont {Hu}\ \emph {et~al.}(2019)\citenamefont {Hu},
  \citenamefont {Zhu}, \citenamefont {Eggert},\ and\ \citenamefont
  {He}}]{Hu2019}%
  \BibitemOpen
  \bibfield  {author} {\bibinfo {author} {\bibfnamefont {S.}~\bibnamefont
  {Hu}}, \bibinfo {author} {\bibfnamefont {W.}~\bibnamefont {Zhu}}, \bibinfo
  {author} {\bibfnamefont {S.}~\bibnamefont {Eggert}},\ and\ \bibinfo {author}
  {\bibfnamefont {Y.-C.}\ \bibnamefont {He}},\ }\bibfield  {title} {\bibinfo
  {title} {Dirac spin liquid on the spin-$1/2$ triangular heisenberg
  antiferromagnet},\ }\href {https://doi.org/10.1103/PhysRevLett.123.207203}
  {\bibfield  {journal} {\bibinfo  {journal} {Phys. Rev. Lett.}\ }\textbf
  {\bibinfo {volume} {123}},\ \bibinfo {pages} {207203} (\bibinfo {year}
  {2019})}\BibitemShut {NoStop}%
\bibitem [{\citenamefont {Itou}\ \emph {et~al.}(2010)\citenamefont {Itou},
  \citenamefont {Oyamada}, \citenamefont {Maegawa},\ and\ \citenamefont
  {Kato}}]{Itou2010}%
  \BibitemOpen
  \bibfield  {author} {\bibinfo {author} {\bibfnamefont {T.}~\bibnamefont
  {Itou}}, \bibinfo {author} {\bibfnamefont {A.}~\bibnamefont {Oyamada}},
  \bibinfo {author} {\bibfnamefont {S.}~\bibnamefont {Maegawa}},\ and\ \bibinfo
  {author} {\bibfnamefont {R.}~\bibnamefont {Kato}},\ }\bibfield  {title}
  {\bibinfo {title} {Instability of a quantum spin liquid in an organic
  triangular-lattice antiferromagnet},\ }\href
  {https://doi.org/10.1038/nphys1715} {\bibfield  {journal} {\bibinfo
  {journal} {Nature Physics}\ }\textbf {\bibinfo {volume} {6}},\ \bibinfo
  {pages} {673} (\bibinfo {year} {2010})}\BibitemShut {NoStop}%
\bibitem [{\citenamefont {Sherman}\ \emph {et~al.}(2023)\citenamefont
  {Sherman}, \citenamefont {Dupont},\ and\ \citenamefont
  {Moore}}]{Sherman2023}%
  \BibitemOpen
  \bibfield  {author} {\bibinfo {author} {\bibfnamefont {N.~E.}\ \bibnamefont
  {Sherman}}, \bibinfo {author} {\bibfnamefont {M.}~\bibnamefont {Dupont}},\
  and\ \bibinfo {author} {\bibfnamefont {J.~E.}\ \bibnamefont {Moore}},\
  }\bibfield  {title} {\bibinfo {title} {Spectral function of the
  ${J}_{1}\ensuremath{-}{J}_{2}$ heisenberg model on the triangular lattice},\
  }\href {https://doi.org/10.1103/PhysRevB.107.165146} {\bibfield  {journal}
  {\bibinfo  {journal} {Phys. Rev. B}\ }\textbf {\bibinfo {volume} {107}},\
  \bibinfo {pages} {165146} (\bibinfo {year} {2023})}\BibitemShut {NoStop}%
\bibitem [{\citenamefont {Czarnik}\ \emph {et~al.}(2019)\citenamefont
  {Czarnik}, \citenamefont {Dziarmaga},\ and\ \citenamefont
  {Corboz}}]{Corboz2019}%
  \BibitemOpen
  \bibfield  {author} {\bibinfo {author} {\bibfnamefont {P.}~\bibnamefont
  {Czarnik}}, \bibinfo {author} {\bibfnamefont {J.}~\bibnamefont {Dziarmaga}},\
  and\ \bibinfo {author} {\bibfnamefont {P.}~\bibnamefont {Corboz}},\
  }\bibfield  {title} {\bibinfo {title} {Time evolution of an infinite
  projected entangled pair state: An efficient algorithm},\ }\href
  {https://doi.org/10.1103/PhysRevB.99.035115} {\bibfield  {journal} {\bibinfo
  {journal} {Phys. Rev. B}\ }\textbf {\bibinfo {volume} {99}},\ \bibinfo
  {pages} {035115} (\bibinfo {year} {2019})}\BibitemShut {NoStop}%
\bibitem [{\citenamefont {Haegeman}\ \emph {et~al.}(2016)\citenamefont
  {Haegeman}, \citenamefont {Lubich}, \citenamefont {Oseledets}, \citenamefont
  {Vandereycken},\ and\ \citenamefont {Verstraete}}]{TDVP2016}%
  \BibitemOpen
  \bibfield  {author} {\bibinfo {author} {\bibfnamefont {J.}~\bibnamefont
  {Haegeman}}, \bibinfo {author} {\bibfnamefont {C.}~\bibnamefont {Lubich}},
  \bibinfo {author} {\bibfnamefont {I.}~\bibnamefont {Oseledets}}, \bibinfo
  {author} {\bibfnamefont {B.}~\bibnamefont {Vandereycken}},\ and\ \bibinfo
  {author} {\bibfnamefont {F.}~\bibnamefont {Verstraete}},\ }\bibfield  {title}
  {\bibinfo {title} {Unifying time evolution and optimization with matrix
  product states},\ }\href {https://doi.org/10.1103/PhysRevB.94.165116}
  {\bibfield  {journal} {\bibinfo  {journal} {Phys. Rev. B}\ }\textbf {\bibinfo
  {volume} {94}},\ \bibinfo {pages} {165116} (\bibinfo {year}
  {2016})}\BibitemShut {NoStop}%
\bibitem [{\citenamefont {Haegeman}\ \emph {et~al.}(2014)\citenamefont
  {Haegeman}, \citenamefont {Mari$\ddot{\rm e}$n}, \citenamefont {Osborne},\
  and\ \citenamefont {Verstraete}}]{MPSManifold2014}%
  \BibitemOpen
  \bibfield  {author} {\bibinfo {author} {\bibfnamefont {J.}~\bibnamefont
  {Haegeman}}, \bibinfo {author} {\bibfnamefont {M.}~\bibnamefont
  {Mari$\ddot{\rm e}$n}}, \bibinfo {author} {\bibfnamefont {T.~J.}\
  \bibnamefont {Osborne}},\ and\ \bibinfo {author} {\bibfnamefont
  {F.}~\bibnamefont {Verstraete}},\ }\bibfield  {title} {\bibinfo {title}
  {Geometry of matrix product states: Metric, parallel transport, and
  curvature},\ }\href {https://doi.org/10.1063/1.4862851} {\bibfield  {journal}
  {\bibinfo  {journal} {J. Math. Phys.}\ }\textbf {\bibinfo {volume} {55}},\
  \bibinfo {pages} {021902} (\bibinfo {year} {2014})}\BibitemShut {NoStop}%
\bibitem [{\citenamefont {Choi}(1972)}]{Choi1972}%
  \BibitemOpen
  \bibfield  {author} {\bibinfo {author} {\bibfnamefont {M.-D.}\ \bibnamefont
  {Choi}},\ }\bibfield  {title} {\bibinfo {title} {Positive linear maps on
  c*-algebras},\ }\href {https://doi.org/10.4153/CJM-1972-044-5} {\bibfield
  {journal} {\bibinfo  {journal} {Canadian Journal of Mathematics}\ }\textbf
  {\bibinfo {volume} {24}},\ \bibinfo {pages} {520} (\bibinfo {year}
  {1972})}\BibitemShut {NoStop}%
\end{thebibliography}%

%
%
%
%
%
%

\end{document}